\documentclass[aps,prx,twocolumn,floatfix,showpacs,superscriptaddress,nofootinbib]{revtex4-2}
\usepackage{amsfonts}
\usepackage{amsmath,bbm, amsfonts}
\usepackage{amssymb}
\usepackage{amsthm}
\usepackage{amsbsy}
\usepackage{bm}
\usepackage{braket}
\usepackage{comment}
\usepackage{dcolumn}
\usepackage{epstopdf}
\usepackage{graphicx}
\usepackage{graphbox}
\usepackage{enumerate}
\usepackage{latexsym}
\usepackage{multirow}
\usepackage{mathtools}
\usepackage{mathrsfs}

\usepackage{subfig}
\usepackage{tabularx}
\usepackage{times}
\usepackage{tikz}
\usepackage{verbatim}
\usepackage{appendix}
\usepackage[normalem]{ulem}
\usepackage[colorlinks, citecolor=blue]{hyperref}
\hypersetup{linkcolor=magenta,urlcolor=blue,citecolor=blue,pdfstartview={FitH},urlcolor=blue}

\begin{document}

\title{Solving Nonequilibrium Dynamics via Influence Matrix Bootstrap: Floquet-PXP Model}

\author{Xiao-Yang Yang}
\affiliation{
Institute for Advanced Study, Tsinghua University, Beijing 100084, China
}

\author{He-Ran Wang}
\email{whr21@tsinghua.org.cn}
\affiliation{
Institute for Advanced Study, Tsinghua University, Beijing 100084, China
}

\author{Zhong Wang}
\email{wangzhongemail@tsinghua.edu.cn}
\affiliation{
Institute for Advanced Study, Tsinghua University, Beijing 100084, China
}

\begin{abstract}
Studies of integrable systems have profoundly deepened the fundamental understanding of quantum many-body physics. While equilibrium properties such as ground states and thermodynamics can often be characterized efficiently, accurately characterizing nonequilibrium integrable dynamics remains a significant challenge. Here, we develop an influence matrix bootstrap approach for exactly solving nonequilibrium dynamics. This approach begins with an ansatz consisting of a set of local tensor relations, termed \textit{generalized zipper conditions}, and uses finite-time data to construct exact solutions of local dynamics that extend to arbitrarily long times.
We demonstrate this approach in  the ``Rule 201'' quantum cellular automaton, an integrable Trotterization of the PXP Hamiltonian. This uncovers a rich landscape of nonequilibrium behavior exhibiting initial-state dependence. As an example, we investigate the fate of persistent oscillating dynamics under local non-integrable perturbations, and present analytical results for non-thermal relaxation constrained by conservation laws. We also obtain numerically exact results for entanglement growth across a broad class of initial states. Furthermore, from an information-theoretic perspective, we identify a refined structure of multitime correlations termed the \textit{hidden Markov order}: the memory encoded in the dynamics separates into finite-length and long-range distributed components, which becomes transparent in an exact split-index matrix-product-state representation of the influence matrix. Our approach enables unified investigations of nonthermalizing and thermalizing regimes of nonequilibrium dynamics within a single analytically tractable model, and can be tested experimentally in state-of-the-art quantum simulators such as Rydberg atom arrays.

\end{abstract}

\maketitle



\section{Introduction}\label{sec:intro}

Nonequilibrium dynamics towards thermalized equilibrium states and the generalized Gibbs ensemble has been the primary focus in the area of quantum many-body physics~\cite{Rigol2007Relaxation,Rigol2008Thermalization,DAlessio2016, Gogolin2016}.
Numerical simulations of generic systems suffer from the curse of dimensionality, particularly manifesting as the rapid growth of entanglement entropy.
Even for integrable systems, the computational overhead can hardly be mitigated: although Bethe ansatz methods enable formally exact solutions of eigenstates and energy spectrum, these are not efficient for calculating time evolution of observables and correlation functions.
The framework of generalized hydrodynamics has been developed to characterize coarse-grained nonequilibrium dynamics through continuity equations of quasiparticles in integrable systems~\cite{Castro2016Emergent, Bertini2016}, while its scope of validity and applicability warrants further investigations~\cite{Doyon2025}.

Rydberg atom arrays have become an attractive experimental platform to study exotic phases of matter and nonequilibrium dynamics, notably including violations of thermalization \cite{Bernien2017Probing}. This phenomenon has been demonstrated by modeling the Rydberg interactions with the PXP Hamiltonian, which hosts quantum many-body scars as atypical and lowly entangled eigenstates embedded within an otherwise thermal spectrum~\cite{Turner2018Weak,Choi2019Emergent}. These scar states account for non-thermalizing dynamics for certain initial states with large overlap with them. 

Furthermore, Refs.~\cite{Iadecola2020,Wilkinson2020,Giudici2024} have promoted the PXP Hamiltonian to Floquet quantum circuits via Trotterization, providing an alternative perspective on the origins of PXP scars and associated non-thermalizing dynamics.
In particular, Ref.~\cite{Wilkinson2020} identified a special parameter point where the Floquet operator induces deterministic evolution on computational-basis states, corresponding to the ``Rule 201'' quantum cellular automaton, which has recently been realized in quantum simulators~\cite{White2026Quantum}.
The model has also been shown to be (super-) integrable, as characterized by ballistic quasiparticles~\cite{Wilkinson2020}.
Although being deterministic on computational-basis states, simulation of the dynamics remains intractable in the presence of quantum perturbations or genuinely quantum initial states.

Recently, the influence matrix framework has been introduced to characterize nonequilibrium dynamics by encoding correlations in the global system into a temporal matrix product state (MPS)~\cite{Banuls2009Matrix,Hermes2012Tensor,Lerose2021Influence,Sonner2021Influence,Ye2021Constructing}.
When the MPS bond dimension grows only subexponentially in time, efficient classical simulation of local observables and correlations becomes feasible.
Such favorable scaling of computational cost has been identified through exact solutions in various settings, including dual-unitary circuits and higher-order generalizations~\cite{Bertini2019Exact, Piroli2020Exact, Yu2023Hierarchical, Bertini2023Exact, Wang2025Temporal, Rampp2026Infinite}, integrable systems~\cite{Giudice2022Temporal,Klobas2021Exact,Klobas2021ExactII,Klobas2021ExactIII}, and hidden Markovian dynamics~\cite{Wang2024Exact}. 

In this work, we present analytical solutions for nonequilibrium dynamics in a quantum version of Rule 201 cellular automaton.
We identify an exact finite-bond-dimension MPS representation of the influence matrix for a class of initial states.
To demonstrate that this solution remains scalable to arbitrarily long evolution times, we develop a set of local graphical rules, which we call \textit{ generalized zipper conditions}, that serve as a sufficient condition enabling exact tensor-network contractions to obtain exact solutions.

At the same time, we provide a practical route for discovering exact influence matrices. Previous exact solutions were obtained either by constructions, as in dual-unitary circuits~\cite{Piroli2020Exact, Yu2023Hierarchical,Bertini2023Exact} and hidden Markovian dynamics~\cite{Wang2024Exact}, or by solving designed tensor-network ansatz, as in Rule 54 and XXZ circuit~\cite{Klobas2021Exact,Giudice2022Temporal}. In contrast, here we adopt a numerical bootstrap strategy. Once the influence matrix is observed numerically to have a finite bond dimension at sufficiently long times, we can transform it into a bulk-shift-invariant MPS form and then extract the building blocks by solving the associated generalized zipper conditions. This method is particularly useful for cases with finite but relatively large bond dimension $\chi$, such as the $\chi=54$ solution found here for certain initial states in Rule 201, which would be difficult to obtain by existing methods.

We further find that the analogous conditions hold in Rule 54 quantum cellular automaton ~\cite{Klobas2021Exact,Klobas2021ExactII,Klobas2021ExactIII}, as well as for several other models with exactly solvable influence matrices~\cite{Bertini2019Exact, Piroli2020Exact, Yu2023Hierarchical,Bertini2023Exact,Klobas2021Exact,Klobas2021ExactII,Klobas2021ExactIII,Wang2024Exact}.
Indeed, among the 256 elementary binary cellular automata in 1D, Rule 201 and Rule 54 are essentially the only cases that admit quantum generalizations and genuine interactions~\cite{Bobenko1993Two}. Therefore, our results fill a notable gap in this context.

In the Rule 201 quantum cellular automaton, one aspect we focus on is the stability of nonthermalizing dynamics under local, nonintegrable perturbations. For a class of computational-basis initial states, the unperturbed dynamics displays short-period revivals reminiscent of quantum many-body scars. 
Leveraging exact solutions for the influence matrix, we show that perturbations generate effective decoherence and drive the local subsystem toward a steady state on a parametrically long relaxation timescale.
Though commonly anticipated in integrable systems, such a relaxation has been established rigorously only in a limited number of cases, most notably for spin-helix states in the XXZ spin chain~\cite{jepsen2020spin,paul2021transverse,jepsen2022long,Cecile2023,Popkov2024}. 


We further characterize thermalizing dynamics through the growth of entanglement entropy after global quantum quenches from coherent initial states. We discover a linear-in-time entanglement growth and relate the rate to a tilt parameter in the initial state. In the slightly tilted regime, we develop a perturbative analysis that quantitatively accounts for these results.

In addition, from an information-theoretic perspective, we identify a refined structure of the Rule 201 influence matrix termed as the \textit{hidden Markov order}.
A classical stochastic process is said to have Markov order $k$ if the conditional probability of a future state depends only on the states within the preceding $k$ time steps. This notion has recently been generalized to quantum processes~\cite{Taranto2019Quantum}. Building on these developments, we introduce the concept of hidden Markov order in terms of the influence matrix. Specifically, the local subsystem can be viewed as being coupled to a finite-dimensional ancilla, such that the non-Markovian evolution of the enlarged system has a finite quantum Markov length.

Our exact solutions of the influence matrix naturally realize this structure, allowing the decomposition of memory in the dynamics into a short-range component, characterized by a finite Markov length, and a long-range component distributed over extended timescales. We further develop an MPS-based algorithm to systematically diagnose hidden Markov order, which is applicable beyond exactly solvable cases.
This perspective differs from our previous work on hidden Markovian dynamics~\cite{Wang2024Exact}, where the virtual bond of the influence matrix was interpreted as an ancilla carrying the full bath memory. In contrast, the present framework enables a more refined separation of memory across different timescales.

The rest of the paper is organized as follows. 
In Sec.~\ref{sec:model}, we set the stage by reviewing the notions of the Rule 201 cellular automaton and Floquet-PXP model.
In Sec.~\ref{sec:solution}, we present our method for solving the Rule 201 influence matrix in diagrammatic forms, together with the numerical bootstrap approach.
We investigate the rich dynamics of the Floquet-PXP model in Sec.~\ref{sec:result}, including the relaxation of the short-orbit dynamics and the entanglement growth from global quench.
In Sec.~\ref{sec:markov_order}, we introduce the hidden Markov order as a framework for uncovering the internal structure of influence matrices.
Finally, we conclude and discuss future directions in Sec.~\ref{sec:conclusion}.

\section{Rule 201 cellular automaton and Floquet-PXP model}\label{sec:model}

\begin{figure*}[ht]
    \hspace{-0.95\linewidth}
    \includegraphics[width=0.95\linewidth]{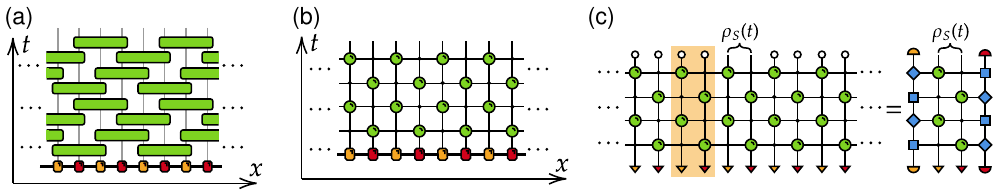}
    \caption{Tensor-network representation of the Rule 201 quantum cellular automaton. Number of time steps $t=2$. (a) Brickwork quantum circuit. (b) Rule 201 quantum cellular automaton represented in interactions round-a-face. (c) Dynamics of a local subsystem on two sites, where the rest of the system is traced out and gives rise to influence matrices, represented as matrix product states on two sides. The shaded region is the spatial transfer matrix.
    }
    \label{fig:exact_dynamics}
\end{figure*}

A three-site $\mathbb{Z}_2$ valued cellular automaton is a one-dimensional deterministic dynamical system of size $N$, where each site $x$ carries a binary variable $s_x \in \{0,1\}$. The state is updated in discrete time according to a global update rule
\begin{equation*}
    (s_1^{t+1},\cdots, s_N^{t+1}) = F(s_1^{t},\cdots, s_N^{t}).
\end{equation*}
The global map $F$ is built from three-site local update rules applied in an alternating pattern:
\begin{equation*}
    F = (\bigotimes_{x\in\text{odd}}f_{x-1,x,x+1}) \circ (\bigotimes_{x\in\text{even}}f_{x-1,x,x+1}).
\end{equation*}
Here $f_{x-1,x,x+1}$ updates only the middle variable $s_x$ according to the values of its two neighboring variables $s_{x-1}$ and $s_{x+1}$, and thus $f$ is a mapping from $\mathbb{Z}_2^3$ to $\mathbb{Z}_2$. 

Such dynamical systems are conventionally named by their Wolfram rule number. For a local function $f$, there are $2^3=8$ possible input configurations, and therefore all the possible outputs of $f$ form an eight-bit binary string. Specifically, writing
\begin{equation*}
(s_1, s'_2, s_3)=f(s_1,s_2,s_3),
\end{equation*}
the output values $s_2'$ associated with eight input configurations, conventionally in the order $[
111,\ 110,\ 101,\ 100,\ 011,\ 010,\ 001,\ 000 
]$, define a binary string. The Wolfram rule number is then given by the decimal value of this binary string.

In this work, we focus on the rule in which the central variable is flipped between $0$ and $1$ only when both its left and right neighboring sites are $0$; otherwise, it is left unchanged. This gives the output string $11001001$, and therefore the corresponding cellular automaton is called Rule 201. The dynamics has been shown to be integrable and exhibits ballistic quasiparticle propagation, together with matrix-product-state representations of equilibrium distributions~\cite{Wilkinson2020}.

The Rule 201 cellular automaton can be extended to quantum dynamics. Consider a one-dimensional spin-$1/2$ chain evolved by a global Floquet operator at each time step:
\begin{align}\label{eq:brickwork}
    \mathbb{U}=\left(\mathop{\bigotimes}\limits_{x \in\text{ odd}}U_{x-1,x,x+1}\right)\left(\mathop{\bigotimes}\limits_{x \in\text{ even}}U_{x-1,x,x+1}\right).
\end{align}
The local three-site unitary is defined as
\begin{align}\label{eq:rule201_def}
     U=\sum_{m,n = 0}^1P^{(m)}\otimes U^{(mn)}\otimes P^{(n)},
\end{align}
where $P^{(m)}=\ket{m}\bra{m}$. For Rule 201 one has $U^{(00)}=X$, $U^{(01)}=U^{(10)}=U^{(11)}=I$, where $X$ is the Pauli-X operator. In addition, for Rule 54 one has $U^{(00)}=I$ and $U^{(01)}=U^{(10)}=U^{(11)}=X$.
The unitary implements a controlled flip of the central qubit conditioned on both neighboring qubits being in $\ket{0}$. Therefore, restricted to computational-basis states, the resulting quantum evolution coincides with the Rule 201 cellular automaton. We hence call it the Rule 201 quantum cellular automaton. 

The model is related to the Trotterized PXP model up to a phase factor. Specifically, consider the exponentiation of the three-site term in the PXP Hamiltonian with Trotter step $\tau$~\cite{Iadecola2020,Giudici2024}:
\begin{align}\label{eq:f-pxp_def}
    U_{x-1,x,x+1}'(\tau)=&e^{-i\tau(P_{x-1}^{(0) }X_xP_{x+1}^{(0)})}\nonumber\\
    =&  I-(1-\cos{\tau})P_{x-1}^{(0)}P_{x+1}^{(0)}\nonumber\\
    &-i\sin{\tau}P_{x-1}^{(0)}X_xP_{x+1}^{(0)}.
\end{align}
At $\tau = \pi/2$, the unitary differs from Eq.~\eqref{eq:rule201_def} only by a phase factor before the $X$ operator. In Appendix~\ref{app:relations}, we demonstrate that this phase does not lead to any qualitative difference in the quantities we consider here, and throughout the main text we focus on the model defined by Eq.~\eqref{eq:rule201_def}.

The brickwork quantum circuit generated by Eq.~\eqref{eq:brickwork} is illustrated in  Fig.~\ref{fig:exact_dynamics}(a), starting from an MPS initial state. For convenience, we fold the unitary evolution together with its complex conjugate, so that each leg in the circuit carries a doubled Hilbert space $\mathcal{H}\otimes\mathcal{H}^*$ of dimension $4$. Physical quantities such as observable expectation values and correlation functions naturally involve both forward and backward time evolution, and can therefore be represented in this folded picture.

The structure of $U$ defined by Eq.~\eqref{eq:rule201_def} allows each layer of the global unitary to be represented as a matrix product operator (MPO), as shown in Fig.~\ref{fig:exact_dynamics}(b). This representation is referred to as the interactions round-a-face (IRF) structure~\cite{Baxter1982exactly, Prosen2021many}. In each layer, qubits on even (odd) sites are updated under the control of qubits on odd (even) sites. The local tensors in the folded pictures are defined by:
\begin{equation}
\begin{tikzpicture}[x=0.75pt,y=0.75pt,yscale=-1,xscale=1]

\draw    (20,31) -- (50,31) ;
\draw    (35,17) -- (35,43) ;
\draw  [fill={rgb, 255:red, 126; green, 211; blue, 33 }  ,fill opacity=1 ] (30,31) .. controls (30,28.24) and (32.24,26) .. (35,26) .. controls (37.76,26) and (40,28.24) .. (40,31) .. controls (40,33.76) and (37.76,36) .. (35,36) .. controls (32.24,36) and (30,33.76) .. (30,31) -- cycle ;
\draw  [draw opacity=0] (35,28) .. controls (35,28) and (35,28) .. (35,28) .. controls (36.66,28) and (38,29.34) .. (38,31) -- (35,31) -- cycle ; \draw   (35,28) .. controls (35,28) and (35,28) .. (35,28) .. controls (36.66,28) and (38,29.34) .. (38,31) ;  
\draw   (206.5,31) .. controls (206.5,30.45) and (206.95,30) .. (207.5,30) .. controls (208.05,30) and (208.5,30.45) .. (208.5,31) .. controls (208.5,31.55) and (208.05,32) .. (207.5,32) .. controls (206.95,32) and (206.5,31.55) .. (206.5,31) -- cycle ;
\draw    (198.5,31) -- (217.5,31) ;
\draw    (207.5,21) -- (207.5,41) ;

\draw (27.8,41.4) node [anchor=north west][inner sep=0.75pt]  [font=\scriptsize]  {$i,i'$};
\draw (3.29,21.67) node [anchor=north west][inner sep=0.75pt]  [font=\scriptsize]  {$j,j'$};
\draw (48.87,21.13) node [anchor=north west][inner sep=0.75pt]  [font=\scriptsize]  {$k,k'$};
\draw (26.93,6) node [anchor=north west][inner sep=0.75pt]  [font=\scriptsize]  {$l,l'$};
\draw (65.5,16.8) node [anchor=north west][inner sep=0.75pt]    {$= U_{li}^{( jk)}\left( U_{l'i'}^{( j'k')}\right)^{*} ,$};
\draw (234.17,24.8) node [anchor=north west][inner sep=0.75pt]    {$=\delta _{ijkl} \delta _{i'j'k'l'} ,$};
\draw (200.51,37.4) node [anchor=north west][inner sep=0.75pt]  [font=\scriptsize]  {$i,i'$};
\draw (182.67,20.33) node [anchor=north west][inner sep=0.75pt]  [font=\scriptsize]  {$j,j'$};
\draw (215.58,20.47) node [anchor=north west][inner sep=0.75pt]  [font=\scriptsize]  {$k,k'$};
\draw (199.31,10) node [anchor=north west][inner sep=0.75pt]  [font=\scriptsize]  {$l,l'$};

\end{tikzpicture}
\end{equation}
where all the indices take values in $\{0,1\}$, and $\delta_{ijkl}$ represents the identity symbol for the four indices.
Finally, we introduce the identity operator on the physical Hilbert space, denoted by hollow dots. In this notation, the unitarity of $U$ reads
\begin{equation}\label{eq:unitary}
\begin{tikzpicture}[x=0.75pt,y=0.75pt,yscale=-1,xscale=1]

\draw    (20,31) -- (50,31) ;
\draw    (35,18) -- (35,44) ;
\draw  [fill={rgb, 255:red, 126; green, 211; blue, 33 }  ,fill opacity=1 ] (30,31) .. controls (30,28.24) and (32.24,26) .. (35,26) .. controls (37.76,26) and (40,28.24) .. (40,31) .. controls (40,33.76) and (37.76,36) .. (35,36) .. controls (32.24,36) and (30,33.76) .. (30,31) -- cycle ;
\draw  [draw opacity=0] (35,28) .. controls (35,28) and (35,28) .. (35,28) .. controls (36.66,28) and (38,29.34) .. (38,31) -- (35,31) -- cycle ; \draw   (35,28) .. controls (35,28) and (35,28) .. (35,28) .. controls (36.66,28) and (38,29.34) .. (38,31) ;  
\draw   (49,31) .. controls (49,30.45) and (49.45,30) .. (50,30) .. controls (50.55,30) and (51,30.45) .. (51,31) .. controls (51,31.55) and (50.55,32) .. (50,32) .. controls (49.45,32) and (49,31.55) .. (49,31) -- cycle ;
\draw    (50,18) -- (50,44) ;
\draw    (20,18) -- (20,44) ;
\draw   (19,31) .. controls (19,30.45) and (19.45,30) .. (20,30) .. controls (20.55,30) and (21,30.45) .. (21,31) .. controls (21,31.55) and (20.55,32) .. (20,32) .. controls (19.45,32) and (19,31.55) .. (19,31) -- cycle ;
\draw  [fill={rgb, 255:red, 255; green, 255; blue, 255 }  ,fill opacity=1 ] (18,18) .. controls (18,16.9) and (18.9,16) .. (20,16) .. controls (21.1,16) and (22,16.9) .. (22,18) .. controls (22,19.1) and (21.1,20) .. (20,20) .. controls (18.9,20) and (18,19.1) .. (18,18) -- cycle ;
\draw  [fill={rgb, 255:red, 255; green, 255; blue, 255 }  ,fill opacity=1 ] (48,18) .. controls (48,16.9) and (48.9,16) .. (50,16) .. controls (51.1,16) and (52,16.9) .. (52,18) .. controls (52,19.1) and (51.1,20) .. (50,20) .. controls (48.9,20) and (48,19.1) .. (48,18) -- cycle ;
\draw  [fill={rgb, 255:red, 255; green, 255; blue, 255 }  ,fill opacity=1 ] (33,18) .. controls (33,16.9) and (33.9,16) .. (35,16) .. controls (36.1,16) and (37,16.9) .. (37,18) .. controls (37,19.1) and (36.1,20) .. (35,20) .. controls (33.9,20) and (33,19.1) .. (33,18) -- cycle ;
\draw    (89,18) -- (89,44) ;
\draw    (101,18) -- (101,44) ;
\draw    (77,18) -- (77,44) ;
\draw  [fill={rgb, 255:red, 255; green, 255; blue, 255 }  ,fill opacity=1 ] (75,18) .. controls (75,16.9) and (75.9,16) .. (77,16) .. controls (78.1,16) and (79,16.9) .. (79,18) .. controls (79,19.1) and (78.1,20) .. (77,20) .. controls (75.9,20) and (75,19.1) .. (75,18) -- cycle ;
\draw  [fill={rgb, 255:red, 255; green, 255; blue, 255 }  ,fill opacity=1 ] (99,18) .. controls (99,16.9) and (99.9,16) .. (101,16) .. controls (102.1,16) and (103,16.9) .. (103,18) .. controls (103,19.1) and (102.1,20) .. (101,20) .. controls (99.9,20) and (99,19.1) .. (99,18) -- cycle ;
\draw  [fill={rgb, 255:red, 255; green, 255; blue, 255 }  ,fill opacity=1 ] (87,18) .. controls (87,16.9) and (87.9,16) .. (89,16) .. controls (90.1,16) and (91,16.9) .. (91,18) .. controls (91,19.1) and (90.1,20) .. (89,20) .. controls (87.9,20) and (87,19.1) .. (87,18) -- cycle ;

\draw (56,26.4) node [anchor=north west][inner sep=0.75pt]    {$=$};

\end{tikzpicture},\qquad\quad
\begin{tikzpicture}[x=0.75pt,y=0.75pt,yscale=-1,xscale=1]

\draw    (20,31) -- (50,31) ;
\draw    (35,18) -- (35,44) ;
\draw  [fill={rgb, 255:red, 126; green, 211; blue, 33 }  ,fill opacity=1 ] (30,31) .. controls (30,28.24) and (32.24,26) .. (35,26) .. controls (37.76,26) and (40,28.24) .. (40,31) .. controls (40,33.76) and (37.76,36) .. (35,36) .. controls (32.24,36) and (30,33.76) .. (30,31) -- cycle ;
\draw  [draw opacity=0] (35,28) .. controls (35,28) and (35,28) .. (35,28) .. controls (36.66,28) and (38,29.34) .. (38,31) -- (35,31) -- cycle ; \draw   (35,28) .. controls (35,28) and (35,28) .. (35,28) .. controls (36.66,28) and (38,29.34) .. (38,31) ;  
\draw   (49,31) .. controls (49,30.45) and (49.45,30) .. (50,30) .. controls (50.55,30) and (51,30.45) .. (51,31) .. controls (51,31.55) and (50.55,32) .. (50,32) .. controls (49.45,32) and (49,31.55) .. (49,31) -- cycle ;
\draw    (50,18) -- (50,44) ;
\draw    (20,18) -- (20,44) ;
\draw   (19,31) .. controls (19,30.45) and (19.45,30) .. (20,30) .. controls (20.55,30) and (21,30.45) .. (21,31) .. controls (21,31.55) and (20.55,32) .. (20,32) .. controls (19.45,32) and (19,31.55) .. (19,31) -- cycle ;
\draw  [fill={rgb, 255:red, 255; green, 255; blue, 255 }  ,fill opacity=1 ] (18,44) .. controls (18,42.9) and (18.9,42) .. (20,42) .. controls (21.1,42) and (22,42.9) .. (22,44) .. controls (22,45.1) and (21.1,46) .. (20,46) .. controls (18.9,46) and (18,45.1) .. (18,44) -- cycle ;
\draw  [fill={rgb, 255:red, 255; green, 255; blue, 255 }  ,fill opacity=1 ] (48,44) .. controls (48,42.9) and (48.9,42) .. (50,42) .. controls (51.1,42) and (52,42.9) .. (52,44) .. controls (52,45.1) and (51.1,46) .. (50,46) .. controls (48.9,46) and (48,45.1) .. (48,44) -- cycle ;
\draw  [fill={rgb, 255:red, 255; green, 255; blue, 255 }  ,fill opacity=1 ] (33,44) .. controls (33,42.9) and (33.9,42) .. (35,42) .. controls (36.1,42) and (37,42.9) .. (37,44) .. controls (37,45.1) and (36.1,46) .. (35,46) .. controls (33.9,46) and (33,45.1) .. (33,44) -- cycle ;
\draw    (89,18) -- (89,44) ;
\draw    (101,18) -- (101,44) ;
\draw    (77,18) -- (77,44) ;
\draw  [fill={rgb, 255:red, 255; green, 255; blue, 255 }  ,fill opacity=1 ] (75,44) .. controls (75,42.9) and (75.9,42) .. (77,42) .. controls (78.1,42) and (79,42.9) .. (79,44) .. controls (79,45.1) and (78.1,46) .. (77,46) .. controls (75.9,46) and (75,45.1) .. (75,44) -- cycle ;
\draw  [fill={rgb, 255:red, 255; green, 255; blue, 255 }  ,fill opacity=1 ] (99,44) .. controls (99,42.9) and (99.9,42) .. (101,42) .. controls (102.1,42) and (103,42.9) .. (103,44) .. controls (103,45.1) and (102.1,46) .. (101,46) .. controls (99.9,46) and (99,45.1) .. (99,44) -- cycle ;
\draw  [fill={rgb, 255:red, 255; green, 255; blue, 255 }  ,fill opacity=1 ] (87,44) .. controls (87,42.9) and (87.9,42) .. (89,42) .. controls (90.1,42) and (91,42.9) .. (91,44) .. controls (91,45.1) and (90.1,46) .. (89,46) .. controls (87.9,46) and (87,45.1) .. (87,44) -- cycle ;

\draw (56,26.4) node [anchor=north west][inner sep=0.75pt]    {$=$};

\end{tikzpicture}.
\end{equation}

\section{Influence matrix and generalized zipper condition}\label{sec:solution}

Although the Rule 201 dynamics is classically tractable when initialized in computational-basis states, generic quantum initial states lead to genuinely quantum evolution that is no longer reducible to deterministic classical dynamics. This motivates the use of alternative methods to characterize the nonequilibrium behavior.

Here, we apply the influence matrix approach to solve the subsystem evolution. 
As shown in Fig.~\ref{fig:exact_dynamics} (c) on the left-hand side, the full system is 
divided into the subsystem $S$ and an effective environment; the subsystem evolution is then obtained by tracing out the environmental degrees of freedom.
The influence of the left and right environments on $S$ are encoded respectively in the left and right multitime MPS, referred to as influence matrices~\cite{Lerose2021Influence}. For the translation-invariant quantum circuits studied here, the influence matrix can be viewed as the leading eigenvector of the spatial transfer matrix, represented by the shaded region in Fig.~\ref{fig:exact_dynamics}(c).

The computational efficiency of the influence-matrix MPS representation is governed by the growth of bond dimension and bipartite entanglement entropy with evolution times. Our first result is that, in the Rule 201 quantum cellular automaton, the influence matrix admits an exact finite-bond-dimension MPS representation for a broad class of initial states, which enables efficient classical simulation of long-time dynamics.

\begin{table*}[ht]
    \renewcommand{\arraystretch}{1.3}
    \begin{tabular}{@{\hspace{10pt}} l @{\hspace{10pt}} c @{\hspace{10pt}} c @{\hspace{10pt}}}
        \hline
        Quantum cellular automaton & Initial state two-site block $\rho \otimes \rho'$ & Influence matrix bond dimension $\chi$ \\
        \hline
        \multirow{3}{*}{Rule 201} & $|0\rangle\langle 0|\otimes |0\rangle\langle 0|$, $|0\rangle\langle 0|\otimes |1\rangle\langle 1|$, $|1\rangle\langle 1|\otimes |0\rangle\langle 0|$ & 12 (Appendix~\ref{sec:rule201,12}) \\
        \cline{2-3}
        & $|0\rangle\langle 0| \otimes \frac{1}{2}(I + a Z)$ & 24 (Appendix~\ref{sec:rule201,24}) \\
        \cline{2-3}
        & $|0\rangle\langle 0| \otimes \rho$ & 54 (Appendix~\ref{app:Hankel}) \\
        \hline
        \multirow{3}{*}{Rule 54} & $|0\rangle\langle 0| \otimes \rho$, $\frac{1}{4}I \otimes I$ & 3 (Ref.~\cite{Klobas2021Exact, Klobas2021ExactII} and Appendix~\ref{sec:rule54,3})\\
        \cline{2-3}
        & $|1\rangle\langle 1| \otimes \frac{1}{2}(I + a Z)$ & 6 (Appendix~\ref{sec:rule54,6}) \\
        \cline{2-3}
        & $|1\rangle\langle 1| \otimes \rho$ & 14 (Appendix~\ref{sec:rule54,14}) \\
        \hline
    \end{tabular}
    \caption{Summary of influence matrix exact solutions for Rule 201 and Rule 54 quantum cellular automata. The initial states are chosen to be two-site shift-invariant product states, with each two-site block of the form $\rho\otimes\rho'$. Explicit expressions of the solutions are reported in Appendices \ref{sec:solution_list} and~\ref{sec:solution_list1}.
    }
    \label{table:solutions}
\end{table*}

Before introducing the recipe for exact solutions, we first summarize our results in Table~\ref{table:solutions}.
It is shown that the bond dimension $\chi$ of the influence matrix can depend on the choice of initial state.
We also include exact solutions for the Rule 54 quantum cellular automaton, which can be obtained using a similar 
method.

As shown in Fig.~\ref{fig:exact_dynamics}(c) on the right-hand side, the MPS representation consists of two rank-three tensors in the bulk together with boundary vectors, depicted as follows:

\begin{equation}\label{eq:building_blocks}
.
\end{equation}
Here, $i = 0, 1, 2, 3$ denotes the basis states of the doubled Hilbert space $\mathcal{H}\otimes \mathcal{H}^*$, and $a = 0, 1, \cdots, \chi-1$ is the basis of the $\chi$-dimensional auxiliary linear space, denoted by $\mathcal{U}$. Accordingly, the top and bottom boundary vectors $|\text{t}), |\text{t}'), |\text{b}), |\text{b}')$ live in $\mathcal{U}$. We further introduce the folded representation of two-site shift-invariant initial states, of which the repeating two-site block is $\rho\otimes\rho'$:
\begin{equation}
%
,
\end{equation}
which can be pure or mixed states.

We find that these local tensors fulfill a set of tensor relations, including relations of bulk tensors:
\begin{equation}\label{eq:bulk}
%
,
\end{equation}
and top boundary conditions:
\begin{equation}\label{eq:top_bc}
%
,
\end{equation}
and finally, bottom boundary conditions:
\begin{equation}\label{eq:btm_bc}
%
.
\end{equation}
Here, we introduce additional tensors:
\begin{equation}
%
,
\end{equation}
where $P_0$ and $P_0'$ are non-orthogonal projectors on the extended Hilbert space $\mathcal{U}\otimes\mathcal{H}\otimes\mathcal{H}^*$, and $P_1$ and $P_1'$ are mappings from $\mathcal{U}\otimes\mathcal{H}\otimes\mathcal{H}^*$ to $\mathcal{U}$.

With these tensor relations, it can be verified that the multitime MPS constructed here is the fixed point of the spatial transfer matrix:
\begin{equation}\label{eq:fp_solution_graphical}
%
.
\end{equation}
The tensor-network contractions are performed as follows.
Using the first two formulas in Eq.~\eqref{eq:btm_bc}, we attach projectors $P_0$ ($P_0'$) to the bottom of the diagram. Then, using the first two relations in Eq.~\eqref{eq:bulk}, these projectors can be propagated through each layer. Next, applying the top-boundary, bulk, and bottom-boundary conditions sequentially, the action of the spatial transfer matrix is reduced, compressing the bond dimension from $4\times\chi$ to $\chi$.
The top-down elimination of tensors is reminiscent of the conventional zipper conditions used to solve MPO-transfer matrix steady states~\cite{Haegeman2017Diag, Klobas2021Exact}, but with additional projection tensors. We therefore refer to these relations as \textit{generalized zipper conditions}.


We present explicit expressions for the local tensors corresponding to different initial states $\rho\otimes\rho'$ in Appendix~\ref{sec:solution_list}. The same diagrammatic relations also hold for Rule 54 influence matrix, and we present the solutions in Appendix~\ref{sec:solution_list1}. That includes both a reproduction of previously reported solutions~\cite{Klobas2021Exact,Klobas2021ExactII} (Appendix~\ref{sec:rule54,3}), and new solutions for different initial states (Appendix~\ref{sec:rule54,6} and~\ref{sec:rule54,14}). Finally, we note that solvable conditions of several influence matrices in brickwork quantum circuits~\cite{Piroli2020Exact, Yu2023Hierarchical,Bertini2023Exact} can be formally recast into a form similar to generalized zipper conditions, as demonstrated in Appendix~\ref{app:generalized_zipper}.

Finally, we note that the practical use of the generalized zipper conditions is somewhat limited if taken as a purely analytical ansatz. 
While these conditions are sufficient to guarantee exact solutions at arbitrary time steps, solving them directly from only the local unitary $U^{(mn)}$ and the initial state can be extremely difficult.
This is because the local tensors to be solved generally have an a priori unknown bond dimension, and the tensor relations generally form cubic equations of variables.

We overcome this hardness through a numerical bootstrap approach. First, we numerically compute the MPS representation of the influence matrix up to a sufficiently large time, where the bond dimension and bipartite entanglement entropy have saturated to finite values.
Starting from this generally non-uniform finite-size MPS, we then use a Hankel-matrix algorithm to bring it into a shift-invariant form, thereby extracting the local tensors $A, B$ and boundary vectors $|\text{t}), |\text{b})$~\cite{Balle2013Spectral}. It follows that only the tensors $P_{0,1}$ and $P'_{0,1}$ remain to be solved in order to complete the generalized zipper conditions. In practice, these equations are generally overdetermined; hence, the existence of solutions provides a consistency check that the generalized zipper conditions are satisfied.

All solutions presented in this work are obtained using this Hankel-matrix algorithm. The method is particularly useful for initial states of the form $\ket{0}\bra{0}\otimes\rho$ in Rule 201, for which the influence matrix has the unexpectedly large but finite bond dimension $\chi=54$. Details of the algorithm, together with those of $\chi=54$ solution, are provided in Appendix~\ref{app:Hankel}. We expect this method to apply whenever a finite-bond-dimension influence matrix is observed numerically at sufficiently long evolution times, thereby facilitating the discovery of new exact solutions.

\section{Nonequilibrium quantum dynamics: Exact results}\label{sec:result}

In this section, we leverage exact solutions of influence matrices to investigate several aspects of quantum many-body dynamics in the Rule 201 quantum cellular automaton. Specifically, we consider different sources of non-classicality that drive the evolution away from deterministic dynamics. In Sec.~\ref{sec:relaxation}, we consider the initial states in the computational basis but introduce local ``quantum defects'' by modifying the gates in a small spatial region away from the cellular automaton. We characterize the relaxation of periodic dynamics supported by computational-basis states. 
In Sec.~\ref{sec:entanglement_growth_global}, we consider initial states that deviate from computational-basis states, parametrized by a tilt parameter. We map out the growth rate of entanglement entropy under time evolution as the tilt parameter is varied.

\subsection{Relaxation of periodic orbits by quantum defects}\label{sec:relaxation}

Ref.~\cite{Wilkinson2020} has developed a quasiparticle picture for the Rule 201 cellular automaton.
In this description, the vacuum is identified with the period-three classical orbit formed by the following computational-basis states:
\begin{align}\label{eq:orbit-3}
    \ket{\phi_0}=\ket{(00)},&& \ket{\phi_1}=\ket{(01)},&& \ket{\phi_2}=\ket{(10)},
\end{align}
where $\ket{(i_1i_2)}$ is a short-hand notation of $\cdots\otimes\ket{i_1}\otimes\ket{i_2}\otimes\ket{i_1}\otimes\ket{i_2}\otimes\cdots$.
The Floquet operator cyclically permutes the three states:
\begin{align}\label{eq:orbit-3-update-rule}
    \ket{\phi_{n+1\text{ mod }{3}}}=\mathbb{U}\ket{\phi_n},&&n=0,1,2.
\end{align}

Correspondingly, quasiparticles are identified as domain walls between two different vacuum configurations. 
For example, consider the initial state \[\ket{\tilde{\phi}_0}=\ket{\cdots0000100000\cdots},\] which is obtained from $\ket{\phi_0}$ by flipping a single qubit from $0$ to $1$. Effectively, two domain walls are created near the flipped qubit.
After an integer multiple of three time steps, the state takes the form
\[\ket{\cdots000000101010\cdots101010000000\cdots}.\]
In other words, the left, right, and central regions correspond to distinct vacuum configurations, and the two domain walls separating them propagate from the initially flipped site with opposite velocities. These ballistically propagating quasiparticles are also referred to as solitons.

\begin{figure}[ht]
    \hspace{-0.9\linewidth}
    \includegraphics[width=0.9\linewidth]{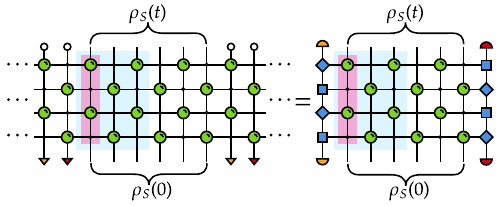}
    \caption{
    Nonequilibrium dynamics in the presence of quantum defects. Local gates in the shaded region are deformed away from the cellular automaton according to Eq.~\eqref{eq:rule201-deform}. The red region corresponds to the period-three initial state, while the light-blue region corresponds to the period-five initial state. Using the exact solution of the influence matrix, the subsystem dynamics can be computed for arbitrarily long times.
    }
    \label{fig:local_dynamics}
\end{figure}

Here, we examine the stability of the vacuum, i.e., the period-three dynamics, in the presence of local quantum defects. 
We deform controlled unitaries acting on the same spatial site at all time steps, as indicated by the red shaded region in Fig.~\ref{fig:local_dynamics}. The deformed unitary reads
\begin{align}\label{eq:rule201-deform}
    U^{(00)}=\left(
    \begin{array}{cc}i\sin{\epsilon} & \cos{\epsilon} \\ \cos{\epsilon} & i\sin{\epsilon}\end{array}
    \right),&&U^{(01)}=U^{(10)}=U^{(11)}=I.
\end{align}
At $\epsilon=0$, it reduces to the cellular automaton.
We take the six sites shown in Fig.~\ref{fig:local_dynamics} as the subsystem.

\begin{figure*}[ht]
    \hspace*{-0.45\linewidth}
    {\includegraphics[width=0.43\linewidth]{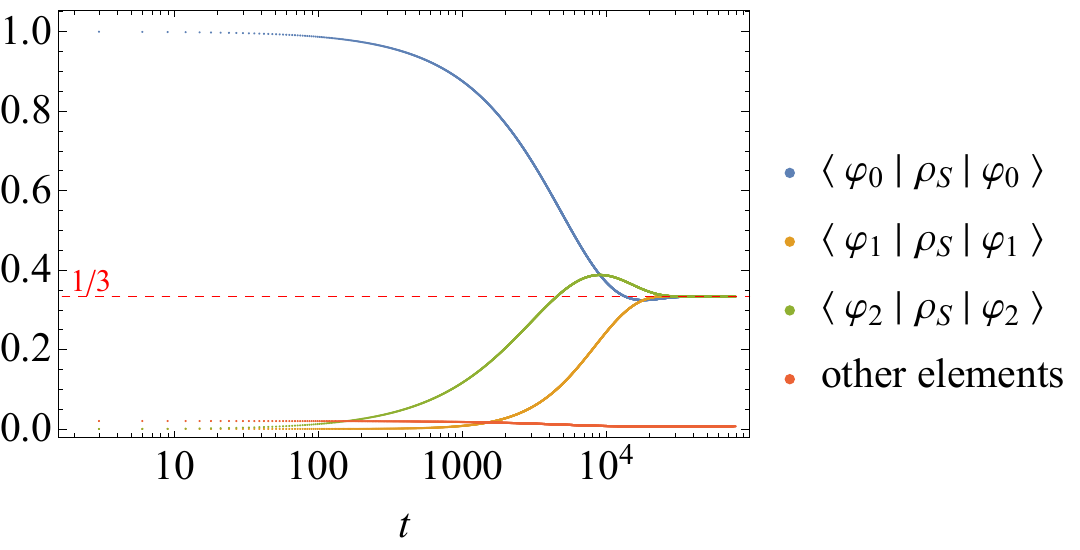}}\qquad
    {\includegraphics[width=0.43\linewidth]{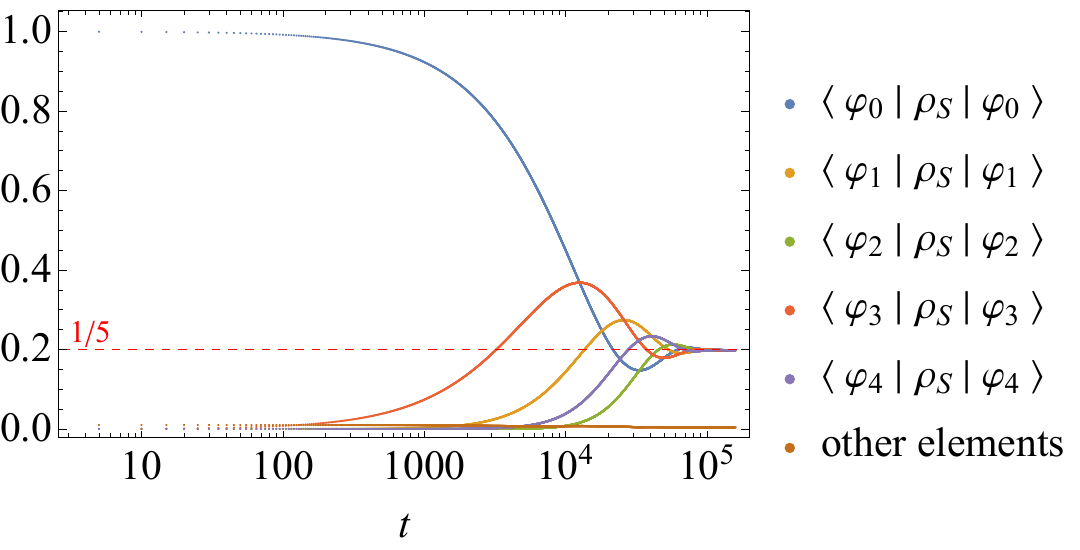}}
    \caption{Time evolution of matrix elements of the subsystem reduced density matrix $\rho_S(t)$ under a quantum deformation of local gates. The deformation parameter $\epsilon = 0.01$. Subsystem states $\ket{\varphi_n}$ are certain classical configuration defined in Eqs.~\eqref{eq:2-translation} and~\eqref{eq:6-translation}. Red dashed lines indicate stationary values of populations with these configurations. The curve labeled ``other elements'' denotes the maximum magnitude among all remaining matrix elements. Left panel: initial state $\ket{(00)}$. Right panel: initial state $\ket{(001001)}$.
    }
    \label{fig:oscillation}
\end{figure*}

Using the influence matrix method, for initial states in the vacuum configurations, the environmental degrees of freedom can be represented by an MPS with bond dimension $\chi=12$. Consequently, evaluating observables from the reduced density matrix $\rho_S(t)$ only involves repeated applications of a finite-dimensional transfer matrix along the time direction.

In the left panel in Fig.~\ref{fig:oscillation}, we show the time evolution of matrix elements of $\rho_S(t)$ up to $80,000$ time steps, starting from $\ket{\phi_0}$.
For convenience, we restrict to times $t\in 3\mathbb{Z}$.
During the first $\sim 100$ time steps, the subsystem remains close to a pure vacuum state. From $\sim 100$ to $\sim 10,000$ time steps, we observe decoherence among the three vacuum configurations, after which the state approaches an equal-weight incoherent mixture. The magnitudes of all other matrix elements of $\rho_S(t)$ are bounded by $\epsilon$. This behavior suggests relaxation toward the maximally mixed state within the zero-soliton sector, consistent with the relaxation induced by weak integrability breaking in integrable systems~\cite{Bertini2015Prethermalization, Durnin2021Non}.

To gain further insight, we perform a perturbative analysis at small $\epsilon$. Taking the initial state $\ket{\phi_0}$, 
after three time steps there is
\begin{align}
    \tilde{\mathbb{U}}^3\ket{\phi_0}=\cos{2\epsilon}\ket{\phi_0}+i\sin{2\epsilon}\ket{\tilde{\phi}_0},
\end{align}
where $\tilde{\mathbb{U}}$ denotes the locally deformed Floquet operator. Iterating this for short time $t\in3\mathbb{Z}$, we obtain
\begin{align}
    \tilde{\mathbb{U}}^t\ket{\phi_0}\approx\left(1-\frac{2\epsilon^2}{3}t\right)\ket{\phi_0}+2i\epsilon\sum_{n=0}^{t/3-1} {\mathbb{U}}^{3n}\ket{\tilde{\phi}_0}.
\end{align}

To evaluate the partial trace, we define vacuum configurations in the subsystem as
\begin{equation}\label{eq:2-translation}
    \ket{\varphi_0}=\ket{000000},\,\ket{\varphi_1}=\ket{010101},\,\ket{\varphi_2}=\ket{101010}.
\end{equation}
From the above analysis, we find that the reduced density matrix of ${\mathbb{U}}^{t}\ket{\tilde{\phi}_0}$ is the configuration $\ket{\varphi_{(t+2)\bmod3}}$. Therefore, the reduced density matrix is
\begin{align}
    \rho_S(t)\approx&\left(1-\frac{4\epsilon^2}{3}t\right)\ket{\varphi_0}\bra{\varphi_0}\notag\\
    &+4\epsilon^2\sum_{n=0}^{t/3-1}\text{Tr}_{\bar{S}}\left({\mathbb{U}}^{3n}\ket{\tilde{\phi}_0}\bra{\tilde{\phi_0}}{\mathbb{U}}^{\dagger3n}\right)\notag\\
    &+2i\epsilon\text{ Tr}_{\bar{S}}\left(\ket{\tilde{\phi}_0}\bra{{\phi_0}}-\ket{{\phi}_0}\bra{\tilde{\phi_0}}\right)\notag\\
    =&\left(1-\frac{4\epsilon^2}{3}t\right)\ket{\varphi_0}\bra{\varphi_0}+4\epsilon^2\left(\frac{t}{3}-2\right)\ket{\varphi_2}\bra{\varphi_2}\notag\\
    &+4\epsilon^2\left(\ket{100000}\bra{100000}+\ket{101000}\bra{101000}\right)\notag\\
    &+2i\epsilon\left(\ket{100000}\bra{000000}-\ket{000000}\bra{100000}\right).
\end{align}
This shows that the population in $\ket{\varphi_0}$ is transferred to $\ket{\varphi_2}$ at a rate $4\epsilon^2/3$. The same rate holds generally for transitions from $\ket{\varphi_n}$ to $\ket{\varphi_{(n+2)\bmod 3}}$. We therefore expect exponential relaxation toward the equal-weight mixture with the same rate $4\epsilon^2/3$, consistent with the parametrically long relaxation time observed in Fig.~\ref{fig:oscillation}. See more numerical results in Appendix~\ref{sec:TEBD}.



The phenomenon is not restricted to the vacuum configuration.
To demonstrate this, we further consider period-five dynamics consisting of the following six-site shift-invariant classical states:
\begin{align}\label{eq:6-translation}
    \ket{\phi_0}=\ket{(001001)},&&\ket{\phi_1}=\ket{(100010)},\notag\\
    \ket{\phi_2}=\ket{(001000)},&&\ket{\phi_3}=\ket{(000001)},\notag\\
    \ket{\phi_4}=\ket{(010100)},&
\end{align}
and their subsystem configuration $\ket{\varphi_n}$ are defined accordingly.
The states evolve as
\[ \ket{\phi_{n+1\text{ mod }{5}}}=\mathbb{U}\ket{\phi_n},\quad n=0,1,2,3,4.\]
In the six-site subsystem, there is one pair of left- and right-moving quasiparticles. 

The corresponding influence matrix solution is not included in Table \ref{table:solutions}; instead, we report the explicit $\chi=19$ MPS representation in Appendix~\ref{sec:rule201,19}. Using this exact solution, we plot the evolution of matrix elements for the system initialized in $\ket{(001001)}$, and locations of quantum defects are indicated by the light-blue shaded region in Fig.~\ref{fig:local_dynamics}. As shown in the right panel in Fig.~\ref{fig:oscillation}, after a parametrically long time the reduced density matrix relaxes to the maximally mixed state within the two-soliton sector, similarly to the period-three case.

\begin{figure*}[ht]
    \hspace*{-0.8\linewidth}
    {\includegraphics[width=0.8\linewidth]{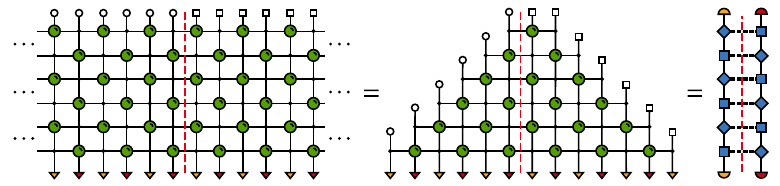}}
    \caption{
        Schematic of the R\'{e}nyi-$n$ entropy calculation. 
        Time steps $t=3$. The red dashed line indicates the location of the bipartition. 
        The black dashed lines on the right-hand side indicate the contraction for the $n$-layer folded tensors $A$ and $B$, as defined in Eqs.~\eqref{eq:renyi_n_1} and~\eqref{eq:renyi_n_2}.
    }
    \label{fig:renyi_n_contraction}
\end{figure*}


\subsection{Entanglement growth from global quench}\label{sec:entanglement_growth_global}

In quantum many-body systems, the entanglement growth starting from shortly correlated initial states is often of great interest, as it provides a diagnostic of ergodicity, scrambling, and thermalization. The Rule 201 quantum cellular automaton does not generate entanglement when evolved from classical configurations. In this section, we therefore consider initial states deformed away from classical configurations by a tilt parameter $\varepsilon$:
\begin{align}\label{eq:ini_state_renyi2}
    \ket{\Phi_{\varepsilon}} = \cdots\otimes\ket{0}\otimes\ket{v_\varepsilon}\otimes\ket{0}\otimes\ket{v_\varepsilon}\otimes\ket{0}\otimes\ket{v_\varepsilon}\otimes\cdots,
\end{align}
where $\ket{v_\varepsilon}=\sqrt{1-\varepsilon^2}\ket{0}+\varepsilon\ket{1}$.
For $\epsilon = 0$ or $1$, it reduces to a classical configuration with period-three dynamics.

\begin{figure*}[ht]
    \hspace*{-0.45\linewidth}
    {\includegraphics[width=0.45\linewidth]{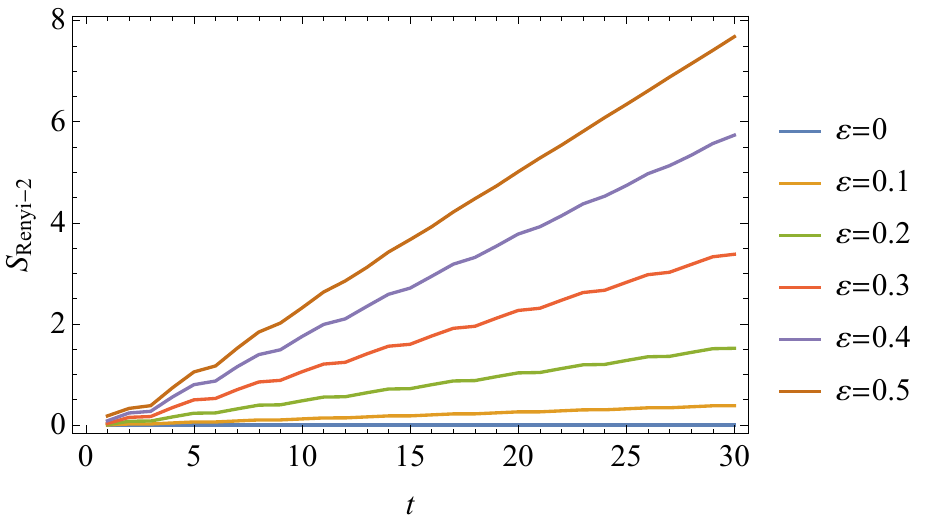}}\qquad\qquad
    {\includegraphics[width=0.39\linewidth]{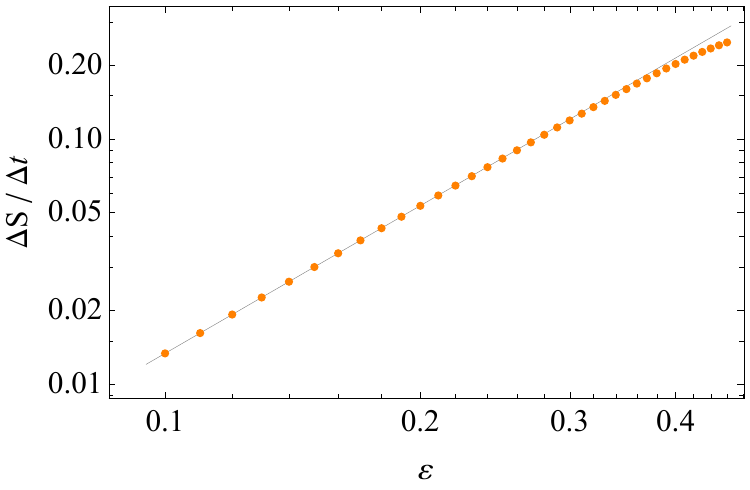}}
    \caption{
        Dependence of the bipartite R\'{e}nyi-2 entropy on time and parameter $\varepsilon$.
        The left panel shows the entropy growth in time.
        The right panel shows the quantity $(S(t=30)-S(t=15))/15$ as a function of $\varepsilon$, with the gray line $4
        \varepsilon^2/3$ shown for reference.
    }
    \label{fig:renyi2}
\end{figure*}

Here, we compute the bipartite R\'enyi entropy to characterize the entanglement growth. Ref.~\cite{Bertini2022Growth} developed a method for calculating R\'enyi entropies using an MPS representation of the influence matrix; here we briefly review the method. 

We consider a bipartition of the system into regions $L$ and $R$, separated by the red dashed line as shown in Fig.~\ref{fig:renyi_n_contraction}. The R\'{e}nyi-$n$ entropy $S_n(t)$ is then defined as 
\begin{equation*}
    S_n(t) = \frac{1}{1-n}\log\text{Tr}_R[ (\text{Tr}_L[\mathbb{U}^t\ket{\Phi_{\varepsilon}}\bra{\Phi_{\varepsilon}}{\mathbb{U}^\dagger}^{t}])^n].
\end{equation*}
To represent $n$ copies of the evolved state, we construct a $2n$-layer tensor network consisting of $n$ conjugate pairs. The different orderings of the partial traces are encoded in the upper boundary tensors, as illustrated in Fig.~\ref{fig:renyi_n_contraction}. Here, hollow circles and hollow squares denote
$\prod_{k=1}^{n}\delta_{i_ki'_k}$ and $\prod_{k=1}^{n}\delta_{i_{(k\text{ mod }n)+1}i_k'}$, respectively. 
On the two sides of the bipartition, the tensor network decomposes into independent sets of $n$ conjugate pairs, each of which can be contracted into $n$ copies of the influence matrix. The two sides differ by a one-layer translation, and therefore the contractions across the bipartition are given by

\begin{align}\label{eq:renyi_n_1}
    \sum_{k_1,k_1',\cdots,k_n,k_n'}\bigotimes_{m=1}^n\left[A^{(k_m,k_m')}\otimes B^{(k'_m,k_{(m\text{ mod }n)+1})}\right]
\end{align}
and
\begin{align}\label{eq:renyi_n_2}
    \sum_{k_1,k_1',\cdots,k_n,k_n'}\bigotimes_{m=1}^n\left[B^{(k_m,k_m')}\otimes A^{(k'_m,k_{(m\text{ mod }n)+1})}\right].
\end{align}
Therefore, once a $\chi$-dimensional influence matrix is obtained, the calculation of $S_n(t)$ only involves the sequential application of a $\chi^{2n}$-dimensional transfer matrix.

Fig.~\ref{fig:renyi2} shows the time evolution of R\'{e}nyi-$2$ entropy starting from  $\ket{\Phi_{\varepsilon}}$.
The left panel demonstrates linear entropy growth in time for different values of $\varepsilon$. In the right panel, we approximate the entanglement velocity $v_n:= \lim_{t\to\infty} S_n(t)/t$ by $\Delta S/\Delta t$ using data at $t=15$ and $t=30$.
For small $\varepsilon$, the result coincides with $4\varepsilon^2/3$.

This relation can be understood perturbatively. For clarity, we place the bipartition between $x=-1$ and $x=0$. From the tensor-network structure shown in Fig.~\ref{fig:renyi_n_contraction}, the initial degrees of freedom outside the inverse light cone, i.e., in the region $|x|>2t$, do not affect the quantity being evaluated.
It is therefore legitimate to replace the initial state by
\begin{align}
    \ket{\Phi'_\varepsilon}&=\ket{\cdots00}\otimes\left(\bigotimes_{k=-t}^{t-1}\ket{0}_{x=2k}\otimes\ket{v_\varepsilon}_{x=2k+1}\right)\otimes\ket{00\cdots}\notag\\
    &\approx(1-\varepsilon^2t)\ket{(00)}+\varepsilon\sum_{k=-t}^{t-1}\ket{\cdots0001_{x=2k+1}000\cdots}.
\end{align}

We now approximate the reduced density matrix of $\mathbb{U}^t\ket{\Phi'_{\varepsilon}}$.
From Sec.~\ref{sec:relaxation}, we know that the configuration $\ket{\cdots000010000\cdots}$ creates a pair of solitons propagating in opposite directions, each moving by two sites every three time steps. After $t\in 3\mathbb{Z}$ time steps, we denote the resulting configuration by
\begin{align*}
    \ket{l(x,t),r(x,t)}=\mathbb{U}^t\ket{\cdots00001_x0000\cdots},
\end{align*}
where $l=x-2t/3$ and $r=x+2t/3$ are the positions of the left- and right-moving solitons. 
Therefore, we have the evolved state
\begin{align}\label{eq:phi_t_renyi2}
    \mathbb{U}^t\ket{\Phi'_\varepsilon}\approx(1-\varepsilon^2t)\ket{(00)}+\varepsilon\sum_{k=-t}^{t-1}\ket{l,r}_{x=2k+1,t}.
\end{align}

Now we compute the reduced density matrix.
The $2t$ terms in the summation $\sum\ket{l,r}$ can be divided into three types, each containing approximately $2t/3$ terms: (i) both solitons lie in $L$; (ii) one soliton lies in $L$ and the other in $R$; and (iii) both solitons lie in $R$. Therefore, in the computational basis, the nonzero block of
$\rho_R(t)=\text{Tr}_L\left[\mathbb{U}^t\ket\phi\bra\phi\left(\mathbb{U}^\dagger\right)^t\right]$ can be expressed as
\begin{align}
    \rho_R(t)=\left(
    \begin{array}{c|ccc|ccc}
         1-4\varepsilon^2t/3&\varepsilon&\cdots&\varepsilon& & & \\
         \hline
         \varepsilon&\varepsilon^2&\cdots&\varepsilon^2& & & \\
         \vdots&\vdots&\ddots&\vdots& & & \\
         \varepsilon&\varepsilon^2&\cdots&\varepsilon^2& & & \\
         \hline
         & & & &\varepsilon^2& &\\
         & & & & &\ddots&\\
         & & & & & &\varepsilon^2\\
    \end{array}\right).
\end{align}
It follows that, to leading order in $\varepsilon$,
\begin{align}
1-\operatorname{Tr}\rho_R^2
\approx
\frac{4\varepsilon^2}{3}t,
\end{align}
and Rényi-$2$ entropy satisfies
\begin{align}
S_2(t)
=
-\log\operatorname{Tr}\rho_R^2
\approx
\frac{4\varepsilon^2}{3}t,
\end{align}
which agrees with numerical results.

\section{Hidden Markov order}\label{sec:markov_order}

\begin{figure*}[ht]
    \hspace*{-\linewidth}
    {\includegraphics[width=\linewidth]{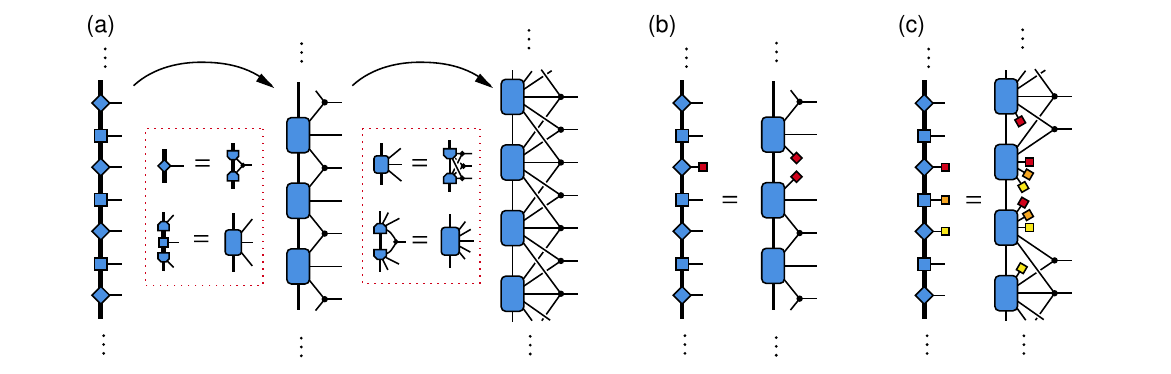}}
    \caption{(a) The transformation from a matrix product state to split-index matrix product states with three and five physical indices in each tensor. 
    (b,c) Measurements can reduce the bond dimension, as captured by the structure of the split-index matrix product states.
    }
    \label{fig:simps}
\end{figure*}

In previous sections, we have used exact representations of the influence matrix to compute observables and physical quantities in nonequilibrium dynamics. In this section, we turn to an information-theoretic perspective and use the exact solution as a concrete example to explore the memory structure encoded in the dynamics.

Ref.~\cite{Taranto2019Quantum} introduced the notion of quantum Markov order, which characterizes the finite memory length of a quantum stochastic process and is particularly well suited to the influence matrix framework. Specifically, one intervenes on the subsystem over an intermediate time window by applying a suitable sequence of instruments, such as unitary operations, measurements, or quantum channels. 
Contracting the corresponding physical legs of the influence matrix with these instruments leaves open legs only in the past and future temporal regions.
If, for a certain intermediate window of length $\ell$, the resulting conditional influence matrix factorizes into the past and the future parts, the process is said to have quantum Markov order $\ell$. 
In this sense, the quantum Markov order quantifies the length of memory encoded in the influence matrix.

We find that finite quantum Markov order can be revealed by converting the representation of influence matrices to
split-index matrix product states (SIMPS)~\cite{stephen2025nononsites}.
The basic idea is to rewrite an MPS of the form $\cdots\mathfrak a^{(l)}\mathfrak a^{(k)}\mathfrak a^{(j)}\mathfrak a^{(i)}\cdots$ to $\cdots\mathfrak m^{(l,k)}\mathfrak m^{(k,j)}\mathfrak m^{(j,i)}\cdots$,
where neighboring physical indices are grouped into one. 
Specifically, consider the MPS representation with an appropriate basis transformation:
\begin{align*}
    \cdots B^{(\nu)}A^{(\mu)}B^{(\lambda)}A^{(\kappa)}B^{(\gamma)}A^{(\beta)}B^{(\alpha)}\cdots
\end{align*}
with bond dimension $\chi$.
For a fixed physical index $\alpha$, $B^{(\alpha)}$ is a $\chi\times\chi$ matrix. If $\chi^{(\alpha)}\equiv\text{rank}\left[B^{(\alpha)}\right]<\chi$ for some $\alpha$, then it can be factorized as $B^{(\alpha)}=U^{(\alpha)}V^{(\alpha)}$ where $U^{(\alpha)}$ is a $\chi\times\chi^{(\alpha)}$ matrix and $V^{(\alpha)}$ is a $\chi^{(\alpha)}\times\chi$ matrix.
This gives the split-index representation
\begin{align}
&\cdots B^{(\nu)}A^{(\mu)}B^{(\lambda)}A^{(\kappa)}B^{(\gamma)}A^{(\beta)}B^{(\alpha)}\cdots\notag\\
&=\cdots M^{(\nu\mu\lambda)}M^{(\lambda\kappa\gamma)}M^{(\gamma\beta\alpha)}\cdots,
\end{align}
where the matrix $M^{(\gamma\beta{\alpha})}=V^{(\gamma)}A^{(\beta)}U^{(\alpha)}$ is of dimension $\chi^{(\gamma)}\times \chi^{(\alpha)}$.


This procedure can be iterated if the split-index tensors have further rank deficiencies. Suppose, for some index tuple $(\gamma,\beta,\alpha)$, there is $\chi^{(\gamma\beta\alpha)}\equiv\text{rank}\left[M^{(\gamma\beta\alpha)}\right]<\min\{\chi^{(\alpha)},\chi^{(\gamma)}\}$, we may then factorize it as $M^{(\gamma\beta\alpha)}=U^{(\gamma\beta\alpha)}V^{(\gamma\beta\alpha)}$, where $U^{(\gamma\beta\alpha)}$ is a $\chi^{(\gamma)}\times\chi^{(\gamma\beta\alpha)}$ matrix and $V^{(\gamma\beta\alpha)}$ is a $\chi^{(\gamma\beta\alpha)}\times\chi^{(\alpha)}$ matrix.
The MPS can then be rewritten as
\begin{align}\label{eq:SIMPS2}
&\cdots M^{(\nu\mu\lambda)}M^{(\lambda\kappa\gamma)}M^{(\gamma\beta\alpha)}\cdots\notag\\
&=\cdots M^{(\nu\mu\lambda\kappa\gamma)}M^{(\lambda\kappa\gamma\beta\alpha)}\cdots,
\end{align}
where $M^{(\lambda\kappa\gamma\beta\alpha)}=V^{(\lambda\kappa\gamma)}U^{(\gamma\beta\alpha)}$.
Whenever additional rank reductions occur, the same factorization step can be repeated, leading to split-index tensors with progressively longer physical indices.

The diagrammatic representation of the above procedure is shown in Fig.~\ref{fig:simps}(a).
The black dots denote the Kronecker delta operator, satisfying the relations
\begin{equation}
    \begin{tikzpicture}[x=0.75pt,y=0.75pt,yscale=-1,xscale=1]

\draw    (36.17,14.17) -- (51.17,29.17) ;
\draw [fill={rgb, 255:red, 255; green, 255; blue, 255 }  ,fill opacity=1 ]   (51.17,29.17) -- (62.17,29.17) ;
\draw  [fill={rgb, 255:red, 0; green, 0; blue, 0 }  ,fill opacity=1 ] (51.17,27.67) .. controls (52,27.67) and (52.67,28.34) .. (52.67,29.17) .. controls (52.67,30) and (52,30.67) .. (51.17,30.67) .. controls (50.34,30.67) and (49.67,30) .. (49.67,29.17) .. controls (49.67,28.34) and (50.34,27.67) .. (51.17,27.67) -- cycle ;
\draw    (36.17,44.17) -- (51.17,29.17) ;
\draw  [fill={rgb, 255:red, 208; green, 2; blue, 27 }  ,fill opacity=1 ] (59.67,26.67) -- (64.67,26.67) -- (64.67,31.67) -- (59.67,31.67) -- cycle ;
\draw    (90.17,14.17) -- (100.17,24.17) ;
\draw    (90.17,44.17) -- (100.17,34.17) ;
\draw  [fill={rgb, 255:red, 208; green, 2; blue, 27 }  ,fill opacity=1 ] (100.17,20.63) -- (103.7,24.17) -- (100.17,27.7) -- (96.63,24.17) -- cycle ;
\draw  [fill={rgb, 255:red, 208; green, 2; blue, 27 }  ,fill opacity=1 ] (100.17,30.63) -- (103.7,34.17) -- (100.17,37.7) -- (96.63,34.17) -- cycle ;

\draw (73.67,25.73) node [anchor=north west][inner sep=0.75pt]    {$=$};

\end{tikzpicture},\qquad
\begin{tikzpicture}[x=0.75pt,y=0.75pt,yscale=-1,xscale=1]

\draw    (41.17,22.17) -- (52.17,33.17) ;
\draw [fill={rgb, 255:red, 255; green, 255; blue, 255 }  ,fill opacity=1 ]   (37.17,33.17) -- (63.17,33.17) ;
\draw  [fill={rgb, 255:red, 0; green, 0; blue, 0 }  ,fill opacity=1 ] (52.17,31.67) .. controls (53,31.67) and (53.67,32.34) .. (53.67,33.17) .. controls (53.67,34) and (53,34.67) .. (52.17,34.67) .. controls (51.34,34.67) and (50.67,34) .. (50.67,33.17) .. controls (50.67,32.34) and (51.34,31.67) .. (52.17,31.67) -- cycle ;
\draw    (41.17,44.17) -- (52.17,33.17) ;
\draw  [fill={rgb, 255:red, 208; green, 2; blue, 27 }  ,fill opacity=1 ] (60.67,30.67) -- (65.67,30.67) -- (65.67,35.67) -- (60.67,35.67) -- cycle ;
\draw    (99.17,18.17) -- (107.17,26.17) ;
\draw [fill={rgb, 255:red, 255; green, 255; blue, 255 }  ,fill opacity=1 ]   (91.17,33.17) -- (102.17,33.17) ;
\draw    (99.17,48.17) -- (107.17,40.17) ;
\draw  [fill={rgb, 255:red, 208; green, 2; blue, 27 }  ,fill opacity=1 ] (107.17,22.63) -- (110.7,26.17) -- (107.17,29.7) -- (103.63,26.17) -- cycle ;
\draw  [fill={rgb, 255:red, 208; green, 2; blue, 27 }  ,fill opacity=1 ] (107.17,36.63) -- (110.7,40.17) -- (107.17,43.7) -- (103.63,40.17) -- cycle ;
\draw  [fill={rgb, 255:red, 208; green, 2; blue, 27 }  ,fill opacity=1 ] (99.67,30.67) -- (104.67,30.67) -- (104.67,35.67) -- (99.67,35.67) -- cycle ;

\draw (74.67,29.73) node [anchor=north west][inner sep=0.75pt]    {$=$};

\end{tikzpicture},
\end{equation}
where the colored square operators denote the basis states of measurements.
As shown in Figs.~\ref{fig:simps}(b) and (c), applying measurements to an intermediate time window induces corresponding reductions in the effective bond dimension of the influence matrix. In particular, measuring a single site reduces the bond dimension from $\chi$ to $\chi^{(\alpha)}$, while measuring three adjacent sites reduces it further to $\chi^{(\gamma\beta\alpha)}$. 

Therefore, if the bond dimension of the SIMPS representation can be reduced to one after grouping finite neighboring indices, it directly implies that the corresponding state factorizes upon measuring a finite number of intermediate sites, giving rise to a finite quantum Markov order. 
We find that the $\chi=3$ solution of Rule 54 realizes this scenario precisely.
Although an analogous SIMPS for the classical Rule 54 cellular automaton was discussed in Ref.~\cite{Klobas2020Matrix}, the present construction provides a quantum example in the sense of Ref.~\cite{Taranto2019Quantum}, with quantum Markov order $\ell=3/2$ time steps.

This suggests a natural generalization to cases where the bond dimension of the SIMPS cannot be reduced any further. We refer to this case as \textit{hidden Markov order}: by regarding the virtual degree of freedom associated with the remaining bond dimension as an ancilla coupled to the subsystem, the joint system of the ancilla and the subsystem has a finite quantum Markov order. 
For instance, for the $\chi=6$ MPS solution of the influence matrix in the Rule 54 quantum cellular automaton, the bond dimension is reduced from $6$ to $2$ at $\Delta t=3/2$, but cannot be further reduced to $1$ by increasing $\Delta t$. See Appendix~\ref{app:hidden_Markov} for more details.
This reveals a finer structure of the influence matrix: the memory effect separates into short-range and long-range components, corresponding respectively to virtual degrees of freedom that vanish under measurements and those that remain. A related structure, organizing memory decomposition by range, can also be found in tree-geometry influence matrices~\cite{Dowling2024Capturing}.


\section{Conclusions and outlook}\label{sec:conclusion}

In summary, we introduced a set of tensor relations that provide sufficient conditions for exactly solving influence matrices. Combined with a numerical bootstrap algorithm, we obtained exact solutions for the Rule 201 quantum cellular automaton. Based on the solutions, we investigated the local nonequilibrium dynamics of the model.
First, we revealed the relaxation of periodic classical dynamics under quantum perturbations. We then studied the entanglement entropy growth following global quenches. Finally, we introduced the notion of hidden Markov order as a characterization of temporal memory encoded in the influence matrix.

It is also instructive to compare the solutions developed here with those obtained from the zipper conditions in Refs.~\cite{Klobas2021Exact,Klobas2021ExactII}. For models where both approaches apply, the resulting solutions are compatible. However, our results suggest that there exist cases in which the conventional zipper conditions are not applicable, whereas the generalized zipper conditions are valid. Establishing a unified framework that clarifies the relation between these two constructions is an important direction for future work.

Another promising direction is to extend the present framework to cases in which the influence-matrix bond dimension grows polynomially with time~\cite{Giudice2022Temporal,Wang2025Temporal}. Such an extension may require the development of new methods for learning MPS representations with a time-dependent bond dimension. This would substantially enlarge the class of models accessible within the present systematic framework.

More broadly, our analytical and numerical methods can be applied to the search for additional solvable influence matrices. Once finite-time numerical data are obtained, the bootstrap method provides both a route for extending them to larger times and a strong consistency check for exact solutions. These methods therefore open a path toward identifying new solvable structures and exploring a broader class of nonequilibrium quantum phenomena.
\\



\textit{Note added}: After the first arXiv posting of this work, we became aware of Ref.~\cite{Klobas2026ExactSubsystem}, which derives exact solutions for classical initial states using ``zipping conditions'' developed in Ref.~\cite{Klobas2021Exact}. Our bootstrap approach additionally applies to quantum initial states and provides a systematic route for discovering exact influence matrices with a priori unknown finite bond dimension.

\begin{acknowledgements}
X.~Y.~Y. thanks Bal\'azs Pozsgay for helpful discussions. This work was supported by the National Natural Science Foundation of China (Grant No. 12125405) and the National Key R$\&$D Program of China (No. 2023YFA1406702).
\end{acknowledgements}

\section*{Data Availability}
The data that support the findings of this manuscript are not publicly available. The data are available from the authors upon reasonable request.

\appendix

\section{Rule 201 quantum cellular automaton solutions}\label{sec:solution_list}

\begin{figure*}[ht]
    \hspace*{-0.8\linewidth}
    {\includegraphics[width=0.8\linewidth]{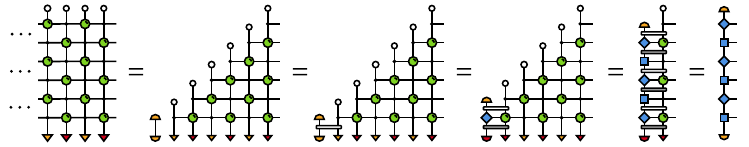}}
    \caption{
        Contractions of the tensor network. Number of time steps $t=3$. 
        The unitary condition, the generalized zipper conditions and Eq.~\eqref{eq:contraction_additional} are applied.
    }
    \label{fig:contractions}
\end{figure*}

In this appendix, we present the explicit form of exact solutions in the Rule 201 quantum cellular automaton.
Fig.~\ref{fig:contractions} illustrates how a semi-infinite tensor network is contracted into an MPS with bond dimension $\chi$.
In the first equality, we use unitarity to contract the circuit to within the light cone, and insert $(\text{t}|\text{b})=1$ on the left side of the network.
We then successively apply the generalized zipper conditions, together with an additional relation
\begin{equation}\label{eq:contraction_additional}
\begin{tikzpicture}[x=0.75pt,y=0.75pt,yscale=-1,xscale=1]

\draw [color={rgb, 255:red, 0; green, 0; blue, 0 }  ,draw opacity=1 ]   (26.61,38.05) -- (26.61,13.39) ;
\draw [color={rgb, 255:red, 0; green, 0; blue, 0 }  ,draw opacity=1 ]   (11.92,37.64) -- (11.92,21.79) ;
\draw  [fill={rgb, 255:red, 245; green, 166; blue, 35 }  ,fill opacity=1 ] (8.25,22.91) .. controls (8.25,22.91) and (8.25,22.91) .. (8.25,22.91) .. controls (8.25,20.89) and (9.89,19.24) .. (11.92,19.24) .. controls (13.95,19.24) and (15.6,20.89) .. (15.6,22.91) -- cycle ;
\draw  [fill={rgb, 255:red, 255; green, 255; blue, 255 }  ,fill opacity=1 ] (24.41,13.39) .. controls (24.41,12.17) and (25.4,11.19) .. (26.61,11.19) .. controls (27.83,11.19) and (28.82,12.17) .. (28.82,13.39) .. controls (28.82,14.61) and (27.83,15.59) .. (26.61,15.59) .. controls (25.4,15.59) and (24.41,14.61) .. (24.41,13.39) -- cycle ;
\draw  [fill={rgb, 255:red, 255; green, 255; blue, 255 }  ,fill opacity=1 ] (9.72,31.29) .. controls (9.72,31.13) and (9.85,30.99) .. (10.01,30.99) -- (28.52,30.99) .. controls (28.69,30.99) and (28.82,31.13) .. (28.82,31.29) -- (28.82,32.17) .. controls (28.82,32.33) and (28.69,32.46) .. (28.52,32.46) -- (10.01,32.46) .. controls (9.85,32.46) and (9.72,32.33) .. (9.72,32.17) -- cycle ;
\draw    (26.61,23.51) -- (35.63,23.49) ;
\draw   (25.88,23.51) .. controls (25.88,23.11) and (26.21,22.78) .. (26.61,22.78) .. controls (27.02,22.78) and (27.35,23.11) .. (27.35,23.51) .. controls (27.35,23.92) and (27.02,24.25) .. (26.61,24.25) .. controls (26.21,24.25) and (25.88,23.92) .. (25.88,23.51) -- cycle ;
\draw [color={rgb, 255:red, 0; green, 0; blue, 0 }  ,draw opacity=1 ]   (60.98,31.68) -- (60.98,13.31) ;
\draw  [fill={rgb, 255:red, 245; green, 166; blue, 35 }  ,fill opacity=1 ] (57.31,13.31) .. controls (57.31,13.31) and (57.31,13.31) .. (57.31,13.31) .. controls (57.31,11.29) and (58.95,9.64) .. (60.98,9.64) .. controls (63.01,9.64) and (64.65,11.29) .. (64.65,13.31) -- cycle ;
\draw    (60.98,24.31) -- (76.61,24.31) ;
\draw [color={rgb, 255:red, 0; green, 0; blue, 0 }  ,draw opacity=1 ]   (68.33,36.93) -- (68.33,31.66) ;
\draw [color={rgb, 255:red, 0; green, 0; blue, 0 }  ,draw opacity=1 ]   (53.63,36.75) -- (53.63,31.65) ;
\draw  [fill={rgb, 255:red, 255; green, 255; blue, 255 }  ,fill opacity=1 ] (51.43,32.41) -- (52.65,30.94) -- (69.3,30.94) -- (70.53,32.41) -- cycle ;
\draw  [fill={rgb, 255:red, 74; green, 144; blue, 226 }  ,fill opacity=1 ] (60.98,20.15) -- (65.13,24.31) -- (60.98,28.46) -- (56.82,24.31) -- cycle ;

\draw (37.39,18.69) node [anchor=north west][inner sep=0.75pt]    {$=$};

\end{tikzpicture}.
\end{equation}

We now present several cases where all tensors in the solution and in the generalized zipper conditions have analytical forms.
We denote the basis of the auxiliary linear space $\mathcal U$ of dimension $\chi$ as
\[
\{|a)=\boldsymbol{e}_a\in\mathbb C^\chi\mid a=0,1,\cdots,\chi-1\,\},
\]
the basis of the folded Hilbert space $\mathcal H\otimes\mathcal H^*$ on one spin-$1/2$ as
\[
\{\,|i\rangle\rangle=\boldsymbol{e}_i\in\mathbb C^4\mid i=0,1,2,3\,\},
\]
and thus the basis of $\mathcal U\otimes\mathcal H\otimes\mathcal H^*$ as
\[
\{\,|a,i)=|a)\otimes|i\rangle\rangle\mid (a,i)=(0,0),(0,1),\cdots,(\chi-1,3)\,\}.
\]
In this appendix, we follow the main text and use the shorthand notation for two-site shift-invariant initial states:
\begin{equation*}
    \ket{(i_1i_2)} = \cdots\otimes\ket{i_1}\otimes\ket{i_2}\otimes\ket{i_1}\otimes\ket{i_2}\otimes\cdots
\end{equation*}

\subsection{$\chi=12$}\label{sec:rule201,12}
The following solutions work for initial states including $\ket{(00)}$, $\ket{(01)}$, and $\ket{(10)}$.

$A$ and $B$ tensors:
\begin{align*}
A^{(00)} =
\left(\begin{array}{cccc}
{p}_1&&&{p}_3\\
&&{p}_2&\\
&{p}_2&&\\
{p}_2&&&\\
\end{array}
\right) ,
\end{align*}
\begin{align*}
A^{(01)} =
\left(\begin{array}{cccc}
{p}_1&&{p}_2&\\
&{p}_1&&{p}_2\\
{p}_2&&&\\
&{p}_2&&\\
\end{array}
\right) ,
\end{align*}
\begin{align*}
A^{(10)} =
\left(\begin{array}{cccc}
{p}_1&{p}_2&&\\
{p}_2&&&\\
&&{p}_1&{p}_2\\
&&{p}_2&\\
\end{array}
\right) ,
\end{align*}
\begin{align*}
A^{(11)} =
\left(\begin{array}{cccc}
{p}_0&&&\\
&{p}_0&&\\
&&{p}_0&\\
&&&{p}_0\\
\end{array}
\right) ,
\end{align*}
\begin{align*}
B^{(00)} =\operatorname{}
\left(\begin{array}{cccc}
I& & & \\
 &O& & \\
 & &O& \\
 & & &O\\
\end{array}
\right) ,&&
B^{(01)} =
\left(\begin{array}{cccc}
O& & & \\
 &I& & \\
 & &O& \\
 & & &O\\
\end{array}
\right) ,
\end{align*}
\begin{align*}
B^{(10)} =
\left(\begin{array}{cccc}
O& & & \\
 &O& & \\
 & &I& \\
 & & &O\\
\end{array}
\right) ,&&
B^{(11)} =
\left(\begin{array}{cccc}
O& & & \\
 &O& & \\
 & &O& \\
 & & &I\\
\end{array}
\right) ,
\end{align*}
where the $3\times3$ matrices are defined as
\begin{align*}
    I=\begin{pmatrix}1&&\\&1&\\&&1\end{pmatrix},&&
    O=\begin{pmatrix}0&&\\&0&\\&&0\end{pmatrix},&&
    {p}_0=\begin{pmatrix}&&1\\1&&\\&1&\end{pmatrix},\\
    {p}_1=\begin{pmatrix}&&0\\1&&\\&0&\end{pmatrix},&&
    {p}_2=\begin{pmatrix}&&1\\0&&\\&1&\end{pmatrix},&&
    {p}_3=\begin{pmatrix}1&&1\\0&&\\&1&\end{pmatrix}.
\end{align*}
$P_0$ and $P_0'$ are the projectors onto the subspaces $\mathcal{V}_0, \mathcal{V}'_0\in\mathcal U\otimes\mathcal H\otimes\mathcal H^*$:
\begin{align*}
    \mathcal{V}_0=\text{span}\{&|0,0),|1,0),|2,0),|3,0),|5,0),|6,0),\\
    &|8,0),|9,0),|0,2),|1,2),|11,0),|0,1),\\
    &|1,1),|2,1),|6,1),|8,1),|2,2),|3,2),\\
    &|5,2),|0,3),|1,3),|2,3)\},
\end{align*}
\begin{align*}
    \mathcal{V}'_0=\text{span}\{&|0,0),|1,0),|3,0),|6,0),|9,0),|11,0),\\
    &|0,1),|1,1),|6,1),|8,1),|0,2),|1,2),\\
    &|3,2),|5,2),|0,3),|1,3),|2,3)\}.
\end{align*}
$P_1$ and $P_1'$ are mappings from $U\otimes\mathcal H\otimes\mathcal H^*$ to $\mathcal U$:
\begin{align*}
    P_1:\,&|0,0)\mapsto|2),&&|1,0)\mapsto|0),&&|2,0)\mapsto|0),\\
    &|9,0)\mapsto|1),&&|11,0)\mapsto|1),&&|0,1)\mapsto|5),\\
    &|1,1)\mapsto|3),&&|8,1)\mapsto|4),&&|0,2)\mapsto|8),\\
    &|1,2)\mapsto|6),&&|5,2)\mapsto|7),&&|0,3)\mapsto|11),\\
    &|1,3)\mapsto|9),&&|2,3)\mapsto|10),&&\text{others}\mapsto 0,\\
\end{align*}
\begin{align*}
    P'_1:\,&|0,0)\mapsto|1),&&|1,0)\mapsto|2),&&|9,0)\mapsto|0),\\
    &|11,0)\mapsto|0),&&|0,1)\mapsto|4),&&|1,1)\mapsto|5),\\
    &|8,1)\mapsto|3),&&|0,2)\mapsto|7),&&|1,2)\mapsto|8),\\
    &|5,2)\mapsto|6),&&|0,3)\mapsto|10),&&|1,3)\mapsto|11),\\
    &|2,3)\mapsto|9),&& \text{others}\mapsto 0.
\end{align*}
Vectors at the top boundary are given by
\begin{align*}
    (\text{t}|=(\text{t}'|=(1, 1, 1, 0, 0, 0, 0, 0, 0, 1, 1, 1).
\end{align*}
Finally, bottom boundary vectors depend on initial states as:

a) for $|\rho\rangle\rangle=(1,0,0,0)^T$, $|\rho'\rangle\rangle=(1,0,0,0)^T$,
\begin{align*}
    |\text{b})=|1),&&|\text{b}')=|0),
\end{align*}

b) for $|\rho\rangle\rangle=(0,0,0,1)^T$, $|\rho'\rangle\rangle=(1,0,0,0)^T$,
\begin{align*}
    |\text{b})=|0),&&|\text{b}')=|11),
\end{align*}

c) for $|\rho\rangle\rangle=(1,0,0,0)^T$, $|\rho'\rangle\rangle=(0,0,0,1)^T$,
\begin{align*}
    |\text{b})=|11),&&|\text{b}')=|1).
\end{align*}

\subsection{ $\chi=24$}\label{sec:rule201,24}
The following solutions work for two-site shift-invariant initial states with the block $|0\rangle\langle 0| \otimes \left(\begin{array}{cc}
    \alpha & 0 \\
    0 & \delta
\end{array}\right)$.

$A$ and $B$ tensors:
\begin{align*}
A^{(00)} =
\left(\begin{array}{cccc}
{q}_1&&&{q}_3\\
&&{q}_2&\\
&{q}_2&&\\
{q}_2'&&&\\
\end{array}
\right) ,
\end{align*}
\begin{align*}
A^{(01)} =
\left(\begin{array}{cccc}
{q}_1&&{q}_2&\\
&{q}_1&&{q}_2\\
{q}_2&&&\\
&{q}_2&&\\
\end{array}
\right) ,
\end{align*}
\begin{align*}
A^{(10)} =
\left(\begin{array}{cccc}
{q}_1&{q}_2&&\\
{q}_2&&&\\
&&{q}_1&{q}_2\\
&&{q}_2&\\
\end{array}
\right) ,
\end{align*}
\begin{align*}
A^{(11)} =
\left(\begin{array}{cccc}
{q}_0'&&&\\
&{q}_0&&\\
&&{q}_0&\\
&&&{q}_0\\
\end{array}
\right) ,
\end{align*}
\begin{align*}
B^{(00)} =
\left(\begin{array}{cccc}
{g}& & & \\
 &\mathcal O& & \\
 & &\mathcal O& \\
 & & &\mathcal O\\
\end{array}
\right) ,&&
B^{(01)} =
\left(\begin{array}{cccc}
\mathcal O& & & \\
 &{g}& & \\
 & &\mathcal O& \\
 & & &\mathcal O\\
\end{array}
\right) ,
\end{align*}
\begin{align*}
B^{(10)} =
\left(\begin{array}{cccc}
\mathcal O& & & \\
 &\mathcal O& & \\
 & &{g}& \\
 & & &\mathcal O\\
\end{array}
\right) ,&&
B^{(11)} =
\left(\begin{array}{cccc}
\mathcal O& & & \\
 &\mathcal O& & \\
 & &\mathcal O& \\
 & & &{g}'\\
\end{array}
\right) ,
\end{align*}
where $6\times6$ matrices are defined as
\begin{align*}
    {q}_0=\begin{pmatrix}&&I\\I&&\\&I&\end{pmatrix},&&
    {q}_0'=\begin{pmatrix}&J&I\\I&&\\&I&\end{pmatrix},&&
    {q}_1=\begin{pmatrix}&&O\\I&&\\&O&\end{pmatrix},\\
    {q}_2=\begin{pmatrix}&&I\\O&&\\&I&\end{pmatrix},&&
    {q}_2'=\begin{pmatrix}&&I\\O&J&\\&I&\end{pmatrix},&&
    {q}_3=\begin{pmatrix}I&&I\\O&&\\&I&\end{pmatrix},    
\end{align*}
\begin{align*}
    \mathcal O=\begin{pmatrix}O&&\\&O&\\&&O\end{pmatrix},
    I=\begin{pmatrix}1&\\&1\end{pmatrix},
    J=\begin{pmatrix}0&\\&1\end{pmatrix},
    O=\begin{pmatrix}0&\\&0\end{pmatrix},
\end{align*}
\begin{align*}
    {g}=\begin{pmatrix}
    0&\alpha\delta/(\alpha+\delta)^2&0&0&0&0\\
    -1&1&0&0&0&0\\
    0&0&0&1&0&0\\
    0&0&1&0&0&0\\
    0&0&0&0&0&1\\
    0&0&0&0&1&0
\end{pmatrix},
\end{align*}
\begin{align*}
    {g}'=\begin{pmatrix}
    0&\alpha\delta/(\alpha+\delta)^2&0&0&0&0\\
    -1&1&0&0&0&0\\
    0&\alpha\delta/(\alpha+\delta)^2&0&1&0&0\\
    0&0&1&0&0&0\\
    0&0&0&0&0&1\\
    0&0&0&0&1&0
\end{pmatrix}.
\end{align*}
\begin{widetext}
$P_0$ and $P'_0$ tensors:
\begin{align*}
(P_0)_{bj,ai} &= 
\begin{cases}
1 & \text{if } (b,j,a,i) \in \\
  & \{(0,0,0,0), (0,1,0,1), (0,2,0,2), (0,3,0,3), (1,0,1,0), (1,1,1,1), (1,2,1,2), (1,3,1,3), \\
  & (2,0,2,0), (2,1,2,1), (2,2,2,2), (2,3,2,3), (3,0,3,0), (3,1,3,1), (3,2,3,2), (3,3,3,3), \\
  & (4,0,4,0), (4,1,4,1), (4,2,4,2), (4,3,4,3), (5,0,5,0), (5,1,5,1), (5,2,5,2), (5,3,5,3), \\
  & (6,0,6,0), (6,2,6,2), (7,0,7,0), (7,2,7,2), (10,0,10,0), (10,2,10,2), (11,0,11,0), \\
  & (11,2,11,2), (12,0,12,0), (12,1,12,1), (13,0,13,0), (13,1,13,1), (16,0,16,0), (16,1,16,1), \\
  & (17,0,17,0), (17,1,17,1), (18,0,20,0), (19,0,19,0), (20,0,20,0), (22,0,22,0), (23,0,23,0)\}, \\
-1 & \text{if } (b,j,a,i) = (18,0,22,0), \\
0 & \text{otherwise},
\end{cases}
\end{align*}
\begin{align*}
(P_0')_{bj,ai} &= 
\begin{cases}
1 & \text{if } (b,j,a,i) \in \\
  & \{(0,0,4,0), (0,1,0,1), (0,2,0,2), (0,3,0,3), (1,0,1,0), (1,1,1,1), (1,2,1,2), (1,3,1,3),\\
  & (2,0,2,0), (2,1,2,1), (2,2,2,2), (2,3,2,3), (3,0,3,0), (3,1,3,1), (3,2,3,2), (3,3,3,3), \\
  & (4,0,4,0), (4,3,4,3), (5,3,5,3), (6,0,6,0), (6,2,6,2), (7,0,7,0), (7,2,7,2), (10,2,10,2),\\
  & (11,2,11,2), (12,0,12,0), (12,1,12,1), (13,0,13,0), (13,1,13,1), (16,1,16,1), (17,1,17,1), \\
  & (18,0,18,0), (19,0,19,0), (21,0,23,0), (22,0,22,0), (23,0,23,0)\}, \\
0 & \text{otherwise}.
\end{cases}
\end{align*}
$P_1$ and $P'_1$ tensors:
\begin{align*}
(P_1)_{b,ai}& = 
\begin{cases}
\dfrac{(\alpha+\delta)^2}{\alpha\delta} & \text{if } (b,a,i) \in  \{(0,3,0), (1,3,0), (4,0,0), (5,0,0), (6,3,1), (7,3,1), (10,0,1), (11,0,1), 
\\&(12,3,2), (13,3,2), (16,0,2), (17,0,2), (18,3,3), (19,3,3), (22,0,3), (23,0,3) \}, \\
1 & \text{if } (b,a,i) \in \{ (0,4,0), (1,5,0), (2,20,0), (3,19,0), (3,23,0), (8,16,1), (9,17,1),\\
&(14,10,2), (15,11,2), (20,4,3), (21,5,3)\}, \\
-1 & \text{if } (b,a,i) \in\{(0,2,0), (5,1,0), (6,2,1), (11,1,1), (12,2,2), (17,1,2), (18,2,3), (23,1,3) \}, \\
0 & \text{otherwise},
\end{cases}
\end{align*}
\begin{align*}
(P_1')_{b,ai} &= 
\begin{cases}
\dfrac{(\alpha+\delta)^2}{\alpha\delta} & \text{if } (b,a,i) \in  \{(4,2,0), (5,2,0), (10,2,1), (11,2,1), (16,2,2), (17,2,2), (22,2,3), (23,2,3) \}, \\
\dfrac{\alpha\delta}{(\alpha+\delta)^2} & \text{if } (b,a,i)=(0, 19, 0), \\
1 & \text{if } (b,a,i) \in \{ (0,23,0), (1,19,0), (1,22,0), (2,1,0), (3,4,0), (6,17,1), (7,16,1), (8,1,1),\\
& (9,0,1), (12,11,2), (13,10,2), (14,1,2), (15,0,2), (18,5,3), (19,4,3), (20,1,3), (21,0,3)\}, \\
-1 & \text{if } (b,a,i) \in\{(1,18,0), (4,3,0), (10,3,1), (16,3,2), (22,3,3) \}, \\
0 & \text{otherwise}.
\end{cases}\notag
\end{align*}
Vectors at the top boundary:
\begin{align*}
    (\text{t}|=(0,1,1,0,-1,1,0,0,0,0,0,0,0,0,0,0,0,0,-1,1,1,-1,-1,1),
\end{align*}
\begin{align*}
    (\text{t}'|=(-1,1,0,1,1,-1,0,0,0,0,0,0,0,0,0,0,0,0,-1,1,-1,1,1,-1).
\end{align*}
Bottom boundary vectors which depend on initial states:

a) for $|\rho\rangle\rangle=(1, 0, 0, 0)^T$, $|\rho'\rangle\rangle=\frac{1}{\alpha+\delta}(\alpha, 0, 0, \delta)^T$,
\begin{align}
    |\text{b})&=\frac{\alpha}{\alpha+\delta}\left(0,0,\frac{\alpha}{\alpha+\delta},0,\frac{\alpha}{\alpha+\delta},1,0,0,0,0,0,0,0,0,0,0,0,0,0,0,\frac{\delta}{\alpha+\delta},0,\frac{\delta}{\alpha+\delta},\frac{\delta}{\alpha}\right)^T,\notag\\
    |\text{b}')&=\frac{\alpha}{\alpha+\delta}\left(0,1,\frac{\delta}{\alpha+\delta},\frac{\delta}{\alpha},0,0,0,0,0,0,0,0,0,0,0,0,0,0,0,0,0,0,0,0\right)^T,\notag
\end{align}

b) for $|\rho\rangle\rangle=\frac{1}{\alpha+\delta}(\alpha, 0, 0, \delta)^T$, $\quad|\rho'\rangle\rangle=(1,0,0,0)^T$,
\begin{align}
    |\text{b})&=\frac{\delta}{\alpha+\delta}\left(\frac{\alpha\delta}{(\alpha+\delta)^2},1,\frac{\alpha}{\delta},\frac{\alpha}{\alpha+\delta},0,0,0,0,0,0,0,0,0,0,0,0,0,0,0,0,0,0,0,0\right)^T,\notag\\
    |\text{b}')&=\frac{\delta}{\alpha+\delta}\left(\frac{\alpha}{\alpha+\delta},\frac{\alpha}{\delta},0,0,\frac{\alpha}{\alpha+\delta},0,0,0,0,0,0,0,0,0,0,0,0,0,\frac{\delta}{\alpha+\delta},1,0,0,\frac{\delta}{\alpha+\delta},0\right)^T.\notag
\end{align}    

\subsection{ $\chi=19$}\label{sec:rule201,19}
In this section, we present exact solutions for six-site shift-invariant initial states defined in Eq.~\eqref{eq:6-translation}.

$A$ and $B$ tensors:
\begin{equation*}
    A^{(00)} = \left(
\begin{array}{ccccccccccccccccccc}
 0 & 0 & 0 & 0 & 0 & 0 & 0 & 0 & 0 & 0 & 0 & 0 & 0 & 0 & 0 & 1 & 1 & 0 & 0 \\
 0 & 0 & 0 & 0 & 0 & 0 & 0 & 0 & 0 & 0 & 0 & 0 & 0 & 0 & 0 & 0 & 0 & 0 & 1 \\
 0 & 0 & 0 & 0 & 0 & 0 & 0 & 0 & 0 & 0 & 0 & 0 & 0 & 0 & 0 & 0 & 0 & 1 & 0 \\
 0 & 1 & 0 & 0 & 0 & 0 & 0 & 0 & 0 & 0 & 0 & 0 & 0 & 0 & 0 & 0 & 0 & 0 & 0 \\
 1 & 0 & 0 & 0 & 0 & 0 & 0 & 0 & 0 & 0 & 0 & 0 & 0 & 0 & 0 & 0 & 0 & 0 & 0 \\
 0 & 0 & 0 & 0 & 0 & 0 & 0 & 0 & 0 & 0 & 0 & 0 & 0 & 0 & 1 & 0 & 0 & 0 & 0 \\
 0 & 0 & 0 & 0 & 0 & 0 & 0 & 0 & 0 & 0 & 0 & 0 & 1 & 0 & 0 & 0 & 0 & 0 & 0 \\
 0 & 0 & 0 & 0 & 0 & 0 & 0 & 0 & 0 & 0 & 0 & 0 & 0 & 1 & 0 & 0 & 0 & 0 & 0 \\
 0 & 0 & 0 & 0 & 0 & 0 & 0 & 0 & 0 & 0 & 0 & 0 & 0 & 0 & 0 & 0 & 0 & 0 & 0 \\
 0 & 0 & 0 & 0 & 0 & 0 & 0 & 0 & 0 & 0 & 0 & 0 & 0 & 0 & 0 & 0 & 0 & 0 & 0 \\
 0 & 0 & 0 & 0 & 0 & 0 & 0 & 0 & 1 & 0 & 0 & 0 & 0 & 0 & 0 & 0 & 0 & 0 & 0 \\
 0 & 0 & 0 & 0 & 0 & 0 & 0 & 0 & 0 & 1 & 0 & 0 & 0 & 0 & 0 & 0 & 0 & 0 & 0 \\
 0 & 0 & 0 & 0 & 0 & 0 & 0 & 0 & 0 & 0 & 0 & 0 & 0 & 0 & 0 & 0 & 0 & 0 & 0 \\
 0 & 0 & 0 & 0 & 0 & 0 & 0 & 0 & 0 & 0 & 0 & 0 & 0 & 0 & 0 & 0 & 0 & 0 & 0 \\
 0 & 0 & 1 & 0 & 1 & 0 & 0 & 0 & 0 & 0 & 0 & 0 & 0 & 0 & 0 & 0 & 0 & 0 & 0 \\
 0 & 0 & 0 & 1 & 0 & 0 & 0 & 0 & 0 & 0 & 0 & 0 & 0 & 0 & 0 & 0 & 0 & 0 & 0 \\
 0 & 0 & 0 & 0 & 0 & 1 & 0 & 0 & 0 & 0 & 0 & 0 & 0 & 0 & 0 & 0 & 0 & 0 & 0 \\
 0 & 0 & 0 & 0 & 0 & 0 & 0 & 0 & 0 & 0 & 0 & 0 & 0 & 0 & 0 & 0 & 0 & 0 & 0 \\
 0 & 0 & 0 & 0 & 0 & 0 & 0 & 0 & 0 & 0 & 0 & 0 & 0 & 0 & 0 & 0 & 0 & 0 & 0 \\
\end{array}
\right) ,
\end{equation*}
\begin{equation*}
    A^{(01)} = \left(
    \begin{array}{ccccccccccccccccccc}
 0 & 0 & 0 & 0 & 0 & 0 & 0 & 0 & 0 & 0 & 1 & 0 & 0 & 0 & 0 & 0 & 0 & 0 & 0 \\
 0 & 0 & 0 & 0 & 0 & 0 & 0 & 0 & 0 & 0 & 0 & 0 & 0 & 1 & 0 & 0 & 0 & 0 & 0 \\
 0 & 0 & 0 & 0 & 0 & 0 & 0 & 0 & 0 & 0 & 0 & 0 & 1 & 0 & 0 & 0 & 0 & 0 & 0 \\
 0 & 1 & 0 & 0 & 0 & 0 & 0 & 0 & 0 & 0 & 0 & 0 & 0 & 0 & 0 & 0 & 0 & 0 & 0 \\
 1 & 0 & 0 & 0 & 0 & 0 & 0 & 0 & 0 & 0 & 0 & 0 & 0 & 0 & 0 & 0 & 0 & 0 & 0 \\
 0 & 0 & 0 & 0 & 0 & 0 & 0 & 0 & 0 & 0 & 0 & 0 & 0 & 0 & 0 & 0 & 0 & 0 & 0 \\
 0 & 0 & 0 & 0 & 0 & 0 & 0 & 0 & 0 & 0 & 0 & 0 & 0 & 0 & 0 & 0 & 0 & 1 & 0 \\
 0 & 0 & 0 & 0 & 0 & 0 & 0 & 0 & 0 & 0 & 0 & 0 & 0 & 0 & 0 & 0 & 0 & 0 & 1 \\
 0 & 0 & 0 & 0 & 0 & 0 & 0 & 1 & 0 & 0 & 0 & 0 & 0 & 0 & 0 & 0 & 0 & 0 & 0 \\
 0 & 0 & 0 & 0 & 0 & 0 & 0 & 0 & 0 & 0 & 0 & 0 & 0 & 0 & 1 & 0 & 0 & 0 & 0 \\
 0 & 0 & 0 & 1 & 0 & 0 & 0 & 0 & 0 & 0 & 0 & 0 & 0 & 0 & 0 & 0 & 0 & 0 & 0 \\
 0 & 0 & 0 & 0 & 0 & 1 & 0 & 0 & 0 & 0 & 0 & 0 & 0 & 0 & 0 & 0 & 0 & 0 & 0 \\
 0 & 0 & 0 & 0 & 0 & 0 & 0 & 0 & 0 & 0 & 0 & 0 & 0 & 0 & 0 & 0 & 0 & 0 & 0 \\
 0 & 0 & 0 & 0 & 0 & 0 & 0 & 0 & 0 & 0 & 0 & 0 & 0 & 0 & 0 & 0 & 0 & 0 & 0 \\
 0 & 0 & 0 & 0 & 0 & 0 & 1 & 0 & 0 & 0 & 0 & 0 & 0 & 0 & 0 & 0 & 0 & 0 & 0 \\
 0 & 0 & 0 & 0 & 0 & 0 & 0 & 0 & 1 & 0 & 0 & 0 & 0 & 0 & 0 & 0 & 0 & 0 & 0 \\
 0 & 0 & 0 & 0 & 0 & 0 & 0 & 0 & 0 & 1 & 0 & 0 & 0 & 0 & 0 & 0 & 0 & 0 & 0 \\
 0 & 0 & 0 & 0 & 0 & 0 & 0 & 0 & 0 & 0 & 0 & 0 & 0 & 0 & 0 & 0 & 0 & 0 & 0 \\
 0 & 0 & 0 & 0 & 0 & 0 & 0 & 0 & 0 & 0 & 0 & 0 & 0 & 0 & 0 & 0 & 0 & 0 & 0 \\
\end{array}
\right) ,
\end{equation*}
\begin{equation*}
    A^{(10)} = \left(
\begin{array}{ccccccccccccccccccc}
 0 & 0 & 0 & 0 & 0 & 0 & 1 & 0 & 0 & 0 & 0 & 0 & 0 & 0 & 0 & 0 & 0 & 0 & 0 \\
 0 & 0 & 0 & 0 & 0 & 0 & 0 & 0 & 0 & 1 & 0 & 0 & 0 & 0 & 0 & 0 & 0 & 0 & 0 \\
 0 & 0 & 0 & 0 & 0 & 0 & 0 & 0 & 1 & 0 & 0 & 0 & 0 & 0 & 0 & 0 & 0 & 0 & 0 \\
 0 & 1 & 0 & 0 & 0 & 0 & 0 & 0 & 0 & 0 & 0 & 0 & 0 & 0 & 0 & 0 & 0 & 0 & 0 \\
 1 & 0 & 0 & 0 & 0 & 0 & 0 & 0 & 0 & 0 & 0 & 0 & 0 & 0 & 0 & 0 & 0 & 0 & 0 \\
 0 & 0 & 0 & 0 & 0 & 0 & 0 & 0 & 0 & 0 & 0 & 0 & 0 & 0 & 0 & 0 & 0 & 0 & 0 \\
 0 & 0 & 0 & 1 & 0 & 0 & 0 & 0 & 0 & 0 & 0 & 0 & 0 & 0 & 0 & 0 & 0 & 0 & 0 \\
 0 & 0 & 0 & 0 & 0 & 1 & 0 & 0 & 0 & 0 & 0 & 0 & 0 & 0 & 0 & 0 & 0 & 0 & 0 \\
 0 & 0 & 0 & 0 & 0 & 0 & 0 & 0 & 0 & 0 & 0 & 0 & 0 & 0 & 0 & 0 & 0 & 0 & 0 \\
 0 & 0 & 0 & 0 & 0 & 0 & 0 & 0 & 0 & 0 & 0 & 0 & 0 & 0 & 0 & 0 & 0 & 0 & 0 \\
 0 & 0 & 0 & 0 & 0 & 0 & 0 & 0 & 0 & 0 & 0 & 0 & 0 & 0 & 0 & 0 & 0 & 1 & 0 \\
 0 & 0 & 0 & 0 & 0 & 0 & 0 & 0 & 0 & 0 & 0 & 0 & 0 & 0 & 0 & 0 & 0 & 0 & 1 \\
 0 & 0 & 0 & 0 & 0 & 0 & 0 & 0 & 0 & 0 & 0 & 1 & 0 & 0 & 0 & 0 & 0 & 0 & 0 \\
 0 & 0 & 0 & 0 & 0 & 0 & 0 & 0 & 0 & 0 & 0 & 0 & 0 & 0 & 1 & 0 & 0 & 0 & 0 \\
 0 & 0 & 0 & 0 & 0 & 0 & 0 & 0 & 0 & 0 & 1 & 0 & 0 & 0 & 0 & 0 & 0 & 0 & 0 \\
 0 & 0 & 0 & 0 & 0 & 0 & 0 & 0 & 0 & 0 & 0 & 0 & 1 & 0 & 0 & 0 & 0 & 0 & 0 \\
 0 & 0 & 0 & 0 & 0 & 0 & 0 & 0 & 0 & 0 & 0 & 0 & 0 & 1 & 0 & 0 & 0 & 0 & 0 \\
 0 & 0 & 0 & 0 & 0 & 0 & 0 & 0 & 0 & 0 & 0 & 0 & 0 & 0 & 0 & 0 & 0 & 0 & 0 \\
 0 & 0 & 0 & 0 & 0 & 0 & 0 & 0 & 0 & 0 & 0 & 0 & 0 & 0 & 0 & 0 & 0 & 0 & 0 \\
\end{array}
\right),
\end{equation*}
\begin{equation*}
    A^{(11)} = \left(
\begin{array}{ccccccccccccccccccc}
 0 & 0 & 1 & 0 & 0 & 0 & 0 & 0 & 0 & 0 & 0 & 0 & 0 & 0 & 0 & 0 & 0 & 0 & 0 \\
 0 & 0 & 0 & 0 & 1 & 1 & 0 & 0 & 0 & 0 & 0 & 0 & 0 & 0 & 0 & 0 & 0 & 0 & 0 \\
 0 & 0 & 0 & 1 & 0 & 0 & 0 & 0 & 0 & 0 & 0 & 0 & 0 & 0 & 0 & 0 & 0 & 0 & 0 \\
 0 & 1 & 0 & 0 & 0 & 0 & 0 & 0 & 0 & 0 & 0 & 0 & 0 & 0 & 0 & 0 & 0 & 0 & 0 \\
 1 & 0 & 0 & 0 & 0 & 0 & 0 & 0 & 0 & 0 & 0 & 0 & 0 & 0 & 0 & 0 & 0 & 0 & 0 \\
 0 & 0 & 0 & 0 & 0 & 0 & 0 & 0 & 0 & 0 & 0 & 0 & 0 & 0 & 0 & 0 & 0 & 0 & 0 \\
 0 & 0 & 0 & 0 & 0 & 0 & 0 & 0 & 1 & 0 & 0 & 0 & 0 & 0 & 0 & 0 & 0 & 0 & 0 \\
 0 & 0 & 0 & 0 & 0 & 0 & 0 & 0 & 0 & 1 & 0 & 0 & 0 & 0 & 0 & 0 & 0 & 0 & 0 \\
 0 & 0 & 0 & 0 & 0 & 0 & 0 & 1 & 0 & 0 & 0 & 0 & 0 & 0 & 0 & 0 & 0 & 0 & 0 \\
 0 & 0 & 0 & 0 & 0 & 0 & 0 & 0 & 0 & 0 & 0 & 0 & 0 & 0 & 0 & 0 & 0 & 0 & 0 \\
 0 & 0 & 0 & 0 & 0 & 0 & 0 & 0 & 0 & 0 & 0 & 0 & 1 & 0 & 0 & 0 & 0 & 0 & 0 \\
 0 & 0 & 0 & 0 & 0 & 0 & 0 & 0 & 0 & 0 & 0 & 0 & 0 & 1 & 0 & 0 & 0 & 0 & 0 \\
 0 & 0 & 0 & 0 & 0 & 0 & 0 & 0 & 0 & 0 & 0 & 1 & 0 & 0 & 0 & 0 & 0 & 0 & 0 \\
 0 & 0 & 0 & 0 & 0 & 0 & 0 & 0 & 0 & 0 & 0 & 0 & 0 & 0 & 0 & 0 & 0 & 0 & 0 \\
 0 & 0 & 0 & 0 & 0 & 0 & 0 & 0 & 0 & 0 & 0 & 0 & 0 & 0 & 0 & 1 & 0 & 0 & 0 \\
 0 & 0 & 0 & 0 & 0 & 0 & 0 & 0 & 0 & 0 & 0 & 0 & 0 & 0 & 0 & 0 & 0 & 1 & 0 \\
 0 & 0 & 0 & 0 & 0 & 0 & 0 & 0 & 0 & 0 & 0 & 0 & 0 & 0 & 0 & 0 & 0 & 0 & 1 \\
 0 & 0 & 0 & 0 & 0 & 0 & 0 & 0 & 0 & 0 & 0 & 0 & 0 & 0 & 0 & 0 & 1 & 0 & 0 \\
 0 & 0 & 0 & 0 & 0 & 0 & 0 & 0 & 0 & 0 & 0 & 0 & 0 & 0 & 1 & 0 & 0 & 0 & 0 \\
\end{array}
\right).
\end{equation*}

\begin{align*}
B^{(00)} =
\left(\begin{array}{cccc}
I_6& & & \\
 &O_4& & \\
 & &O_4& \\
 & & &O_5\\
\end{array}
\right) ,&&
B^{(01)} =
\left(\begin{array}{cccc}
O_6& & & \\
 &I_4& & \\
 & &O_4& \\
 & & &O_5\\
\end{array}
\right) ,
\end{align*}
\begin{align*}
B^{(10)} =
\left(\begin{array}{cccc}
O_6& & & \\
 &O_4& & \\
 & &I_4& \\
 & & &O_5\\
\end{array}
\right) ,&&
B^{(11)} =
\left(\begin{array}{cccc}
O_6& & & \\
 &O_4& & \\
 & &O_4& \\
 & & &I_5\\
\end{array}
\right) ,
\end{align*}
where
\begin{align*}
I_{k}=\left(\begin{array}{ccc}
     1  \\ &\ddots\\ &&1
\end{array}\right)_{k\times k},&&
O_{k}=\left(\begin{array}{ccc}
     0  \\ &\ddots\\ &&0
\end{array}\right)_{k\times k}.
\end{align*}
$P_0$ and $P_0'$ are the projectors onto the subspaces $\mathcal{V}_0, \mathcal{V}'_0\in\mathcal U\otimes\mathcal H\otimes\mathcal H^*$ respectively, where
\begin{align*}
    \mathcal{V}_0=\text{span}\{&|0,0),|0,1),|0,2),|0,3),|1,0),|1,1),|1,2),|1,3),|2,0),|2,1),|2,2),|2,3),|3,0),\\&|3,1),|3,2),|3,3),|4,0),|4,1),|4,2),|4,3),|5,0),|5,1),|5,2),|5,3),|6,0),|6,2),\\&|7,0),|7,2),|10,0),|10,1),|11,0),|11,1),|14,0),|15,0),|16,0)\},\
\end{align*}
\begin{align*}
     \mathcal{V}'_0=\text{span}\{&|0,0),|0,1),|0,2),|0,3),|1,3),|2,3),|3,0),|3,1),|3,2),|3,3),|4,0),|4,1),|4,2),\\&|4,3),|5,0),|6,2),|7,2),|10,1),|11,1),|14,0),|15,0),|16,0)\}.
\end{align*}
$P_1$ and $P_1'$ are mappings from $U\otimes\mathcal H\otimes\mathcal H^*$ to $\mathcal U$:
\begin{align*}
    P_1:\,&|2,0)\mapsto|0),&&|4,0)\mapsto|0),&&|3,0)\mapsto|1),&&|5,0)\mapsto|1),&&|0,0)\mapsto|2),&&|15,0)\mapsto|3),&&|16,0)\mapsto|3),\\
    &|14,0)\mapsto|4),&&|1,0)\mapsto|5),&&|0,1)\mapsto|6),&&|3,1)\mapsto|7),&&|10,1)\mapsto|8),&&|1,1)\mapsto|9),&&|0,2)\mapsto|10),\\
    &|3,2)\mapsto|11), &&|6,2)\mapsto|12),&&|1,2)\mapsto|13),&&|4,3)\mapsto|14),&&|5,3)\mapsto|14),&&|0,3)\mapsto|15),&&|3,3)\mapsto|16),\\
    &|2,3)\mapsto|17),&&|1,3)\mapsto|18),&&\,\text{others}\mapsto 0,
\end{align*}
\begin{align*}
    P'_1:\,&|14,0)\mapsto|0),&&|15,0)\mapsto|1),&&|16,0)\mapsto|1),&&|4,0)\mapsto|2),&&|0,0)\mapsto|3),&&|5,0)\mapsto|4),&&|3,0)\mapsto|5),\\
    &|4,1)\mapsto|6),&&|10,1)\mapsto|7),&&|0,1)\mapsto|8),&&|3,1)\mapsto|9),&&|4,2)\mapsto|10),&&|6,2)\mapsto|11),&&|0,2)\mapsto|12),\\
    &|3,2)\mapsto|13),&&|1,3)\mapsto|14),&&|4,3)\mapsto|15),&&|2,3)\mapsto|16),&&|0,3)\mapsto|17),&&|3,3)\mapsto|18),&&\text{others}\mapsto 0.
\end{align*}
Vectors at the top boundary:
\begin{align*}
(\text{t}|=(\text{t}'|=(1, 1, 1, 1, 1, 1, 0, 0, 0, 0, 0, 0, 0, 0, 1, 1, 1, 1, 1).
\end{align*}

\end{widetext}
Bottom boundary conditions:
\begin{align*}
    &P_1:|v_{2n})\otimes|\rho_{2n})\mapsto|v_{2n+1}),\\
    &P_1':|v_{2n+1})\otimes|\rho_{2n+1})\mapsto|v_{(2n+2)\bmod 6}), &&n=0,1,2.
\end{align*}

Bottom boundary vectors and corresponding initial states:

a)\quad for $\rho_0=|0\rangle\langle0|$, $\rho_1=|0\rangle\langle0|$, $\rho_2=|1\rangle\langle1|$, $\rho_3=|0\rangle\langle0|$, $\rho_4=|0\rangle\langle0|$, $\rho_5=|1\rangle\langle1|$, 
\begin{align*}
    &|v_0)=|16), &&|v_1)=|3),&& |v_2)=|5), \\
    &|v_3)=|14),&& |v_4)=|0), &&|v_5)=|2),
\end{align*}

b)\quad for $\rho_0=|1\rangle\langle1|$, $\rho_1=|0\rangle\langle0|$, $\rho_2=|0\rangle\langle0|$, $\rho_3=|0\rangle\langle0|$, $\rho_4=|1\rangle\langle1|$, $\rho_5=|0\rangle\langle0|$, 
\begin{align*}
    &|v_0)=|0), &&|v_1)=|15),&& |v_2)=|1),\\
    &|v_3)=|5),&& |v_4)=|4), &&|v_5)=|14),
\end{align*}

c)\quad for $\rho_0=|0\rangle\langle0|$, $\rho_1=|0\rangle\langle0|$, $\rho_2=|1\rangle\langle1|$, $\rho_3=|0\rangle\langle0|$, $\rho_4=|0\rangle\langle0|$, $\rho_5=|0\rangle\langle0|$, 
\begin{align*}
    &|v_0)=|4), &&|v_1)=|0),&& |v_2)=|3),\\
    &|v_3)=|16),&& |v_4)=|1), &&|v_5)=|5),
\end{align*}

d)\quad for $\rho_0=|0\rangle\langle0|$, $\rho_1=|0\rangle\langle0|$, $\rho_2=|0\rangle\langle0|$, $\rho_3=|0\rangle\langle0|$, $\rho_4=|0\rangle\langle0|$, $\rho_5=|1\rangle\langle1|$, 
\begin{align*}
    &|v_0)=|14), &&|v_1)=|4),&& |v_2)=|2),\\
    &|v_3)=|0),&& |v_4)=|3), &&|v_5)=|1),
\end{align*}

e)\quad  for $\rho_0=|0\rangle\langle0|$, $\rho_1=|1\rangle\langle1|$, $\rho_2=|0\rangle\langle0|$, $\rho_3=|1\rangle\langle1|$, $\rho_4=|0\rangle\langle0|$, $\rho_5=|0\rangle\langle0|$, 
\begin{align*}
    &|v_0)=|5), &&|v_1)=|1),&& |v_2)=|14),\\
    &|v_3)=|4),&& |v_4)=|15), &&|v_5)=|3).
\end{align*}

\section{Rule 54 quantum cellular automaton solutions}\label{sec:solution_list1}

The generalized zipper conditions also work for exact solutions in the Rule 54 quantum cellular automaton. In this appendix, we present the solutions and corresponding initial states.

\subsection{ $\chi=3$}\label{sec:rule54,3}
The following solutions work for two-site shift-invariant initial states with block $|0\rangle\langle 0| \otimes \left(\begin{array}{cc}
    \alpha & \beta \\
    \gamma & \delta
\end{array}\right)$. Solutions obtained by a different approach have been reported in Ref.~\cite{Klobas2021Exact}.

$A$ and $B$ tensors:
\begin{align*}
A^{(00)} =
\left(\begin{array}{ccc}
&1&\\
1&1&\\
&&1\\
\end{array}
\right) ,&&
A^{(11)} =
\left(\begin{array}{ccc}
&1&\\
1&&1\\
&1&\\
\end{array}
\right) ,
\end{align*}
\begin{align*}
A^{(01)} =A^{(10)} =
\left(\begin{array}{ccc}
&1&\\
1&&\\
&&0\\
\end{array}
\right) ,
\end{align*}
\begin{align*}
B^{(00)} =
\left(\begin{array}{ccc}
&&0\\
&\frac{\alpha}{\alpha+\delta}&\\
0&&\\
\end{array}
\right) ,&&
B^{(11)} =
\left(\begin{array}{ccc}
&&1\\
&0&\\
\frac{\delta}{\alpha}&&\\
\end{array}
\right),
\end{align*}
\begin{align*}
B^{(01)} =
B^{(10)} =
\left(\begin{array}{ccc}
&&0\\
&0&\\
0&&\\
\end{array}
\right) .
\end{align*}
$P_0$ and $P'_0$ tensors:
\begin{align*}
(P_0)_{bj,ai}=
\left(\begin{array}{cccccccccccc}
1&0&0&0&0&0&0&0&0&0&0&0\\
0&0&0&0&0&0&0&0&0&0&0&0\\
0&0&0&0&0&0&0&0&0&0&0&0\\
0&0&0&1&0&0&0&0&0&0&0&0\\
0&0&0&0&1&0&0&0&0&0&0&0\\
0&0&0&0&0&1&0&0&0&0&0&0\\
0&0&0&0&0&0&1&0&0&0&0&0\\
0&0&0&0&0&0&0&1&0&0&0&0\\
\frac{\delta}{\alpha}&0&0&0&0&0&0&0&0&0&0&0\\
0&0&0&0&0&0&0&0&0&1&0&0\\
0&0&0&0&0&0&0&0&0&0&1&0\\
0&0&0&0&0&0&0&0&0&0&0&1\\
\end{array}
\right)_{bj,ai},
\end{align*}
\begin{align*}
(P_0')_{bj,ai}=
\left(\begin{array}{cccccccccccc}
1&0&0&0&0&0&0&0&0&0&0&0\\
0&1&0&0&0&0&0&0&0&0&0&0\\
0&0&1&0&0&0&0&0&0&0&0&0\\
0&0&0&1&0&0&0&0&0&0&0&0\\
1&0&0&0&0&0&0&0&\frac{\alpha}{\delta}&0&0&0\\
0&0&0&0&0&0&0&0&0&0&0&0\\
0&0&0&0&0&0&0&0&0&0&0&0\\
0&0&0&0&0&0&0&1&0&0&0&0\\
0&0&0&0&0&0&0&0&1&0&0&0\\
0&0&0&0&0&0&0&0&0&0&0&0\\
0&0&0&0&0&0&0&0&0&0&0&0\\
0&0&0&1&0&0&0&0&0&0&0&0\\
\end{array}
\right)_{bj,ai}.
\end{align*}
$P_1$ and $P'_1$ tensors:
\begin{align*}
(P_1)_{b,ai}=
\left(\begin{array}{cccccccccccc}
0&0&0&\frac{\alpha}{\delta}&0&0&0&0&0&0&0&\frac{\alpha}{\delta}\\
\frac{\alpha+\delta}{\alpha}&0&0&0&\frac{\alpha+\delta}{\alpha}&0&0&0&0&0&0&0\\
0&0&0&0&0&0&0&\frac{\alpha+\delta}{\alpha}&0&0&0&0\\
\end{array}
\right)_{b,ai},
\end{align*}
\begin{align*}
(P_1')_{b,ai}
=
\left(\begin{array}{cccccccccccc}
0&0&0&\frac{\alpha+\delta}{\alpha}&0&0&0&0&0&0&0&0\\
1&0&0&0&0&0&0&0&\frac{\alpha}{\delta}&0&0&0\\
0&0&0&0&0&0&0&1&0&0&0&0\\
\end{array}
\right)_{b,ai}.
\end{align*}
Vectors at the top boundary:
\begin{align*}
    (\text{t}|=\left(1, \frac{\alpha+\delta}{\alpha}, 1\right),&&(\text{t}'|=\left(\frac\delta\alpha,1,1\right).
\end{align*}
Initial states and bottom boundary vectors:

a) for $|\rho\rangle\rangle=\frac{1}{\alpha+\delta}(\alpha,\beta,\gamma,\delta)^T$, $|\rho'\rangle\rangle=(1, 0, 0, 0)^T$,
\begin{align*}
    |\text{b})=\left(0,\frac{\alpha}{\alpha+\delta},0\right)^T,&&|\text{b}')=\left(0,\frac{\alpha}{\alpha+\delta},\frac{\delta}{\alpha+\delta}\right)^T,
\end{align*}

b) for $|\rho\rangle\rangle=(1, 0, 0, 0)^T$, $|\rho'\rangle\rangle=\frac{1}{\alpha+\delta}(\alpha,\beta,\gamma,\delta)^T$, we have modified bottom boundary conditions
\begin{equation}\label{eq:btm_bc_mod}
\begin{tikzpicture}[x=0.75pt,y=0.75pt,yscale=-1,xscale=1]

\draw [color={rgb, 255:red, 0; green, 0; blue, 0 }  ,draw opacity=1 ]   (65.74,35.02) -- (65.74,20.37) ;
\draw    (65.74,31.96) -- (87.02,31.96) ;
\draw  [fill={rgb, 255:red, 245; green, 166; blue, 35 }  ,fill opacity=1 ] (63.22,30.45) .. controls (63.22,29.9) and (63.67,29.45) .. (64.23,29.45) -- (67.25,29.45) .. controls (67.8,29.45) and (68.25,29.9) .. (68.25,30.45) -- (68.25,33.47) .. controls (68.25,34.03) and (67.8,34.48) .. (67.25,34.48) -- (64.23,34.48) .. controls (63.67,34.48) and (63.22,34.03) .. (63.22,33.47) -- cycle ;
\draw [color={rgb, 255:red, 0; green, 0; blue, 0 }  ,draw opacity=1 ]   (78.32,39.84) -- (78.32,20.05) ;
\draw  [fill={rgb, 255:red, 245; green, 166; blue, 35 }  ,fill opacity=1 ] (78.32,41.42) -- (75.8,38.27) -- (80.83,38.27) -- cycle ;
\draw   (77.69,31.96) .. controls (77.69,31.61) and (77.97,31.33) .. (78.32,31.33) .. controls (78.66,31.33) and (78.94,31.61) .. (78.94,31.96) .. controls (78.94,32.31) and (78.66,32.59) .. (78.32,32.59) .. controls (77.97,32.59) and (77.69,32.31) .. (77.69,31.96) -- cycle ;
\draw [color={rgb, 255:red, 0; green, 0; blue, 0 }  ,draw opacity=1 ]   (102.21,34.7) -- (102.21,19.42) ;
\draw    (102.21,31.96) -- (123.5,31.96) ;
\draw  [fill={rgb, 255:red, 245; green, 166; blue, 35 }  ,fill opacity=1 ] (99.7,30.45) .. controls (99.7,29.9) and (100.15,29.45) .. (100.7,29.45) -- (103.72,29.45) .. controls (104.28,29.45) and (104.73,29.9) .. (104.73,30.45) -- (104.73,33.47) .. controls (104.73,34.03) and (104.28,34.48) .. (103.72,34.48) -- (100.7,34.48) .. controls (100.15,34.48) and (99.7,34.03) .. (99.7,33.47) -- cycle ;
\draw [color={rgb, 255:red, 0; green, 0; blue, 0 }  ,draw opacity=1 ]   (114.79,39.84) -- (114.79,19.51) ;
\draw  [fill={rgb, 255:red, 245; green, 166; blue, 35 }  ,fill opacity=1 ] (114.79,41.42) -- (112.27,38.27) -- (117.3,38.27) -- cycle ;
\draw   (114.16,31.96) .. controls (114.16,31.61) and (114.44,31.33) .. (114.79,31.33) .. controls (115.14,31.33) and (115.42,31.61) .. (115.42,31.96) .. controls (115.42,32.31) and (115.14,32.59) .. (114.79,32.59) .. controls (114.44,32.59) and (114.16,32.31) .. (114.16,31.96) -- cycle ;
\draw  [fill={rgb, 255:red, 155; green, 155; blue, 155 }  ,fill opacity=1 ] (100.5,23.77) .. controls (100.5,23.63) and (100.61,23.52) .. (100.75,23.52) -- (116.6,23.52) .. controls (116.74,23.52) and (116.85,23.63) .. (116.85,23.77) -- (116.85,24.52) .. controls (116.85,24.66) and (116.74,24.77) .. (116.6,24.77) -- (100.75,24.77) .. controls (100.61,24.77) and (100.5,24.66) .. (100.5,24.52) -- cycle ;

\draw (84.64,24.28) node [anchor=north west][inner sep=0.75pt]    {$=$};

\end{tikzpicture},\quad
\begin{tikzpicture}[x=0.75pt,y=0.75pt,yscale=-1,xscale=1]

\draw [color={rgb, 255:red, 0; green, 0; blue, 0 }  ,draw opacity=1 ]   (36.36,32.27) -- (36.36,15.86) ;
\draw [color={rgb, 255:red, 0; green, 0; blue, 0 }  ,draw opacity=1 ]   (23.06,31.56) -- (23.06,15.47) ;
\draw  [fill={rgb, 255:red, 208; green, 2; blue, 27 }  ,fill opacity=1 ] (36.36,33.47) -- (33.23,29.56) -- (39.48,29.56) -- cycle ;
\draw  [fill={rgb, 255:red, 208; green, 2; blue, 27 }  ,fill opacity=1 ] (26.97,29.53) .. controls (26.97,29.53) and (26.97,29.53) .. (26.97,29.53) .. controls (26.97,31.69) and (25.22,33.44) .. (23.06,33.44) .. controls (20.9,33.44) and (19.15,31.69) .. (19.15,29.53) -- cycle ;
\draw [color={rgb, 255:red, 0; green, 0; blue, 0 }  ,draw opacity=1 ]   (75.46,33.48) -- (75.46,17.08) ;
\draw [color={rgb, 255:red, 0; green, 0; blue, 0 }  ,draw opacity=1 ]   (59.82,32.78) -- (59.82,16.68) ;
\draw  [fill={rgb, 255:red, 208; green, 2; blue, 27 }  ,fill opacity=1 ] (75.46,34.69) -- (72.33,30.78) -- (78.59,30.78) -- cycle ;
\draw  [fill={rgb, 255:red, 208; green, 2; blue, 27 }  ,fill opacity=1 ] (63.73,30.75) .. controls (63.73,30.75) and (63.73,30.75) .. (63.73,30.75) .. controls (63.73,32.91) and (61.98,34.66) .. (59.82,34.66) .. controls (57.66,34.66) and (55.91,32.91) .. (55.91,30.75) -- cycle ;
\draw  [fill={rgb, 255:red, 155; green, 155; blue, 155 }  ,fill opacity=1 ] (57.43,22.1) .. controls (57.43,21.92) and (57.57,21.78) .. (57.74,21.78) -- (77.45,21.78) .. controls (77.62,21.78) and (77.76,21.92) .. (77.76,22.1) -- (77.76,23.03) .. controls (77.76,23.21) and (77.62,23.35) .. (77.45,23.35) -- (57.74,23.35) .. controls (57.57,23.35) and (57.43,23.21) .. (57.43,23.03) -- cycle ;

\draw (41.67,21.61) node [anchor=north west][inner sep=0.75pt]    {$=$};

\end{tikzpicture},\quad
\begin{tikzpicture}[x=0.75pt,y=0.75pt,yscale=-1,xscale=1]

\draw [color={rgb, 255:red, 0; green, 0; blue, 0 }  ,draw opacity=1 ]   (28.17,26.38) -- (28.17,20.54) ;
\draw [color={rgb, 255:red, 0; green, 0; blue, 0 }  ,draw opacity=1 ]   (21.88,34.74) -- (21.88,26.38) ;
\draw    (21.88,34.2) -- (43.16,34.2) ;
\draw  [fill={rgb, 255:red, 245; green, 166; blue, 35 }  ,fill opacity=1 ] (19.37,32.69) .. controls (19.37,32.13) and (19.82,31.68) .. (20.37,31.68) -- (23.39,31.68) .. controls (23.95,31.68) and (24.4,32.13) .. (24.4,32.69) -- (24.4,35.71) .. controls (24.4,36.26) and (23.95,36.71) .. (23.39,36.71) -- (20.37,36.71) .. controls (19.82,36.71) and (19.37,36.26) .. (19.37,35.71) -- cycle ;
\draw [color={rgb, 255:red, 0; green, 0; blue, 0 }  ,draw opacity=1 ]   (34.46,42.08) -- (34.46,26.06) ;
\draw  [fill={rgb, 255:red, 245; green, 166; blue, 35 }  ,fill opacity=1 ] (34.46,43.65) -- (31.94,40.51) -- (36.97,40.51) -- cycle ;
\draw   (33.83,34.2) .. controls (33.83,33.85) and (34.11,33.57) .. (34.46,33.57) .. controls (34.81,33.57) and (35.09,33.85) .. (35.09,34.2) .. controls (35.09,34.54) and (34.81,34.83) .. (34.46,34.83) .. controls (34.11,34.83) and (33.83,34.54) .. (33.83,34.2) -- cycle ;
\draw  [fill={rgb, 255:red, 155; green, 155; blue, 155 }  ,fill opacity=1 ] (19.99,27.01) -- (21.04,25.75) -- (35.3,25.75) -- (36.35,27.01) -- cycle ;
\draw [color={rgb, 255:red, 0; green, 0; blue, 0 }  ,draw opacity=1 ]   (60.73,42.67) -- (60.73,23.43) ;
\draw    (60.73,32.88) -- (71.28,32.88) ;
\draw  [fill={rgb, 255:red, 74; green, 144; blue, 226 }  ,fill opacity=1 ] (60.73,29.32) -- (64.29,32.88) -- (60.73,36.44) -- (57.17,32.88) -- cycle ;
\draw  [fill={rgb, 255:red, 208; green, 2; blue, 27 }  ,fill opacity=1 ] (63.88,41.68) .. controls (63.88,41.68) and (63.88,41.68) .. (63.88,41.68) .. controls (63.88,41.68) and (63.88,41.68) .. (63.88,41.68) .. controls (63.88,43.42) and (62.47,44.83) .. (60.73,44.83) .. controls (58.99,44.83) and (57.59,43.42) .. (57.59,41.68) -- cycle ;

\draw (41.41,26.51) node [anchor=north west][inner sep=0.75pt]    {$=$};

\end{tikzpicture},\quad
\begin{tikzpicture}[x=0.75pt,y=0.75pt,yscale=-1,xscale=1]

\draw    (23.29,37.14) -- (44.28,37.14) ;
\draw [color={rgb, 255:red, 0; green, 0; blue, 0 }  ,draw opacity=1 ]   (35.88,46.26) -- (35.88,28.48) ;
\draw  [fill={rgb, 255:red, 208; green, 2; blue, 27 }  ,fill opacity=1 ] (35.88,47.83) -- (33.36,44.68) -- (38.39,44.68) -- cycle ;
\draw [color={rgb, 255:red, 0; green, 0; blue, 0 }  ,draw opacity=1 ]   (23.29,45.89) -- (23.29,28.77) ;
\draw  [fill={rgb, 255:red, 74; green, 144; blue, 226 }  ,fill opacity=1 ] (23.29,33.58) -- (26.84,37.14) -- (23.29,40.69) -- (19.73,37.14) -- cycle ;
\draw  [fill={rgb, 255:red, 208; green, 2; blue, 27 }  ,fill opacity=1 ] (26.43,44.68) .. controls (26.43,46.42) and (25.02,47.83) .. (23.29,47.83) .. controls (21.55,47.83) and (20.14,46.42) .. (20.14,44.68) -- cycle ;
\draw [color={rgb, 255:red, 0; green, 0; blue, 0 }  ,draw opacity=1 ]   (61.8,35.88) -- (61.8,26.88) ;
\draw    (61.8,35.88) -- (71.33,35.88) ;
\draw  [fill={rgb, 255:red, 245; green, 166; blue, 35 }  ,fill opacity=1 ] (59.28,34.37) .. controls (59.28,33.81) and (59.73,33.36) .. (60.29,33.36) -- (63.31,33.36) .. controls (63.86,33.36) and (64.31,33.81) .. (64.31,34.37) -- (64.31,37.39) .. controls (64.31,37.94) and (63.86,38.4) .. (63.31,38.4) -- (60.29,38.4) .. controls (59.73,38.4) and (59.28,37.94) .. (59.28,37.39) -- cycle ;
\draw  [fill={rgb, 255:red, 126; green, 211; blue, 33 }  ,fill opacity=1 ] (32.73,37.13) .. controls (32.73,35.4) and (34.14,33.99) .. (35.88,33.99) .. controls (37.61,33.99) and (39.02,35.4) .. (39.02,37.13) .. controls (39.02,38.87) and (37.61,40.28) .. (35.88,40.28) .. controls (34.14,40.28) and (32.73,38.87) .. (32.73,37.13) -- cycle ;
\draw  [draw opacity=0] (35.9,35.56) .. controls (36.75,35.58) and (37.43,36.26) .. (37.45,37.11) -- (35.88,37.13) -- cycle ; \draw   (35.9,35.56) .. controls (36.75,35.58) and (37.43,36.26) .. (37.45,37.11) ;  

\draw [color={rgb, 255:red, 0; green, 0; blue, 0 }  ,draw opacity=1 ]   (29.1,27.78) -- (29.1,22.97) ;
\draw  [fill={rgb, 255:red, 255; green, 255; blue, 255 }  ,fill opacity=1 ] (21.24,28.97) -- (22.29,27.71) -- (36.54,27.71) -- (37.59,28.97) -- cycle ;

\draw (42.97,29.51) node [anchor=north west][inner sep=0.75pt]    {$=$};

\end{tikzpicture},
\end{equation}
with another tensor
\begin{equation}
    \begin{tikzpicture}[x=0.75pt,y=0.75pt,yscale=-1,xscale=1]

\draw    (27,33.28) -- (40.28,33.28) ;
\draw [color={rgb, 255:red, 0; green, 0; blue, 0 }  ,draw opacity=1 ]   (27,33.28) -- (27,19.97) ;
\draw  [fill={rgb, 255:red, 245; green, 166; blue, 35 }  ,fill opacity=1 ] (23,30.85) .. controls (23,29.97) and (23.72,29.25) .. (24.6,29.25) -- (29.4,29.25) .. controls (30.28,29.25) and (31,29.97) .. (31,30.85) -- (31,35.65) .. controls (31,36.54) and (30.28,37.25) .. (29.4,37.25) -- (24.6,37.25) .. controls (23.72,37.25) and (23,36.54) .. (23,35.65) -- cycle ;

\draw (-36.67,17.51) node [anchor=north west][inner sep=0.75pt]  [font=\normalsize]  {$C^{(i,i')}_a=$};
\draw (23.17,9.89) node [anchor=north west][inner sep=0.75pt]  [font=\scriptsize]  {$a$};
\draw (43,28.68) node [anchor=north west][inner sep=0.75pt]  [font=\scriptsize]  {$i,i'$};
\end{tikzpicture}
\end{equation}
at the bottom boundary, where
\begin{align*}
    C^{(00)}=\frac{1}{\alpha+\delta}\left(0,\alpha,\delta\right)^T,&&C^{(01)}=\frac{1}{\alpha+\delta}\left(0,\beta,\gamma\right)^T,\\
    C^{(10)}=\frac{1}{\alpha+\delta}\left(0,\gamma,\beta\right)^T,&&C^{(11)}=\frac{1}{\alpha+\delta}\left(0,\delta,\alpha\right)^T,
\end{align*}
and $|\text{b}')=\left(0,1,0\right)^T$.

\subsection{$\chi=6$}\label{sec:rule54,6}
The solutions work for two-site shift-invariant initial states with block $|1\rangle\langle 1| \otimes \left(\begin{array}{cc}
    \alpha & 0 \\
    0 & \delta
\end{array}\right)$.

$A$ and $B$ tensors:
\begin{align*}
A^{(00)}=\left(\begin{array}{ccc}&I&\\I&J&\\&&I\\\end{array}\right),&&
A^{(11)}=\left(\begin{array}{ccc}&I&\\I&&I\\&J&\\\end{array}\right),
\end{align*}
\begin{align*}
A^{(01)}=A^{(10)}=\left(\begin{array}{ccc}&I&\\I&&\\&&O\\\end{array}\right),
\end{align*}
\begin{align*}
B^{(00)}=\left(\begin{array}{ccc}&&O\\&Q&\\O&&\\\end{array}\right),&&
B^{(11)}=\left(\begin{array}{ccc}&&I\\&O&\\S&&\\\end{array}\right),
\end{align*}
\begin{align*}
B^{(01)}=B^{(10)}=\left(\begin{array}{ccc}&&O\\&O&\\O&&\\\end{array}\right),
\end{align*}
where $2\times2$ matrices are defined as
\begin{align*}
    I=\begin{pmatrix}1&\\&1\end{pmatrix},&&
    J=\begin{pmatrix}0&\\&1\end{pmatrix},&&
    O=\begin{pmatrix}0&\\&0\end{pmatrix},
\end{align*}
\begin{align*}
    S=\begin{pmatrix}1&1\\0&-1\end{pmatrix},&&
    Q=\begin{pmatrix}0&\alpha\delta/{(\alpha+\delta)^2}\\1&0\end{pmatrix}.
\end{align*}
$P_0$ and $P'_0$ tensors:
\begin{align*}
(P_0)_{bj,ai} = 
\begin{cases}
1 & \text{if } (b,j,a,i) \in\\
  &\{(0,3,0,3),(1,0,1,0),(1,3,1,3),\\
  &\;(2,0,2,0),(2,1,2,1),(2,2,2,2),\\
  &\;(2,3,2,3),(3,0,3,0),(3,1,3,1),\\
  &\;(3,2,3,2),(3,3,3,3),(4,0,4,0),\\
  &\;(4,1,4,1),(4,2,4,2),(4,3,4,3),\\
  &\;(5,1,5,1),(5,2,5,2),(5,3,5,3)\}, \\
-1 & \text{if } (b,j,a,i) = (5,0,1,0), \\
0 & \text{otherwise},
\end{cases}
\end{align*}
\begin{align*}
(P_0')_{bj,ai}= 
\begin{cases}
1 & \text{if } (b,j,a,i) \in \\
  &\{(0,0,0,0),(0,1,0,1),(0,2,0,2),\\
  &\;(0,3,0,3),(1,0,1,0),(1,1,1,1),\\
  &\;(1,2,1,2), (1,3,1,3),(2,3,2,3),\\
  &\;(3,0,1,0),(3,3,3,3),(4,0,4,0),\\
  &\;(5,0,5,0),(5,3,1,3)\}, \\
-1 & \text{if } (b,j,a,i) = (3,0,5,0), \\
0 & \text{otherwise}.
\end{cases}
\end{align*}
$P_1$ and $P'_1$ tensors:
\begin{align*}
(P_1)_{b,ai}= 
\begin{cases}
\dfrac{(\alpha+\delta)^2}{\alpha\delta} & \text{if } (b,a,i) \in\{(3,2,0),(3,4,0),(5,2,3)\}, \\
-\dfrac{(\alpha+\delta)^2}{\alpha\delta} & \text{if } (b,a,i) =(3,1,0), \\
1 & \text{if } (b,a,i) \in \{(0,0,3),(0,1,3),(0,4,3),\\
&\;(0,5,3),(2,1,0),(2,3,0),(4,3,3)\}, \\
-1 & \text{if } (b,a,i) \in\{(1,1,3),(1,5,3)\}, \\
0 & \text{otherwise},
\end{cases}  
\end{align*}
\begin{align*}
(P_1')_{b,ai}= 
\begin{cases}
\dfrac{(\alpha+\delta)^2}{\alpha\delta} & \text{if } (b,a,i)=(1,0,3) \\
1 & \text{if } (b,a,i) \in \{ (0,1,3),(2,0,0),(2,4,0),\\
&\;(2,5,0),(3,1,0),(4,2,3),(5,3,3)\}, \\
-1 & \text{if } (b,a,i)=(3,5,0), \\
0 & \text{otherwise}.\notag
\end{cases}
\end{align*}
Vectors at the top boundary:
\begin{align*}
    (\text{t}|&=\left(1,\frac{\alpha\delta}{(\alpha+\delta)^2},1,1,1,\frac{\alpha\delta}{(\alpha+\delta)^2}\right),\\
    (\text{t}'|&=\left(1,1-\frac{\alpha\delta}{(\alpha+\delta)^2},1,\frac{\alpha\delta}{(\alpha+\delta)^2},1,\frac{\alpha\delta}{(\alpha+\delta)^2}\right).
\end{align*}
Solvable initial states and bottom boundary vectors:

a)\quad for $|\rho\rangle\rangle=(0, 0, 0, 1)^T$, $|\rho'\rangle\rangle=\frac{1}{\alpha+\delta}(\alpha, 0, 0, \delta)^T$,
\begin{align*}
    |\text{b})&=\left(1,-\frac{\alpha+\delta}{\delta},\frac{\alpha}{\alpha+\delta},0,0,0\right)^T,\\
    |\text{b}')&=\left(-\frac{\alpha}{\delta},\frac{\alpha+\delta}{\delta},0,0,0,\frac{\alpha+\delta}{\delta}\right)^T,
\end{align*}

b)\quad for $|\rho\rangle\rangle=\frac{1}{\alpha+\delta}(\alpha, 0, 0, \delta)^T$, $|\rho'\rangle\rangle=(0,0,0,1)^T$,
\begin{align*}
    |\text{b})&=\left(0,\frac{\alpha+\delta}{\alpha},0,0,1,-\frac{\alpha+\delta}{\alpha}\right)^T,\\
    |\text{b}')&=\left(\frac{\delta}{\alpha+\delta},0,1,-\frac{\alpha+\delta}{\alpha},0,0\right)^T.
\end{align*}

\begin{widetext}
\subsection{$\chi=14$}\label{sec:rule54,14}

The solutions work for two-site shift-invariant initial states with block $|1\rangle\langle 1| \otimes \left(\begin{array}{cc}
    \alpha & \beta \\
    \gamma & \delta
\end{array}\right)$.

$A$ and $B$ tensors:
\begin{align*}
A^{(00)}=\left(\begin{array}{cccccccccccccc}
0 & 0 & 0 & 0 & 0 & 0 & 0 & 0 & \frac{\delta}{\alpha + \delta} & 0 & 0 & 0 & 0 & 0\\
1 & 0 & 0 & 0 & 0 & 0 & 0 & 0 & 0 & 0 & 1 & 0 & 0 & 0\\
1 & 0 & 0 & 0 & 0 & 0 & 0 & 0 & 0 & 0 & 0 & 0 & 0 & 0\\
0 & 0 & 0 & 0 & 0 & 0 & 0 & 1 & 0 & 0 & 0 & 0 & 0 & 0\\
0 & 0 & 0 & 0 & 0 & 0 & 0 & 0 & 0 & 1 & 0 & 0 & 0 & 0\\
1 & 0 & 0 & 0 & 0 & 0 & 0 & 0 & 0 & 0 & 0 & 0 & 0 & 0\\
0 & 0 & 0 & 0 & 1 & 0 & 0 & 0 & 0 & 0 & 0 & 0 & 0 & 0\\
0 & 0 & 0 & 0 & 0 & 0 & 0 & 0 & 0 & 0 & 0 & 1 & 0 & 0\\
0 & 0 & 0 & 0 & 0 & 0 & 0 & 0 & 0 & 0 & 0 & 0 & 1 & 1\\
0 & 0 & 0 & 0 & 0 & 0 & \frac{\beta}{\alpha + \delta} & 0 & 0 & 0 & 0 & 0 & 0 & 0\\
0 & 0 & 0 & 0 & 0 & 0 & 0 & 0 & 0 & 0 & 0 & 0 & 0 & 1\\
0 & 0 & 0 & \frac{\gamma}{\alpha + \delta} & 0 & 0 & 0 & 0 & 0 & 0 & 0 & 0 & 0 & 0\\
1 & 0 & 0 & 0 & 0 & 0 & 0 & 0 & 0 & 0 & 0 & 0 & 0 & 0\\
0 & \frac{\alpha}{\alpha + \delta} & 0 & 0 & 0 & 0 & 0 & 0 & 0 & 0 & 0 & 0 & 0 & 0
\end{array}\right),
\end{align*}
\begin{align*}
A^{(01)}=\left(\begin{array}{cccccccccccccc}
0 & 0 & 0 & 0 & 0 & 0 & 0 & 0 & \frac{\delta}{\alpha + \delta} & 0 & 0 & 0 & 0 & 0\\
0 & 0 & 0 & \frac{\gamma}{\alpha + \delta} & 0 & 0 & 0 & 0 & 0 & 0 & 1 & 0 & 0 & 0\\
1 & 0 & 0 & 0 & 0 & 0 & 0 & 0 & 0 & 0 & 0 & 0 & 0 & 0\\
0 & 0 & 0 & 0 & 0 & 0 & 0 & 1 & 0 & 0 & 0 & 0 & 0 & 0\\
0 & 0 & 0 & 0 & 0 & 0 & 0 & 0 & 0 & 0 & 0 & 0 & 0 & 0\\
0 & 0 & 0 & \frac{\gamma}{\alpha + \delta} & 0 & 0 & 0 & 0 & 0 & 0 & 0 & 0 & 0 & 0\\
0 & 0 & 1 & 0 & 1 & 0 & 0 & 0 & 0 & 0 & 0 & 0 & 0 & 0\\
0 & 0 & 0 & 0 & 0 & 0 & 0 & 0 & 0 & 0 & 0 & 1 & 0 & 0\\
0 & 0 & 0 & 0 & 0 & 0 & 0 & 0 & 0 & 0 & 0 & 0 & 0 & 0\\
0 & 0 & 0 & 0 & 0 & 0 & \frac{\beta}{\alpha + \delta} & 0 & 0 & 0 & 0 & 0 & 0 & 0\\
0 & 0 & 0 & 0 & 0 & 0 & \frac{\beta}{\alpha + \delta} & 0 & 0 & 0 & 0 & 0 & 0 & 0\\
1 & 0 & 0 & 0 & 0 & 0 & 0 & 0 & 0 & 0 & 0 & 0 & 0 & 0\\
1 & 0 & 0 & 0 & 0 & 0 & 0 & 0 & 0 & 0 & 0 & 0 & 0 & 0\\
0 & \frac{\alpha}{\alpha + \delta} & 0 & 0 & 0 & 0 & 0 & 0 & 0 & 0 & 0 & 0 & 0 & 0
\end{array}\right),
\end{align*}
\begin{align*}
A^{(10)}=\left(\begin{array}{cccccccccccccc}
0 & 0 & 0 & 0 & 0 & 0 & 0 & 0 & \frac{\delta}{\alpha + \delta} & 0 & 0 & 0 & 0 & 0\\
0 & 0 & 0 & 0 & 0 & 0 & \frac{\beta}{\alpha + \delta} & 0 & 0 & 0 & 1 & 0 & 0 & 0\\
0 & 0 & 0 & 0 & 0 & 0 & \frac{\beta}{\alpha + \delta} & 0 & 0 & 0 & 0 & 0 & 0 & 0\\
0 & 0 & 0 & 0 & 0 & 1 & 0 & 1 & 0 & 0 & 0 & 0 & 0 & 0\\
0 & 0 & 0 & 0 & 0 & 0 & 0 & 0 & 0 & 1 & 0 & 0 & 0 & 0\\
1 & 0 & 0 & 0 & 0 & 0 & 0 & 0 & 0 & 0 & 0 & 0 & 0 & 0\\
0 & 0 & 0 & 0 & 1 & 0 & 0 & 0 & 0 & 0 & 0 & 0 & 0 & 0\\
0 & 0 & 0 & 0 & 0 & 0 & 0 & 0 & 0 & 0 & 0 & 0 & 0 & 0\\
0 & 0 & 0 & 0 & 0 & 0 & 0 & 0 & 0 & 0 & 0 & 0 & 0 & 0\\
1 & 0 & 0 & 0 & 0 & 0 & 0 & 0 & 0 & 0 & 0 & 0 & 0 & 0\\
0 & 0 & 0 & \frac{\gamma}{\alpha + \delta} & 0 & 0 & 0 & 0 & 0 & 0 & 0 & 0 & 0 & 0\\
0 & 0 & 0 & \frac{\gamma}{\alpha + \delta} & 0 & 0 & 0 & 0 & 0 & 0 & 0 & 0 & 0 & 0\\
1 & 0 & 0 & 0 & 0 & 0 & 0 & 0 & 0 & 0 & 0 & 0 & 0 & 0\\
0 & \frac{\alpha}{\alpha + \delta} & 0 & 0 & 0 & 0 & 0 & 0 & 0 & 0 & 0 & 0 & 0 & 0
\end{array}\right),
\end{align*}
\begin{align*}
A^{(11)}=\left(\begin{array}{cccccccccccccc}
0 & 0 & 0 & 0 & 0 & 0 & 0 & 0 & \frac{\delta}{\alpha + \delta} & 0 & 0 & 0 & \frac{\delta}{\alpha + \delta} & \frac{\delta}{\alpha + \delta}\\
0 & 0 & 0 & 0 & 0 & 0 & 0 & 0 & 0 & 0 & 1 & 0 & 0 & 1\\
0 & 0 & 0 & 0 & 0 & 0 & \frac{\beta}{\alpha + \delta} & 0 & 0 & 0 & 0 & 0 & 0 & 0\\
0 & 0 & 0 & 0 & 0 & 1 & 0 & 1 & 0 & 0 & 0 & 0 & 0 & 0\\
0 & 0 & 0 & 0 & 0 & 0 & 0 & 0 & 0 & 0 & 0 & 0 & 0 & 0\\
0 & 0 & 0 & \frac{\gamma}{\alpha + \delta} & 0 & 0 & 0 & 0 & 0 & 0 & 0 & 0 & 0 & 0\\
0 & 0 & 1 & 0 & 1 & 0 & 0 & 0 & 0 & 0 & 0 & 0 & 0 & 0\\
0 & 0 & 0 & 0 & 0 & 0 & 0 & 0 & 0 & 0 & 0 & 0 & 0 & 0\\
0 & 0 & 0 & 0 & 0 & 0 & 0 & 0 & 0 & 0 & 0 & 0 & 0 & 0\\
1 & 0 & 0 & 0 & 0 & 0 & 0 & 0 & 0 & 0 & 0 & 0 & 0 & 0\\
1 & 0 & 0 & 0 & 0 & 0 & 0 & 0 & 0 & 0 & 0 & 0 & 0 & 0\\
1 & 0 & 0 & 0 & 0 & 0 & 0 & 0 & 0 & 0 & 0 & 0 & 0 & 0\\
1 & 0 & 0 & 0 & 0 & 0 & 0 & 0 & 0 & 0 & 0 & 0 & 0 & 0\\
0 & \frac{\alpha}{\alpha + \delta} & 0 & 0 & 0 & 0 & 0 & 0 & 0 & 0 & 0 & 0 & 0 & 0
\end{array}\right),
\end{align*}
\begin{align*}
B^{(00)} =
\left(\begin{array}{cccc}
I_2& & & \\
 &O_3& & \\
 & &O_3& \\
 & & &O_6\\
\end{array}
\right),&&
B^{(01)} =
\left(\begin{array}{cccc}
O_2& & & \\
 &I_3& & \\
 & &O_3& \\
 & & &O_6\\
\end{array}
\right),
\end{align*}
\begin{align*}
B^{(10)} =
\left(\begin{array}{cccc}
O_2& & & \\
 &O_3& & \\
 & &I_3& \\
 & & &O_6\\
\end{array}
\right),&&
B^{(11)} =
\left(\begin{array}{cccc}
O_2& & & \\
 &O_3& & \\
 & &O_3& \\
 & & &I_6\\
\end{array}
\right).
\end{align*}
$P_0$ and $P'_0$ tensors:
\begin{align*}
(P_0)_{bj,ai} &=
\begin{cases}
1 & \text{if } (b,j,a,i) \in \\
  & \{(0,0,0,0),(0,1,0,1),(0,2,0,2),(0,3,0,3),(1,0,1,0),(1,1,1,1),(1,2,1,2),(1,3,1,3),  \\
  & (2,0,2,0),(2,1,2,1),(2,2,2,2),(2,3,2,3),(3,0,3,0),(3,1,3,1),(3,2,3,2),(3,3,3,3),\\
  & (4,1,4,1),(4,3,4,3),(5,0,5,0),(5,1,5,1),(5,2,5,2),(5,3,5,3),(6,0,6,0),(6,1,6,1),\\
  & (6,2,6,2),(6,3,6,3),(7,2,7,2),(7,3,7,3),(8,3,8,3),(9,0,9,0),(9,1,9,1),(9,2,9,2),\\
  & (9,3,9,3),(10,0,9,0),(10,1,10,1),(10,2,9,2),(10,3,10,3),(11,0,9,0),(11,1,10,1),\\
  & (11,2,11,2),(11,3,11,3),(12,0,9,0),(12,1,9,1),(12,2,11,2),(12,3,12,3),(13,0,13,0),\\
  & (13,1,13,1),(13,2,13,2),(13,3,13,3)\},\\
0 & \text{otherwise},
\end{cases}
\end{align*}
\begin{align*}
(P_0')_{bj,ai} &= 
\begin{cases}
1 & \text{if } (b,j,a,i)\in\\
 & \{(0,3,0,3),(1,0,1,0),(1,1,1,1),(1,2,1,2),(1,3,1,3),(2,0,2,0),(2,1,2,1),(2,2,2,2),\\
 & (2,3,2,3),(3,2,3,2),(3,3,3,3),(4,0,1,0),(4,2,1,2),(5,0,2,0),(5,1,5,1),(5,2,5,2),\\
 & (5,3,5,3),(6,1,6,1),(6,3,6,3),(7,0,1,0),(7,1,1,1),(8,0,8,0),(9,0,9,0),(9,1,9,1),\\
 & (9,2,9,2),(9,3,9,3),(10,0,2,0),(10,0,8,0),(10,1,9,1),(10,2,10,2),(10,3,9,3),(11,0,11,0),\\
 & (11,1,2,1),(11,2,10,2),(11,3,9,3),(12,0,2,0),(12,1,2,1),(12,2,9,2),(12,3,9,3),(13,0,13,0),\\
 & (13,1,13,1),(13,2,13,2),(13,3,13,3)\},\\
-1 & \text{if } (b,j,a,i)\in\{(4,0,2,0),(4,2,2,2),(7,0,2,0),(7,1,5,1),(10,0,1,0)\},\\
-\alpha/\gamma & \text{if } (b,j,a,i)\in\{(10,0,5,2),(12,0,5,2)\},\\
\alpha/\gamma & \text{if } (b,j,a,i)\in\{(10,0,9,2),(12,0,9,2)\},\\
-\beta/\gamma & \text{if } (b,j,a,i)\in\{(11,1,5,2),(12,1,5,2)\},\\
\beta/\gamma & \text{if } (b,j,a,i)\in\{(11,1,9,2),(12,1,9,2)\},\\
0 & \text{otherwise}.
\end{cases}
\end{align*}
$P_1$ and $P'_1$ tensors:
\begin{align*}
(P_1)_{b,ai}& = 
\begin{cases}
1 & \text{if } (b,a,i)\in\\
& \{(0,0,0),(0,9,0),(1,1,0),(2,2,1),(2,4,1),(3,3,1),(4,6,1),(5,5,2),\\
& (5,7,2),(6,6,2),(7,3,2),(9,2,3),(9,4,3),(10,1,3),(11,5,3),(11,7,3),\\
& (12,8,3),(12,12,3),(13,10,3),(13,13,3)\},\\
(\alpha+\delta)/\alpha & \text{if } (b,a,i)=(1,13,0),\\
(\alpha+\delta)/\beta & \text{if } (b,a,i)=(6,9,2),\\
(\alpha+\delta)/\gamma & \text{if } (b,a,i)=(3,10,1),\\
(\alpha+\delta)/\delta & \text{if } (b,a,i)=(8,0,3),\\
-1 & \text{if } (b,a,i)\in\{(12,10,3)\},\\
0 & \text{otherwise},
\end{cases}
\end{align*}
\begin{align*}
(P_1')_{b,ai} &= 
\begin{cases}
1 & \text{if } (b,a,i)\in\{(0,1,0),(2,6,1),(5,3,2),(9,6,3),(11,3,3),(13,1,3)\},\\
-(\alpha+\delta)/\alpha & \text{if } (b,a,i)=(1,1,0),\\
(\alpha+\delta)/\alpha & \text{if } (b,a,i)\in\{(1,2,0),(1,8,0),(1,13,0),(10,13,3)\},\\
-(\alpha+\delta)/\gamma & \text{if } (b,a,i)=(1,5,2),\\
(\alpha+\delta)/\beta & \text{if } (b,a,i)\in\{(4,9,1),(6,1,2)\},\\
(\alpha+\delta)/\gamma & \text{if } (b,a,i)\in\{(1,9,2),(3,1,1),(7,10,2)\},\\
(\alpha+\delta)/\delta & \text{if } (b,a,i)\in\{(8,9,3),(12,0,3)\},\\
-1 & \text{if } (b,a,i)=(12,1,3),\\
0 & \text{otherwise}.
\end{cases}
\end{align*}
Vectors at the top boundary:
\begin{align*}
(\mathrm t|=(\mathrm t'|=\left(1,\frac{\alpha}{\alpha+\delta},0,0,0,0,0,0,\frac{\delta}{\alpha+\delta},0,\frac{\alpha}{\alpha+\delta},0,\frac{\delta}{\alpha+\delta},1\right).
\end{align*}
Bottom boundary vectors for $|\rho\rangle\rangle=(0,0,0,1)^{\mathsf T}$, $|\rho'\rangle\rangle=\frac{1}{\alpha+\delta}(\alpha,\beta,\gamma,\delta)^{\mathsf T},$:
\begin{align*}
|\mathrm b)=\left(0,1,0,0,1,0,0,1,1,0,0,0,0,0\right)^{\mathsf T},&&|\mathrm b')=\left(0,0,0,0,0,0,0,0,0,1,1,1,1,0\right)^{\mathsf T}.
\end{align*}

\end{widetext}

\section{Relations between the Rule 201 quantum cellular automaton and Trotterized PXP model}\label{app:relations}

The local unitary in the Rule 201 quantum cellular automaton and in the Trotterized PXP model at Trotter step $\tau=\pi/2$ differ by a phase factor $i$.
Here, we show that such phases do not affect the solvability.

To this end, we introduce the parameterization
\begin{align}
    U(\psi,\phi)&=\sum_{m,n = 0}^1P^{(m)}\otimes U^{(mn)}(\psi, \phi)\otimes P^{(n)},
\end{align}
where $P^{(m)}=\ket{m}\bra{m}$, $U^{(01)}=U^{(10)}=U^{(11)}=I$ and
\begin{align}
    U^{(00)}(\psi, \phi)=
    \left(\begin{array}{cc}
        0&e^{i\psi}\\e^{i\phi}&0\\
    \end{array}\right).
\end{align}
At $\psi=\phi=0$, the model reduces to the Rule 201 automaton, while at $\psi=\phi=-\pi/2$ it becomes the Trotterized PXP model.

Taking the case in Sec.~\ref{sec:rule201,12} as an example, we show that exact solutions work for arbitrary parameters $\psi$ and $\phi$. Specifically, 
generalized zipper conditions can be satisfied by deforming tensors $A$ and $B$ as
\begin{align*}
A^{(00)} =
\left(\begin{array}{cccc}
{p}_1&&&{p}_3\\
&&e^{i(\psi-\phi)}{p}_2&\\
&e^{i(\phi-\psi)}{p}_2&&\\
{p}_2&&&\\
\end{array}
\right),
\end{align*}
\begin{align*}
A^{(01)} =
\left(\begin{array}{cccc}
{p}_1&&e^{i\psi}{p}_2&\\
&{p}_1&&e^{i\psi}{p}_2\\
e^{i\phi}{p}_2&&&\\
&e^{i\phi}{p}_2&&\\
\end{array}
\right),
\end{align*}
\begin{align*}
A^{(10)} =
\left(\begin{array}{cccc}
{p}_1&e^{-i\psi}{p}_2&&\\
e^{-i\phi}{p}_2&&&\\
&&{p}_1&e^{-i\psi}{p}_2\\
&&e^{-i\psi}{p}_2&\\
\end{array}
\right),
\end{align*}
\begin{align*}
A^{(11)} =
\left(\begin{array}{cccc}
{p}_0&&&\\
&{p}_0&&\\
&&{p}_0&\\
&&&{p}_0\\
\end{array}
\right),\\
\end{align*}
\begin{align*}
B^{(00)} =
\left(\begin{array}{cccc}
I& & & \\
 &O& & \\
 & &O& \\
 & & &O\\
\end{array}
\right),
B^{(01)} =
\left(\begin{array}{cccc}
O& & & \\
 &I'& & \\
 & &O& \\
 & & &O\\
\end{array}
\right),
\end{align*}
\begin{align*}
B^{(10)} =
\left(\begin{array}{cccc}
O& & & \\
 &O& & \\
 & &(I')^\dagger& \\
 & & &O\\
\end{array}
\right),
B^{(11)} =
\left(\begin{array}{cccc}
O& & & \\
 &O& & \\
 & &O& \\
 & & &I\\
\end{array}
\right),
\end{align*}
where 
\begin{align*}
    I'=\begin{pmatrix}e^{i\phi}&&\\&e^{i\psi}&\\&&1\end{pmatrix},
\end{align*}
and all other tensors defined in Sec.~\ref{sec:rule201,12} remain unchanged.

\section{Generalized zipper conditions}\label{app:generalized_zipper}

In this appendix, we demonstrate the extension of generalized zipper conditions to quantum circuits with other architectures, in particular the brickwork quantum circuits. We also show how these conditions hold for several previously known exactly solvable influence matrices.

Formally, consider a transfer matrix in an MPO form, which exhibits shift-invariance in the bulk:
\begin{align}
    \mathfrak w^{(r,s)}\mathfrak u^{(m_{l},n_{l})}\mathfrak u^{(m_{l-1},n_{l-1})}\cdots \mathfrak u^{(m_{1},n_{1})}\mathfrak v^{(p,q)}.
\end{align}
The spatial transfer matrix associated with a quantum circuit generally admits such a representation.

Suppose that the leading eigenvector admits an MPS representation that is likewise shift-invariant in the bulk:
\begin{equation}
    \mathfrak c^{(k)}\mathfrak a^{(i_l)}\mathfrak a^{(i_{l-1})}\cdots \mathfrak a^{(i_1)}\mathfrak b^{(j)}.
\end{equation}
Now the eigenvector condition can be expressed as
\begin{align}
    &\mathfrak f^{(k)}\mathfrak d^{(i_l)}\mathfrak d^{(i_{l-1})}\cdots \mathfrak d^{(i_1)}\mathfrak e^{(j)}\notag\\
    &=\mathfrak c^{(k)}\mathfrak a^{(i_l)}\mathfrak a^{(i_{l-1})}\cdots \mathfrak a^{(i_1)}\mathfrak b^{(j)},&&\forall\,i_1,\cdots,i_l,j,k,
\end{align}
where
\begin{align}
    \mathfrak d^{(i)}&=\sum_{m}\mathfrak a^{(m)} \otimes\mathfrak u^{(m,i)},\notag\\ 
    \mathfrak e^{(i)}&=\sum_{m}\mathfrak b^{(m)} \otimes\mathfrak v^{(m,i)},\notag\\ 
    \mathfrak f^{(i)}&=\sum_{m}\mathfrak c^{(m)} \otimes\mathfrak w^{(m,i)}.
\end{align}
Here, for any $(i,m,n)$, $\mathfrak a^{(i)}$ and $\mathfrak u^{(m,n)}$ are maps on the linear space $V_a$ and $V_u$ respectively, $\mathfrak b^{(i)}$, $\mathfrak c^{(i)}$ are vectors in $V_a$, and $\mathfrak e^{(i)}$, $\mathfrak f^{(i)}$ are vectors in $V_u$.

In this setting, the generalized zipper conditions read:
\begin{align}\label{eq:gzc}
    \begin{dcases}
        \mathfrak e^{(i)}=P_0\mathfrak e^{(i)}, \\
        \mathfrak d^{(i)}P_0=P_0\mathfrak d^{(i)}P_0,\\ 
        \mathfrak f^{(i)}P_0=\mathfrak c^{(i)}P_1,\\
        P_1\mathfrak d^{(i)}P_0=\mathfrak a^{(i)}P_1,\\
        P_1\mathfrak e^{(i)}=\mathfrak b^{(i)},
    \end{dcases}
\end{align}
where $P_0:V_a\otimes V_u\to V_a\otimes V_u$ and $P_1:V_a\otimes V_u\to V_a$. The tensor contractions are represented graphically as
\begin{equation*}
.
\end{equation*}
Therefore, the generalized zipper conditions provide a sufficient condition for the eigenvector relation to hold.

For exact solutions in Rule 201 and Rule 54 quantum cellular automata, the building blocks in the MPS and MPO representations are identified as
\begin{equation*}
%
.
\end{equation}
Here, $\mathfrak b$ and $\mathfrak v$ are written in the above forms to maintain consistency with both Eq.~\eqref{eq:btm_bc} and Eq.~\eqref{eq:btm_bc_mod}. In many cases, however, they reduce to the simpler forms appearing in Eq.~\eqref{eq:btm_bc}: $\mathfrak b=|b)$, $\mathfrak v=|\rho\rangle\rangle\otimes|\rho'\rangle\rangle$.
Moreover, this bottom boundary condition can be generalized from the product initial state $\rho\otimes\rho'$ to a local MPS.

Next, we consider several exactly solvable quantum circuits with brickwork architecture. We show that the solvability of influence matrices can be recast as generalized zipper conditions in a certain form, with $P_0$ equal to the identity matrix. They can therefore be expressed in the form of Eq.~\eqref{eq:gzc}.

Specifically, dual-unitary circuits satisfy~\cite{Bertini2019Exact, Piroli2020Exact}
\begin{equation}

\end{equation*}
Second-level dual-unitary circuits satisfy~\cite{Yu2023Hierarchical}
\begin{equation}
%
\end{equation*}
Quantum circuits with exact hidden Markovian dynamics satisfy~\cite{Wang2024Exact}
\begin{equation}
%
.
\end{equation*}
In these cases, the tensors $\mathfrak c$ and $\mathfrak w$ are obtained from $\mathfrak a$ and $\mathfrak u$, respectively, by attaching trace operators at the top.
When the boundary tensors $\mathfrak b$ and $\mathfrak e$ are chosen such that Eq.~\eqref{eq:gzc} is satisfied, the corresponding quantum circuits yield exactly solvable influence matrices. This unified formulation demonstrates the broad applicability of the generalized zipper conditions.

\section{Numerical bootstrap method}\label{app:Hankel}

In this appendix, we present the bootstrap procedure for solving influence matrices. As discussed in the main text, although the generalized zipper conditions are broadly applicable, they remain difficult to construct and solve directly for a given quantum circuit. The bootstrap approach circumvents this difficulty by using finite-time numerical data to reconstruct exact solutions.

To sketch the procedure, we begin with numerically obtained influence matrices at finite time steps. We then impose the assumption that the solution extends to arbitrarily long times as an MPS with finite bond dimension. Under this assumption, we develop an algorithm to extract the local building blocks (i.e., $\mathfrak a$, $\mathfrak b$ and $\mathfrak c$ defined in Appendix~\ref{app:generalized_zipper}) from the numerical data. The construction is validated by the consistency with generalized zipper conditions.

The ingredients are numerically exact MPS representations of influence matrices at finite time steps $t$ and $t+1$, respectively shown as 
\begin{align*}
    \mathcal C^{(i_t)}\mathcal A_{t-1}^{(i_{t-1})}\mathcal A_{t-2}^{(i_{t-2})}\cdots\mathcal A_{2}^{(i_2)}\mathcal B^{(i_1)},
\end{align*}
and 
\begin{align*}
     {\mathcal C'}^{(i_{t+1})}{\mathcal A'_{t}}^{(i_{t})}{\mathcal A'_{t-1}}^{(i_{t-1})}\cdots{\mathcal A_{2}'}^{(i_2)}\mathcal {B'}^{(i_1)},
\end{align*}
where $i_{1,2,\cdots,t, t+1}$ denote the open indices. For each $i$, $\mathcal B^{(i)}, {\mathcal B'}^{(i)}$ and $\mathcal C^{(i)}, {\mathcal C'}^{(i)}$ are vectors in the auxiliary linear space, and $\mathcal A_{\tau}^{(i)}, {\mathcal A_{\tau}'}^{(i)}$ are matrices.
These MPS representations are generally non-uniform and therefore do not exhibit shift-invariance. Moreover, the local tensors in the two MPS representations are generally different.

We now impose shift-invariance in the bulk. This assumption is natural because the spatial transfer matrix is itself shift-invariant in the bulk. The leading eigenvector, namely the influence matrix, is therefore expected to approach a bulk shift-invariant form, particularly in the long-time limit. Moreover, such an MPS representation can be extended to arbitrarily long times and is well suited for determining asymptotic dynamical properties.

Motivated by these considerations, we adopt the bootstrap ansatz that the two numerically obtained MPS admit equivalent representations with a common shift-invariant bulk tensor:
\begin{align}\label{eq:hankelt}
 &\mathcal C^{(i_t)}\mathcal A_{t-1}^{(i_{t-1})}\mathcal A_{t-2}^{(i_{t-2})}\cdots\mathcal A_{3}^{(i_3)}\mathcal A_{2}^{(i_2)}\mathcal B^{(i_1)}\notag\\
 &=\mathfrak c^{(i_t)}\mathfrak a^{(i_{t-1})}\mathfrak a^{(i_{t-2})}\cdots\mathfrak a^{(i_3)}\mathfrak a^{(i_2)}\mathfrak b^{(i_1)},
\end{align}
and
\begin{align}\label{eq:hankelt1}
     &{\mathcal C'}^{(i_{t+1})}{\mathcal A'_{t}}^{(i_{t})}{\mathcal A'_{t-1}}^{(i_{t-1})}\cdots{\mathcal A_{3}'}^{(i_3)}{\mathcal A_{2}'}^{(i_2)}\mathcal {B'}^{(i_1)}\notag\\
 &=\mathfrak c^{(i_{t+1})}\mathfrak a^{(i_{t})}\mathfrak a^{(i_{t-1})}\cdots\mathfrak a^{(i_3)}\mathfrak a^{(i_2)}\mathfrak b^{(i_1)}.
\end{align}
Here, the bulk tensor $\mathfrak a^{(i)}$ is uniform in time, and tensors are the same between two time steps. We aim to reconstruct $\mathfrak a$, $\mathfrak b$ and $\mathfrak c$ from the non-uniform numerical representations. To this end, we apply the Hankel matrix method developed in the learning theory of hidden Markov models~\cite{Balle2013Spectral}. The central idea is to encode information of the many-body state into a low-dimensional Hankel matrix.

To begin, choose an intermediate time step $1<\tau<t$ at which the bond dimension reaches the maximal value $\chi$. We reshape the time-$t$ influence matrix into a matrix whose column index is the past string ($\boldsymbol p$) $i_1, i_2, \cdots, i_\tau$ and row index is the future string ($\boldsymbol f$) $i_{\tau+1}, \cdots, i_t$. Because the MPS has been compressed to the minimal form, 
this matrix has rank $\chi$. It follows that one can select $\chi$ past and future strings,
\begin{align*}
    \boldsymbol p_\alpha&=(i_{\tau}^{\alpha},i_{\tau-1}^{\alpha},\cdots,i_{1}^{\alpha}), \\
    \boldsymbol f_\beta&=(i_{t}^{\beta},i_{t-1}^{\beta},\cdots,i_{\tau+1}^{\beta}),
\end{align*} 
where $\alpha, \beta = 1,\cdots,\chi$, such that the 
corresponding $\chi$-dimensional submatrix is full-rank. This submatrix is defined as the Hankel matrix:
\begin{align}
    H_{\beta\alpha}\equiv  \mathcal C^{(i_t^\beta)}\mathcal A_{t-1}^{(i_{t-1}^\beta)}\cdots\mathcal A_{\tau+1}^{(i_{\tau+1}^\beta)}\mathcal A_{\tau}^{(i_\tau^\alpha)}\cdots\mathcal A_{2}^{(i_2^\alpha)}\mathcal B^{(i_1^\alpha)}.
\end{align}
All entries of $H$ can be evaluated directly from the numerical MPS representation, with a computational cost governed primarily by the bond dimension $\chi$ and scaling polynomially with $t$.

Similarly, using the time-$t+1$ influence matrix, we  construct a family of Hankel matrices $H^{\prime(i)}$ with an additional index $i$, from the same subsets of past and future strings:
\begin{align}
    &H_{\beta\alpha}^{\prime(i)}\notag\\  &\equiv{\mathcal C'}^{(i_t^\beta)}{\mathcal A_{t}'}^{(i_{t-1}^\beta)}\cdots{\mathcal A'}_{\tau+2}^{(i_{\tau+1}^\beta)}{\mathcal A'}^{(i)}_{\tau+1}{\mathcal A'_{\tau}}^{(i_\tau^\alpha)}\cdots{\mathcal A'_{2}}^{(i_2^\alpha)}{\mathcal B'}^{(i_1^\alpha)}.
\end{align}

Next, we express the Hankel matrices in terms of the tensors $\mathfrak a$, $\mathfrak b$ and $\mathfrak c$. For convenience, we define matrices $\mathfrak L$ and $\mathfrak R$:
\begin{align}\label{eq:hankel_lr}
    \mathfrak L_{\beta a}&\equiv \left(\mathfrak c^{(i_t^\beta)}\mathfrak a^{(i_{t-1}^\beta)}\cdots \mathfrak a^{(i_{\tau+1}^\beta)}\right)_a,\notag\\
    \mathfrak R_{a\alpha}&\equiv\left(\mathfrak a^{(i_\tau^\alpha)}\cdots\mathfrak a ^{(i_2^\alpha)}\mathfrak b^{(i_1^\alpha)}\right)_a.
\end{align}
Using Eqs.~(\ref{eq:hankelt}, \ref{eq:hankelt1}), the Hankel matrices can then be factorized as
\begin{align}
    H&=\mathfrak L\mathfrak R,&&
    H^{\prime(i)}=\mathfrak L\mathfrak a^{(i)}\mathfrak R.
\end{align}
Since $H$ is nonsingular, both $\mathfrak L$ and $\mathfrak R$ are invertible.
It therefore follows that 
\begin{equation}
     \tilde{ \mathfrak a}^{(i)} = \mathfrak R^{-1}\mathfrak a^{(i)}\mathfrak R = H^{-1}H^{\prime (i)}.
\end{equation}
Thus, the bulk tensor $\mathfrak a^{(i)}$ can be reconstructed from numerically accessible Hankel matrices up to a similarity transformation. Such a transformation is precisely an MPS gauge freedom.  Consequently, $ \tilde{ \mathfrak a}^{(i)}$ provides an equivalent representation of the bulk tensor in a gauge fixed by the chosen subsets of past and future strings. Notably, the transformation matrix $\mathfrak R$ is introduced only to establish this gauge equivalence; its explicit form is neither available nor required for reconstructing the uniform MPS.

Finally, we determine boundary tensors in the gauge fixed by $ \tilde{ \mathfrak a}^{(i)}$, which are
\begin{align}
    \tilde{\mathfrak b}^{(i)}=\mathfrak R^{-1}\mathfrak b^{(i)},&&\tilde{\mathfrak c}^{(i)}=\mathfrak c^{(i)}\mathfrak R.
\end{align}
These tensors can be solved from linear equations involving only $\tilde{\mathfrak a}^{(i)}$ and Hankel matrices:
\begin{align}
    \left(\tilde{\mathfrak a}^{(i_\tau^\alpha)}\cdots\tilde{\mathfrak a} ^{(i_2^\alpha)}\tilde{\mathfrak b}^{(i_1^\alpha)}\right)_\beta=& \left(\mathfrak R^{-1}{\mathfrak a}^{(i_\tau^\alpha)}\cdots{\mathfrak a} ^{(i_2^\alpha)}{\mathfrak b}^{(i_1^\alpha)}\right)_\beta\nonumber\\
    =&\sum_{a}(\mathfrak R^{-1})_{\beta a} \left ( {\mathfrak a}^{(i_\tau^\alpha)}\cdots{\mathfrak a} ^{(i_2^\alpha)}{\mathfrak b}^{(i_1^\alpha)}\right )_{a \alpha}\nonumber\\
    =&
    \sum_{a}(\mathfrak R^{-1})_{\beta a}\mathfrak R_{a\alpha}=\delta_{\beta\alpha}
\end{align}
and
\begin{align}
\left(\tilde{\mathfrak c}^{(i_t^\beta)}\tilde{\mathfrak a}^{(i_{t-1}^\beta)}\cdots \tilde{\mathfrak a}^{(i_{\tau+1}^\beta)}\right)_\alpha=&
    \left(\mathfrak c^{(i_t^\beta)}\mathfrak a^{(i_{t-1}^\beta)}\cdots \mathfrak a^{(i_{\tau+1}^\beta)}\mathfrak R\right)_\alpha\nonumber\\
    =&\sum_a 
    \left(\mathfrak c^{(i_t^\beta)}\mathfrak a^{(i_{t-1}^\beta)}\cdots \mathfrak a^{(i_{\tau+1}^\beta)}\right)_a\mathfrak R_{a\alpha}\nonumber\\
    =&\sum_a\mathfrak L_{\beta a}\mathfrak R_{a\alpha}=H_{\beta\alpha}.
\end{align}
As before, the matrices $\mathfrak R$ and $\mathfrak L$ are introduced only as conceptual tools in the derivation, while the explicit forms are not required.

Alternatively, the boundary tensors can be determined jointly from the bilinear equations
\begin{align*}
    H_{\beta\alpha}\equiv  {\mathfrak c}^{(i_t^\beta)}{\mathfrak a}^{(i_{t-1}^\beta)}\cdots{\mathfrak a}^{(i_{\tau+1}^\beta)}{\mathfrak a}^{(i_\tau^\alpha)}\cdots{\mathfrak a}^{(i_2^\alpha)}{\mathfrak b}^{(i_1^\alpha)},&&\forall\,\alpha,\beta,
\end{align*}
or from Eq.~\eqref{eq:hankelt}.
These equations are generally overdetermined. In all exactly solvable cases considered here, however, they admit self-consistent solutions.



As an illustrative example, we apply the bootstrap procedure to the Rule 201 quantum cellular automaton with two-site shift-invariant initial states with the block $\ket0\bra0\otimes\rho$. Because of the large bond dimension
$\chi=54$, determining a closed form analogous to those presented in Appendix~\ref{sec:solution_list} is impractical. Nevertheless, the bootstrap method allows us to obtain a numerically exact representation and to compute long-time dynamics. 

First, we calculate the influence matrix utilizing the light-cone growth algorithm (LCGA)~\cite{Lerose2023Overcoming}.
The algorithm reads:
\begin{equation*}
.
\end{equation*}
The first equation contracts the tensor network into the light cone, while the second gives the iterative update equation for the MPS for longer time steps.
Figure~\ref{fig:rule201_54} shows the growth of the entanglement entropy and the bond dimension during the update.
The bond dimension reaches $\chi=54$ at around $t=10$ and persists over the subsequent time steps.
We therefore use $t = 12$ and $t = 13$ influence matrices to perform the bootstrap procedure.


\begin{figure}[ht]
    \hspace*{-0.9\linewidth}
    {\includegraphics[width=0.9\linewidth]{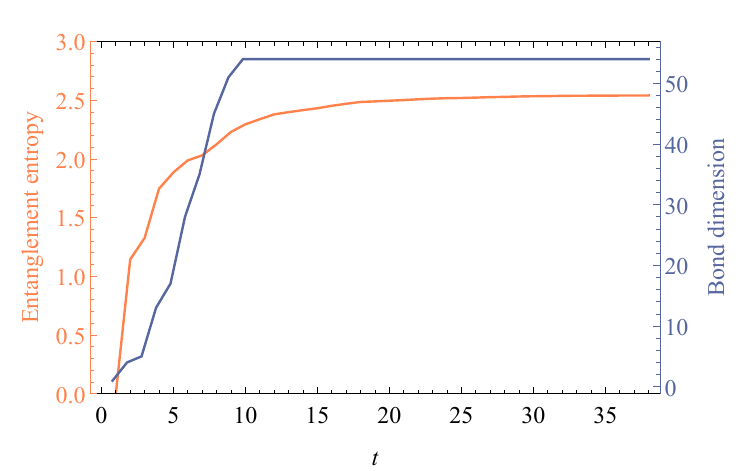}}
    \caption{The growth of entanglement entropy and bond dimension of the influence matrix in the Rule 201 quantum cellular automaton, with the initial state $\ket 0\bra0\otimes\ket+\bra+$.
    The bond dimension saturates at around $t=10$, and the entanglement entropy saturates at around $t=30$.
    }
    \label{fig:rule201_54}
\end{figure}


Here, we present one choice of the past and future string subsets for which the resulting Hankel matrix $H$ has full rank:
{
\allowdisplaybreaks[1]
\begin{align*}
\{\boldsymbol{p}_\alpha\}=\{&(6,3,1,0,15,14),&&(3,3,1,0,15,14),\\
&(8,3,1,0,15,14),&&(0,3,1,0,15,14),\\
&(9,3,1,0,15,14),&&(4,3,1,0,15,14),\\
&(12,3,1,0,15,14),&&(13,6,1,0,15,14),\\
&(2,6,1,0,15,14),&&(7,6,1,0,15,14),\\
&(8,6,1,0,15,14),&&(5,12,1,0,15,14),\\
&(0,12,1,0,15,14),&&(10,12,1,0,15,14),\\
&(15,12,1,0,15,14),&&(14,9,1,0,15,14),\\
&(1,9,1,0,15,14),&&(11,9,1,0,15,14),\\
&(4,9,1,0,15,14),&&(12,2,6,0,15,14),\\
&(8,2,6,0,15,14),&&(6,2,6,0,15,14),\\
&(2,2,6,0,15,14),&&(9,2,6,0,15,14),\\
&(3,2,6,0,15,14),&&(7,5,6,0,15,14),\\
&(8,5,6,0,15,14),&&(13,5,6,0,15,14),\\
&(2,5,6,0,15,14),&&(9,1,3,0,15,14),\\
&(12,1,3,0,15,14),&&(3,1,3,0,15,14),\\
&(6,1,3,0,15,14),&&(5,4,8,0,15,14),\\
&(2,4,8,0,15,14),&&(15,15,12,0,15,14),\\
&(0,15,12,0,15,14),&&(5,15,12,0,15,14),\\
&(10,15,12,0,15,14),&&(1,0,12,0,15,14),\\
&(14,11,9,0,15,14),&&(9,2,5,15,15,14),\\
&(3,2,5,15,15,14),&&(0,7,5,15,15,14),\\
&(7,7,5,15,15,14),&&(15,13,5,15,15,14),\\
&(11,8,5,15,15,14),&&(10,10,15,15,15,14),\\
&(4,1,10,15,15,14),&&(7,4,10,15,15,14),\\
&(15,14,11,10,10,14),&&(5,14,11,10,10,14),\\
&(15,13,4,10,10,14),&&(10,13,4,10,10,14)\},
\end{align*}
\begin{align*}
\{\boldsymbol{f}_\beta\}=\{&(13,7,4,11,11,10),&&(13,7,4,11,11,9),\\
&(13,7,4,11,11,11),&&(13,7,4,11,11,8),\\
&(13,7,4,11,9,3),&&(13,7,4,11,9,0),\\
&(13,7,4,11,9,1),&&(13,7,4,11,9,2),\\
&(13,7,4,11,8,4),&&(13,7,4,11,8,7),\\
&(13,7,4,11,8,1),&&(13,7,4,11,8,0),\\
&(13,7,4,11,10,14),&&(13,7,4,11,10,10),\\
&(13,7,4,11,10,9),&&(13,7,4,11,10,8),\\
&(13,7,4,11,10,11),&&(13,7,4,11,10,15),\\
&(13,7,4,11,10,13),&&(13,7,4,11,10,12),\\
&(13,7,4,10,12,2),&&(13,7,4,10,12,3),\\
&(13,7,4,10,12,1),&&(13,7,4,10,12,0),\\
&(13,7,4,10,15,13),&&(13,7,4,10,15,14),\\
&(13,7,4,10,14,10),&&(13,7,4,10,14,9),\\
&(13,7,4,0,8,6),&&(13,7,4,0,11,10),\\
&(13,7,4,0,11,8),&&(13,7,4,0,11,11),\\
&(13,7,4,0,15,13),&&(13,7,4,0,15,14),\\
&(13,7,4,0,15,15),&&(13,7,4,0,15,12),\\
&(13,7,4,0,10,15),&&(13,7,4,0,10,14),\\
&(13,7,4,1,10,10),&&(13,7,4,1,10,8),\\
&(13,7,4,3,1,0),&&(13,7,4,8,0,5),\\
&(13,7,4,8,0,4),&&(13,7,7,5,12,3),\\
&(13,7,7,5,14,9),&&(13,7,7,5,14,10),\\
&(13,7,7,5,13,7),&&(13,7,5,14,8,5),\\
&(13,6,1,11,10,15),&&(13,6,1,11,10,13),\\
&(13,6,1,0,13,5),&&(13,4,8,1,3,1),\\
&(13,4,8,1,3,3),&&(13,4,8,1,1,10)\}.
\end{align*}
}
Comparing the notation with that of Eq.~\eqref{eq:building_blocks}, we identify
\begin{align*}
    \mathfrak a^{(k)}=B^{(i,i')}A^{(j,j')},
\end{align*}
and we flatten the binary multi-index $(i,i',j,j')$ into a single index according to $k=8i+4i'+2j+j'$, such that $k = 0, 1, \cdots, 15$. We then reconstruct the local tensors using the bootstrap method described above. The results are used to compute the long-time dynamics shown in Fig.~\ref{fig:renyi2}.

\section{More details of local dynamics with quantum defects}\label{sec:TEBD}

\begin{figure*}[ht]
    \hspace*{-0.45\linewidth}
    {\includegraphics[width=0.47\linewidth]{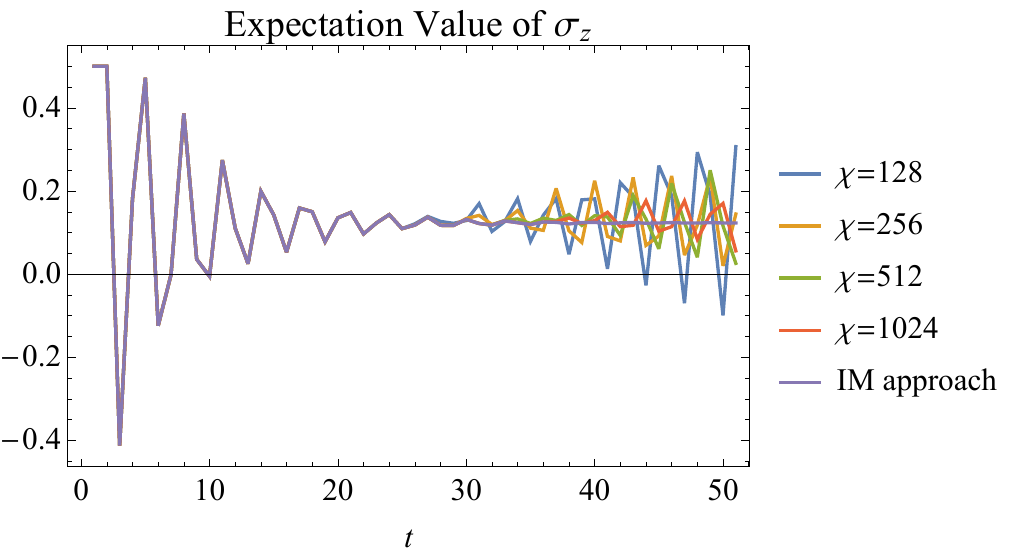}}\qquad\qquad
    {\includegraphics[width=0.42\linewidth]{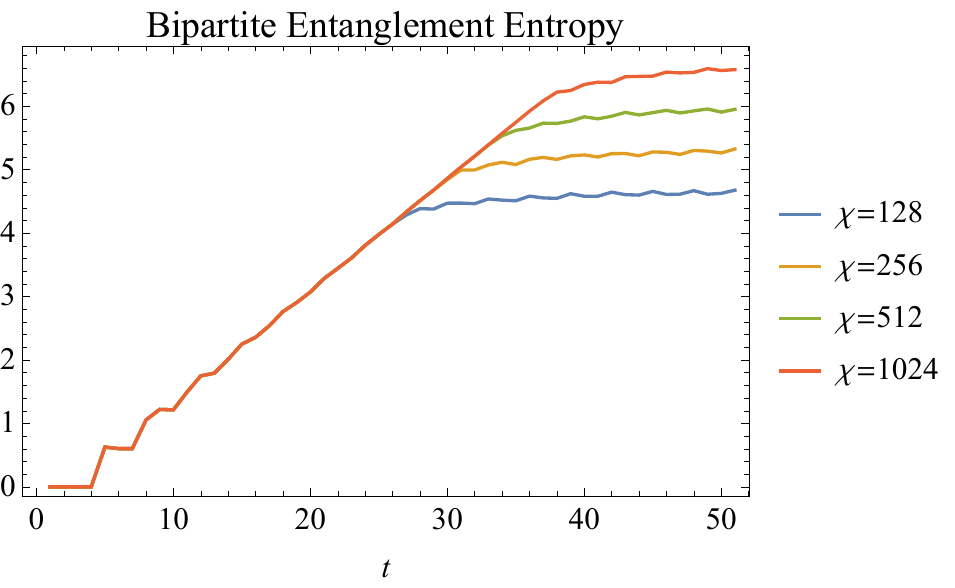}}
    \caption{
    Subsystem dynamics computed using the time-evolving block decimation method.
    Left panel: time evolution of the $\sigma_z$ expectation value on the single-site subsystem, in comparison with the exact solution obtained via the influence matrix approach. 
    Right panel: time evolution of the maximum entanglement entropy over all bipartitions in the TEBD simulation. 
    $\chi$ denotes the truncated bond dimension. 
    }
    \label{fig:tebd}
\end{figure*}

In this appendix, we demonstrate the advantages of the influence matrix approach over conventional tensor-network methods for computing subsystem dynamics.
Specifically, we compare it with the time-evolving block decimation (TEBD), which represents the full time-evolved many-body state as an MPS with a truncated bond dimension.

More specifically, we again consider the local deformation defined in Eq.~\eqref{eq:rule201-deform}:
\begin{align*}
    U^{(00)}=\left(
    \begin{array}{cc}i\sin{\epsilon} & \cos{\epsilon} \\ \cos{\epsilon} & i\sin{\epsilon}\end{array}
    \right),&&U^{(01)}=U^{(10)}=U^{(11)}=I,
\end{align*}
which is applied to the gates in the red shaded region shown in Fig.~\ref{fig:local_dynamics}. 
Within the influence-matrix approach, the dynamics of the single-site subsystem is governed by a finite-dimensional transfer matrix:
\begin{equation*}
\begin{tikzpicture}[x=0.75pt,y=0.75pt,yscale=-1,xscale=1]

\draw  [draw opacity=0][fill={rgb, 255:red, 255; green, 105; blue, 180 }  ,fill opacity=0.52 ] (20.41,8.13) -- (32.88,8.13) -- (32.88,38.13) -- (20.41,38.13) -- cycle ;
\draw [color={rgb, 255:red, 0; green, 0; blue, 0 }  ,draw opacity=1 ]   (26.64,41.46) -- (26.64,4.79) ;
\draw [color={rgb, 255:red, 0; green, 0; blue, 0 }  ,draw opacity=1 ]   (10.64,41.46) -- (10.64,4.79) ;
\draw    (10.68,30.92) -- (44.44,30.92) ;
\draw    (10.68,15.33) -- (44.44,15.33) ;
\draw   (25.86,30.92) .. controls (25.86,30.51) and (26.19,30.18) .. (26.6,30.18) .. controls (27.01,30.18) and (27.34,30.51) .. (27.34,30.92) .. controls (27.34,31.33) and (27.01,31.66) .. (26.6,31.66) .. controls (26.19,31.66) and (25.86,31.33) .. (25.86,30.92) -- cycle ;
\draw  [fill={rgb, 255:red, 126; green, 211; blue, 33 }  ,fill opacity=1 ] (22.96,15.35) .. controls (22.96,13.3) and (24.62,11.64) .. (26.67,11.64) .. controls (28.72,11.64) and (30.38,13.3) .. (30.38,15.35) .. controls (30.38,17.4) and (28.72,19.06) .. (26.67,19.06) .. controls (24.62,19.06) and (22.96,17.4) .. (22.96,15.35) -- cycle ;
\draw  [draw opacity=0] (26.7,13.5) .. controls (27.7,13.51) and (28.51,14.32) .. (28.52,15.32) -- (26.67,15.35) -- cycle ; \draw   (26.7,13.5) .. controls (27.7,13.51) and (28.51,14.32) .. (28.52,15.32) ;  

\draw  [fill={rgb, 255:red, 74; green, 144; blue, 226 }  ,fill opacity=1 ] (10.68,11.13) -- (14.88,15.33) -- (10.68,19.53) -- (6.48,15.33) -- cycle ;
\draw  [fill={rgb, 255:red, 74; green, 144; blue, 226 }  ,fill opacity=1 ] (7.71,27.95) -- (13.65,27.95) -- (13.65,33.89) -- (7.71,33.89) -- cycle ;
\draw [color={rgb, 255:red, 0; green, 0; blue, 0 }  ,draw opacity=1 ]   (42.64,41.46) -- (42.64,4.79) ;
\draw  [fill={rgb, 255:red, 74; green, 144; blue, 226 }  ,fill opacity=1 ] (42.68,11.13) -- (46.88,15.33) -- (42.68,19.53) -- (38.48,15.33) -- cycle ;
\draw  [fill={rgb, 255:red, 74; green, 144; blue, 226 }  ,fill opacity=1 ] (39.71,27.95) -- (45.65,27.95) -- (45.65,33.89) -- (39.71,33.89) -- cycle ;

\end{tikzpicture}.
\end{equation*}
The dimension is $4\times\chi^2$ where $\chi$ depends on the initial state. Here we consider $\ket{\cdots000000\cdots}$ where $\chi = 12$. Using the exact solution, we can solve the steady state of single-site reduced density matrix for finite $\epsilon$:
\begin{align*}
    \rho_{S}(t\to\infty)=\frac12(I+a_x\sigma_x+a_y\sigma_y+a_z\sigma_z),
\end{align*}
where
\begin{align*}
    \begin{dcases}
        a_x=\dfrac{8\cos(5\epsilon)-8\cos(3\epsilon)}{38-11\cos(2\epsilon)-2\cos(4\epsilon)-\cos(6\epsilon)},\\
        a_y=\dfrac{2\sin(2\epsilon)+6\sin(4\epsilon)-2\sin(6\epsilon)}{38-11\cos(2\epsilon)-2\cos(4\epsilon)-\cos(6\epsilon)},\\
        a_z=\dfrac{16-17\cos(2\epsilon)+12\cos(4\epsilon)-3\cos(6\epsilon)}{38-11\cos(2\epsilon)-2\cos(4\epsilon)-\cos(6\epsilon)}.
    \end{dcases}
\end{align*}
The limit of $\epsilon\to 0$ is consistent with perturbative analysis in Sec.~\ref{sec:relaxation}, where $a_x = a_y = 0$ and $a_z = 1/3$.
Here, we compare the results at larger $\epsilon$ to those obtained by TEBD.

In Fig.~\ref{fig:tebd}, we set $\epsilon=0.3$ and compare the numerical results with the analytical solution.
The left panel shows the expectation value of $\sigma_z$ on the single site.
For $t<26$, the TEBD results with different bond-dimension truncations agree well with the analytical solution.
At later times, however, clear truncation error can be viewed.
The exact solution approaches a stationary value, whereas the TEBD results exhibit spurious revivals of the oscillations, which become increasingly pronounced as the bond-dimension truncation is reduced.

The effects of truncation are also evident in the right panel, showing the time evolution of the maximum entanglement entropy over all bipartitions of the many-body state MPS representation.
Deviations from the expected linear growth become visible at
$t \sim 30$, 
signaling a substantial effect caused by truncations.
This example illustrates a fundamental limitation of conventional TEBD simulations of local dynamics: accurate representation of the full many-body state may require a rapidly growing bond dimension even when only a local observable is of interest.

\section{Examples of hidden Markov order}\label{app:hidden_Markov}

In this appendix, we elaborate on the hidden Markov order encoded in the $\chi=3$ and $\chi = 6$ MPS representations of the influence matrices in the Rule 54 quantum cellular automaton. The MPS representations have been presented in Appendix~\ref{sec:solution_list1}.

For convenience, we relabel the physical indices of the local tensors as $\{A^{(\alpha)}\mid\alpha=0,1,2,3\}=\{A^{(0,0)},A^{(0,1)},A^{(1,0)},A^{(1,1)}\}$ and $\{B^{(\alpha)}\mid\alpha=0,1,2,3\}=\{B^{(0,0)},B^{(0,1)},B^{(1,0)},B^{(1,1)}\}$.
In the following, we introduce a set of matrices that serve as the local building blocks of a SIMPS representation, which satisfy relations:
\begin{align}
    \begin{cases}
        B^{(\alpha)}=U^{(\alpha)}V^{(\alpha)},\\
        M^{(\gamma\beta\alpha)}=V^{(\gamma)}A^{(\beta)}U^{(\alpha)},\\
        M^{(\gamma\beta\alpha)}=U^{(\gamma\beta\alpha)}V^{(\gamma\beta\alpha)},\\
        M^{(\lambda\kappa\gamma\beta\alpha)}=V^{(\lambda\kappa\gamma)}U^{(\gamma\beta\alpha)}.\\
    \end{cases}
\end{align}

In the $\chi=3$ case, the non-zero matrices are given by
\begin{align*}
&U^{(0)} =
\left(\begin{array}{ccc}
0\\\frac{\alpha}{\alpha+\delta}\\0\\
\end{array}
\right) ,&&
U^{(3)} =
\left(\begin{array}{cc}
0&1\\0&0\\\frac{\delta}{\alpha}&0\\
\end{array}
\right),\\
&V^{(0)} =
\left(\begin{array}{ccc}
0&1&0\\
\end{array}
\right) ,&&
V^{(3)} =
\left(\begin{array}{ccc}
1&0&0\\0&0&1\\
\end{array}
\right),
\end{align*}
\begin{align*}
&U^{(000)}=U^{(003)}=U^{(013)}=U^{(023)}=U^{(033)}=1,\\
&U^{(300)}=U^{(310)}=U^{(320)}=(1,0)^T,\\
&U^{(303)}=(0,1)^T,\qquad U^{(330)}=(1,1)^T,\\
&V^{(000)}=V^{(300)}=V^{(310)}=V^{(320)}=V^{(330)}=\frac{\alpha}{\alpha+\delta},\\
&V^{(003)}=V^{(013)}=V^{(023)}=(0,1),\\
&V^{(303)}=(\delta/\alpha,0), \qquad V^{(033)}=(\delta/\alpha,1).
\end{align*}
It follows that $M^{(\lambda\kappa\gamma\beta\alpha)}$  is a scalar, therefore the corresponding SIMPS has bond dimension one.

In the $\chi=6$ case, we have
\begin{align*}
&U^{(0)} =
\left(\begin{array}{ccc}
O\\Q\\O\\
\end{array}
\right) ,&&
U^{(3)} =
\left(\begin{array}{cc}
O&I\\O&O\\S&O\\
\end{array}
\right),\\
&V^{(0)} =
\left(\begin{array}{ccc}
O&I&O\\
\end{array}
\right) ,&&
V^{(3)} =
\left(\begin{array}{ccc}
I&O&O\\O&O&I\\
\end{array}
\right),
\end{align*}
\begin{align*}
&U^{(000)}=(0,1)^T,\\
&U^{(003)}=U^{(013)}=U^{(023)}=U^{(033)}=I,\\
&U^{(300)}=U^{(310)}=U^{(320)}=(I,O)^T,\\
&U^{(303)}=(O,I)^T,\qquad U^{(330)}=(I,J)^T,\\
&V^{(000)}=(1,0),\\
&V^{(300)}=V^{(310)}=V^{(320)}=V^{(330)}=Q,\\
&V^{(003)}=V^{(013)}=V^{(023)}=(O,I),\\
&V^{(303)}=(S,O), \qquad V^{(033)}=(S,I).
\end{align*}
Here, the two-dimensional matrices $I,J,O,S,Q$ follow the same definition in Appendix~\ref{sec:rule54,6}.
It follows that the maximal dimension of $M^{(\lambda\kappa\gamma\beta\alpha)}$ is two.
Thus, a nontrivial bond degree of freedom remains after the physical indices are grouped, and the dimension cannot be reduced further.

\bibliography{Heran_circuit}

\end{document}